\documentclass[twocolumn]{aastex63}

\usepackage{natbib}
\usepackage{color}
\usepackage{mathrsfs}
\usepackage{comment}
\usepackage{longtable}

\newcommand{\HI}{\mbox{H\,{\sc i}}}

\received{June 1, 2019}
\revised{January 10, 2019}
\accepted{\today}
\submitjournal{ApJ}

\shorttitle{Molecular gas in Fornax galaxies}
\shortauthors{Morokuma-Matsui et al.}

\begin{document}

\title{CO($J$=1-0) mapping survey of 64 galaxies in the Fornax cluster with the ALMA Morita array}

\correspondingauthor{Kana Morokuma-Matsui}
\email{kanamoro@ioa.s.u-tokyo.ac.jp}

\author[0000-0003-3932-0952]{Kana Morokuma-Matsui}
\altaffiliation{JSPS Fellow}
\affiliation{Institute of Astronomy, Graduate School of Science, The University of Tokyo, 2-21-1 Osawa, Mitaka, Tokyo 181-0015, Japan}
\affiliation{Institute of Space and Astronautical Science, Japan Aerospace Exploration Agency, 3-1-1 Yoshinodai, Chuo-ku, Sagamihara, Kanagawa 252-5210, Japan}

\author[0000-0001-6163-4726]{Kenji Bekki}
\affiliation{International Centre for Radio Astronomy Research (ICRAR), M468, University of Western Australia, 35 Stirling Hwy, Crawley, WA 6009, Australia}

\author[0000-0002-6593-8820]{Jing Wang}
\affiliation{Kavli Institute for Astronomy and Astrophysics, Peking University, Beijing 100871, People's Republic of China}

\author[0000-0001-5965-252X]{Paolo Serra}
\affiliation{INAF-Osservatorio Astronomico di Cagliari, Via della Scienza 5, 09047, Selargius, Italy}

\author[0000-0002-0479-3699]{Yusei Koyama}
\affiliation{Subaru Telescope, National Astronomical Observatory of Japan, 650 North A'ohoku Place, Hilo, HI 96720, USA}

\author[0000-0001-7449-4814]{Tomoki Morokuma}
\affiliation{Planetary Exploration Research Center, Chiba Institute of Technology, 2-17-1, Tsudanuma, Narashino, Chiba 275-0016, Japan}
\affiliation{Institute of Astronomy, Graduate School of Science, The University of Tokyo, 2-21-1 Osawa, Mitaka, Tokyo 181-0015, Japan}

\author[0000-0002-1639-1515]{Fumi Egusa}
\affiliation{Institute of Astronomy, Graduate School of Science, The University of Tokyo, 2-21-1 Osawa, Mitaka, Tokyo 181-0015, Japan}

\author[0000-0002-0196-5248]{Bi-Qing For}
\affiliation{International Centre for Radio Astronomy Research, University of Western Australia, 35 Stirling Hwy, Crawley, WA 6009, Australia}
\affiliation{ARC Centre of Excellence for All Sky Astrophysics in 3 Dimensions (ASTRO 3D)}

\author[0000-0002-6939-0372]{Kouichiro Nakanishi}
\affiliation{National Astronomical Observatory of Japan, 2-21-1 Osawa, Mitaka, Tokyo 181-8588, Japan}
\affiliation{Department of Astronomy, School of Science, The Graduate University for Advanced Studies, SOKENDAI, Mitaka, Tokyo 181-8588, Japan}

\author[0000-0003-4351-993X]{B\"arbel S. Koribalski}
\affiliation{Australia Telescope National Facility, CSIRO Astronomy and Space Science, P.O. Box 76, Epping, NSW 1710, Australia}
\affiliation{School of Science, Western Sydney University, Locked Bag 1797, Penrith, NSW 2751, Australia}

\author[0000-0003-0137-2490]{Takashi Okamoto}
\affiliation{Faculty of Science, Hokkaido University, N10 W8 Kitaku, Sapporo, 060-0810, Japan}

\author[0000-0002-2993-1576]{Tadayuki Kodama}
\affiliation{Graduate School of Science, Tohoku University, 6-3 Aramaki Aza-Aoba, Sendai, Miyagi 980-8578, Japan}

\author[0000-0002-3810-1806]{Bumhyun Lee}
\affiliation{Korea Astronomy and Space Science Institute, 776 Daedeokdae-ro, Daejeon 34055, Republic of Korea}
\affiliation{Kavli Institute for Astronomy and Astrophysics, Peking University, Beijing 100871, People's Republic of China}

\author[0000-0002-9930-1844]{Filippo M. Maccagni}
\affiliation{INAF-Osservatorio Astronomico di Cagliari, Via della Scienza 5, 09047 Selargius, Italy; Netherlands Institute for Radio Astronomy, Oude Hoogeveensedijk 4, 7991 PD Dwingeloo, The Netherlands}

\author[0000-0001-8187-7856]{Rie E. Miura}
\affiliation{Departamento de F\'{i}sica Te\'{o}rica y del Cosmos, Campus de Fuentenueva, Universidad de Granada, E-18071 Granada, Spain}
\affiliation{National Astronomical Observatory of Japan, National Institutes of Natural Sciences, 2-21-1 Osawa, Mitaka, Tokyo 181-8588, Japan}

\author[0000-0002-8726-7685]{Daniel Espada}
\affiliation{Departamento de F\'{i}sica Te\'{o}rica y del Cosmos, Campus de Fuentenueva, Universidad de Granada, E-18071 Granada, Spain}

\author[0000-0001-8416-7673]{Tsutomu T.\ Takeuchi}
\affiliation{Division of Particle and Astrophysical Science, Nagoya University, Furo-cho, Chikusa-ku, Nagoya 464--8602, Japan}
\affiliation{The Research Center for Statistical Machine Learning, the Institute of Statistical Mathematics, 10-3 Midori-cho, Tachikawa, Tokyo 190---8562, Japan}

\author[0000-0002-5679-3447]{Dong Yang}
\affiliation{Kavli Institute for Astronomy and Astrophysics, Peking University, Beijing 100871, People’s Republic of China}
\affiliation{Department of Astronomy, School of Physics, Peking University, Beijing 100871, People’s Republic of China}

\author[0000-0002-2419-3068]{Minju M. Lee}
\affiliation{Cosmic Dawn Center (DAWN)} 
\affiliation{DTU-Space, Technical University of Denmark, Elektrovej 327, DK2800 Kgs. Lyngby, Denmark}

\author{Masaki Ueda}
\affiliation{Department of Physics, Tokyo University of Science, 1-3 Kagurazaka, Shinjuku-ku, Tokyo 162-8601, Japan}

\author[0000-0003-2907-0902]{Kyoko Matsushita}
\affiliation{Department of Physics, Tokyo University of Science, 1-3 Kagurazaka, Shinjuku-ku, Tokyo 162-8601, Japan}



\begin{abstract}

We conduct a $^{12}$C$^{16}$O($J$=1-0) (hereafter CO) mapping survey of 64 galaxies in the Fornax cluster using the ALMA Morita array in cycle 5. 
CO emission is detected from 23 out of the 64 galaxies.
Our sample includes dwarf, spiral and elliptical galaxies with stellar masses of $M_{\rm star}\sim10^{6.3-11.6}$~M$_\odot$.
The achieved beam size and sensitivity are $15''\times8''$ and $\sim12$~mJy~beam$^{-1}$ at the velocity resolution of $\sim10$~km~s$^{-1}$, respectively.
We study the cold-gas (molecular- and atomic-gas) properties of 38 subsamples with $M_{\rm star}>10^9$~M$_\odot$ combined with literature \HI~data.
We find that:
(1) the low star-formation (SF) activity in the Fornax galaxies is caused by the decrease in the cold-gas mass fraction with respect to stellar mass (hereafter, gas fraction) rather than the decrease of the SF efficiency from the cold gas;
(2) the atomic-gas fraction is more heavily reduced than the molecular-gas fraction of such galaxies with low SF activity.
A comparison between the cold-gas properties of the Fornax galaxies and their environmental properties suggests that the atomic gas is stripped tidally and by the ram pressure, which leads to the molecular gas depletion with an aid of the strangulation and consequently SF quenching.
Pre-processes in the group environment would also play a role in reducing cold-gas reservoirs in some Fornax galaxies.

\end{abstract}

\keywords{Galaxy environments (2029), Galaxy clusters (584), Molecular gas (1073)}


\section{Introduction} \label{sec:intro}

The cosmic star-formation rate (SFR) density decreases in the last $\sim10$~Gyr \citep[e.g.,][]{Madau:2014fk}.
Studying how the star formation in galaxies has been quenched is essential to understand galaxy evolution.
Especially, the star-formation quenching due to the galaxy environment becomes important in the current universe since the large-scale structure of the universe develops with time \citep[e.g.,][]{Peng:2010eq,Darvish:2016gj}.
Indeed, various observations find that the galaxies in dense regions are gas poor and passive \citep[e.g,][]{Dressler:1980yl,Gomez:2003hb,Balogh:2004vf,Kauffmann:2004rj,Hogg:2004oj,Baldry:2006kd}

The environmental effects on the star formation in the galaxies can be roughly classified into two categories: gravitational interactions and hydrodynamical interactions.
One of the major differences between the two categories is that gravitational interactions act in the same way on all components of a galaxy (dark matter, stars and gas) while hydrodynamic interactions only affect the gas component in the galaxy (circumgalactic medium, CGM and interstellar medium, ISM).
The former includes the galaxy-galaxy interaction \citep{Merritt:1983cv} and the galaxy-cluster interaction \citep{Byrd:1990pd}.
The multiple high-speed galaxy encounter is particularly called ``galaxy harassment'' \citep{Moore:1996mv,Moore:1999hw}, which is expected to be more important in the core than in the outskirt of the cluster.
Although the slow galaxy-galaxy encounter is rare in the cluster core, it is important in the group environment.
The hydrodynamical interactions include the ram-pressure stripping of ISM in galaxies by the intracluster medium (ICM) \citep{Gunn:1972kc} and the strangulation \citep{Larson:1980ok}.
The typical timescale of the ISM ram-pressure stripping is $\lesssim500$~Myr \citep{Abadi:1999id,Vollmer:2001ir,Roediger:2006kb}.
In the process of the strangulation, the hot coronal gas of galaxies is stripped by the ram pressure and/or tidally but the star formation continues until the remaining ISM is consumed.
The typical timescale of the strangulation is estimated to be $\sim4$~Gyr from the stellar metallicity difference between quiescent and star forming galaxies \citep{Peng:2015kx}.

It is essential to investigate the molecular gas (H$_2$) properties of cluster galaxies to understand the quenching processes in the galaxy clusters.
This is because the H$_2$ molecule is the dominant molecular species in the universe and is the raw material for star formation.
However the H$_2$ molecule does not have a dipole moment and does not emit under cold environment with temperature of $\sim10$~K where stars generally form in the local universe.
Astronomers have made use of $^{12}$C$^{16}$O($J=1-0$) (hereafter CO) as a tracer of molecular gas since it is the most abundant molecule after H$_2$ in galaxies.


The Fornax cluster is the second nearest cluster from the Milky Way at the distance of $20$~Mpc \citep{Blakeslee:2009nf}, after the Virgo cluster at $16.5$~Mpc \citep{Mei:2007wt}.
The total mass enclosed at the virial radius of $0.7$~Mpc is estimated to be $5\times10^{13}$~M$_\odot$ \citep{Drinkwater:2001so}.
From the X-ray observation of its ICM, the Fornax cluster is a cool-core cluster, in which the cooling timescale at the cluster core is less than 1~Gyr, while the Fornax cluster is considered to be in the process of forming a cool core \citep[i.e., ``nascent'' core,][]{Burns:2008dj} unlike the typical cool-core clusters that generally have a well-established core \citep{Hudson:2010uu}.
The member galaxies have been also studied in various projects in mostly optical wavelengths such as The Fornax Cluster Catalog \citep[FCC,][]{Ferguson:1989eo}, The Fornax spectroscopic survey \citep{Drinkwater:2000gq}, the ACS Fornax Cluster Survey \citep{Jordan:2007cx}, Next Generation Fornax Survey \citep{Munoz:2015us}, The Fornax Deep Survey \citep{Venhola:2018zi}, The Fornax3D project \citep{Sarzi:2018el}, and The Fornax Cluster VLT Spectroscopic Survey \citep{Pota:2018tr}.
Furthermore, they are observed in infrared \citep[The Herschel Fornax Cluster Survey,][]{Davies:2013pq}, \HI~and radio continuum \citep{Serra:2019mh,Maccagni:2020fe,Kleiner:2021bx,Loni:2021qe}.

CO emission has been also searched for in the Fornax galaxies.
\cite{Horellou:1995iq} conducted CO single/multiple pointing observations toward 21 Fornax spiral and lenticular galaxies with the 15m Swedish-ESO Submillimeter Telescope (SEST) whose beam size was $44''$ which corresponds to $4.3$~kpc at $20$~Mpc.
They detected CO emission from 11 galaxies and found that CO emission of the Fornax galaxies is $\sim10$ times lower than controlled field galaxies with similar far-infrared luminosity.
They considered that CO deficiency in the Fornax galaxies is attributed to tidal interactions rather than ram pressure stripping provided the weak X-ray emission and the high number density of galaxies in the cluster. 
Recently, the CO deficiency in the Fornax galaxies has been confirmed by the ALMA Fornax Cluster Survey (AlFoCS) project \citep{Zabel:2019ne}.
They observed 30 galaxies with stellar mass of $10^{8.5-11}$~M$_\odot$ only with the main array of the Atacama Large Millimeter/submillimeter Array (ALMA) and detected CO emission from 15 galaxies.
Among the 30 galaxies, ten galaxies were also observed with the Mopra telescope.
With the high resolution ALMA observations ($2''-3''\sim0.2-0.3$~kpc at $20$~Mpc), they also revealed that the CO-deficient galaxies with low stellar masses with $<3\times10^9$~M$_\odot$ show disturbed CO morphologies.
This suggests that the molecular gas in such less massive galaxies could be affected by the ram pressure from the ICM in the Fornax cluster.

We conduct CO mapping observations toward the 64~galaxies in the Fornax cluster using the Morita array (a.k.a., the Atacama Compact Array, ACA) of the ALMA to measure the spatial distribution and the total flux in CO of the galaxies, making it by far the largest CO survey of the Fornax galaxies so far.
With the data of one of the sample galaxies, NGC~1316 (a.k.a., Fornax~A), obtained in the survey, we have already presented the complex distribution and kinematics of molecular gas \citep{Morokuma-Matsui:2019hu}, the recurrent nuclear activity \citep{Maccagni:2020fe} and the physical properties of the multiphase gas in NGC~1316 \citep{Maccagni:2021li}.
The cold-gas properties of the Fornax-A group have been also presented in \cite{Kleiner:2021bx}.

The structure of this paper is as follows.
We briefly introduce the ALMA observations and data in section~\ref{sec:sampledata}.
The ancillary data (stellar mass, star-formation rate, and atomic gas mass) used in this study are described in section~\ref{sec:ancillary}.
With the data set, we present the galaxy-integrated cold (molecular and atomic) gas properties of the Fornax galaxies as a function of the clustocentric distance, local galaxy number density, and accretion phase to the cluster that is defined on the projected phase-space diagram (PSD) in section~\ref{sec:coldgassfproperties}.
Since the various properties of galaxies depend on their stellar masses and the mass segregation is often observed in galaxy clusters, we subtract the mass dependence of the quantities explored in this study before the investigation.
We compare our results with previous studies and discuss the possible dominant quenching process in the Fornax cluster in section~\ref{sec:discussion}.
Finally, we summarize our study in section~\ref{sec:summary}.

We adopt the following parameters of the Fornax cluster throughout the paper: the coordinates of the cluster center $({\rm R.A.}, {\rm Decl.})=(54.6^\circ, -35.5^\circ)$ , the distance of $20$~Mpc \citep{Blakeslee:2009nf}, the virial radius of $2.0$~degree ($\sim0.7$~Mpc at $20$~Mpc) \citep{Drinkwater:2001so}, the velocity dispersion of $318$~km~s$^{-1}$, and the systemic velocity of $1442$~km~s$^{-1}$ \citep{Maddox:2019mm}.

\section{ALMA observations and data} \label{sec:sampledata}

\subsection{Sample}

\begin{figure*}[]
\begin{center}
\includegraphics[width=\textwidth, bb=0 0 2095 2095]{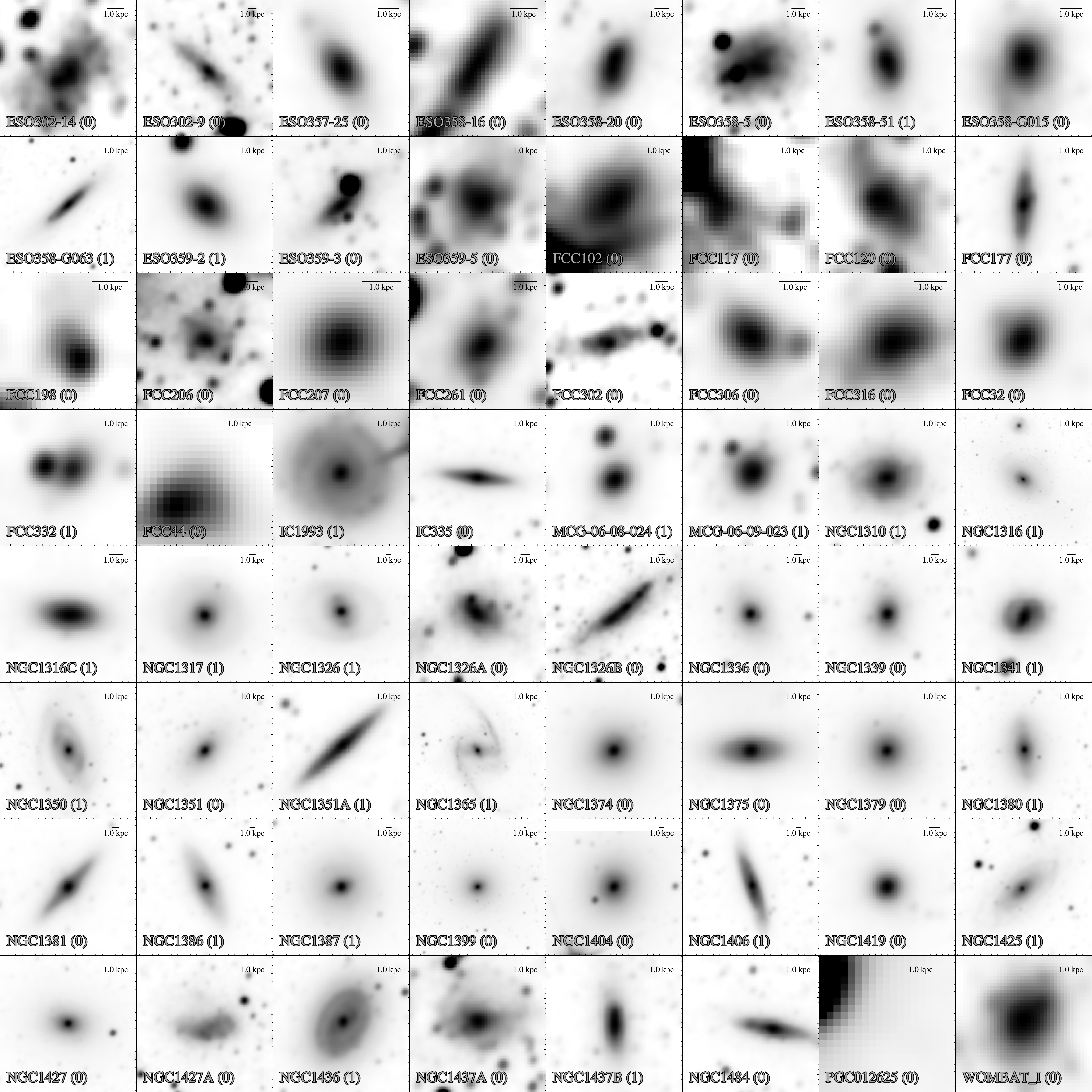}
\end{center}
\caption{
WISE $3.4~\mu$m images of our sample galaxies (North-up and East-left).
A bar on the upper right corner in each panel indicates $1$~kpc.
The integer in the brackets after the galaxy name indicates the detection (1) or non-detection (0) of the ALMA CO observation.
FCC~117 and PGC~012625 are not clearly detected in the WISE band.
}
\label{fig:wise}
\end{figure*}

\begin{figure*}[]
\begin{center}
\includegraphics[width=.49\textwidth, bb=0 0 638 599]{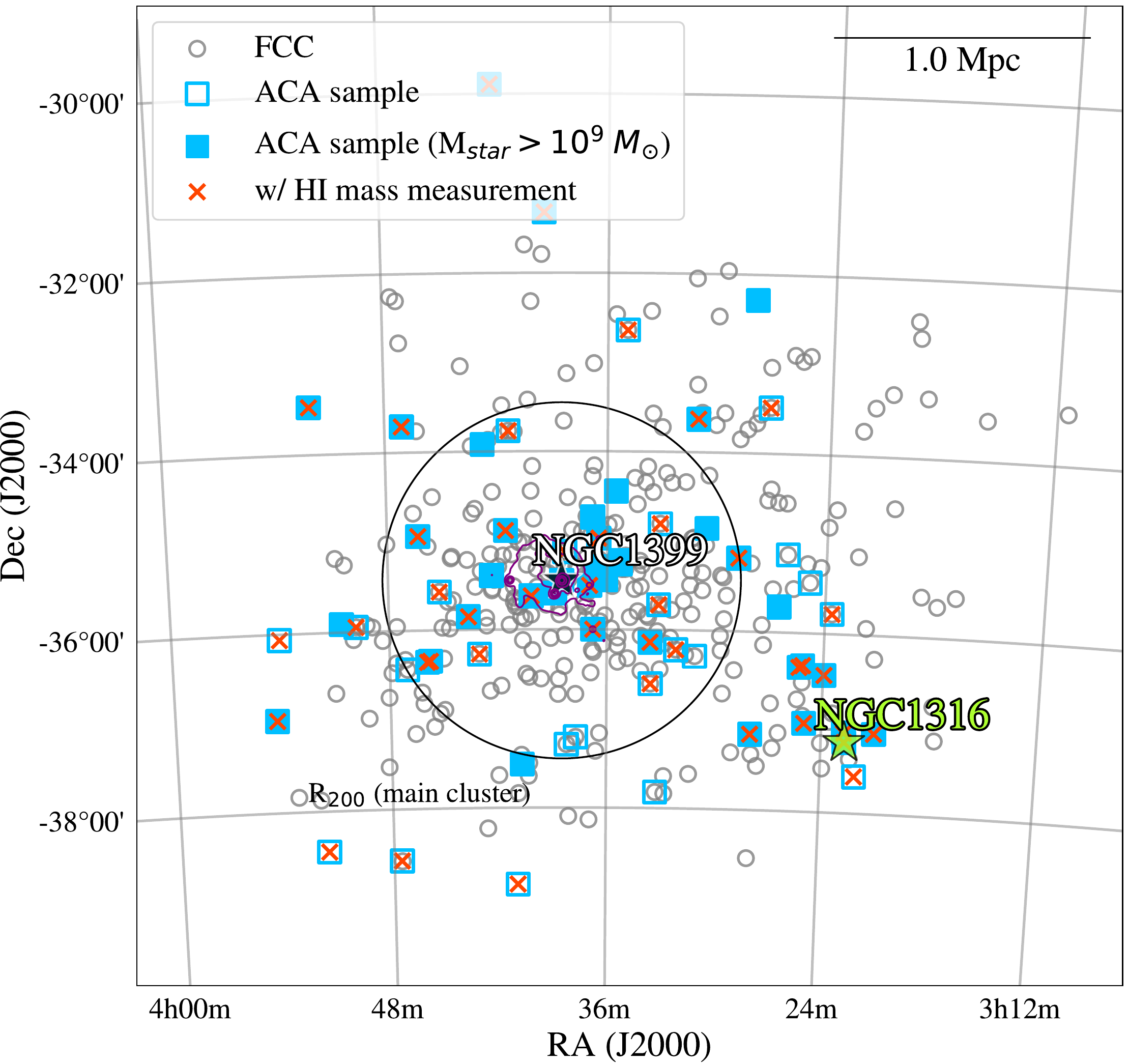}
\includegraphics[width=.48\textwidth, bb=0 0 626 606]{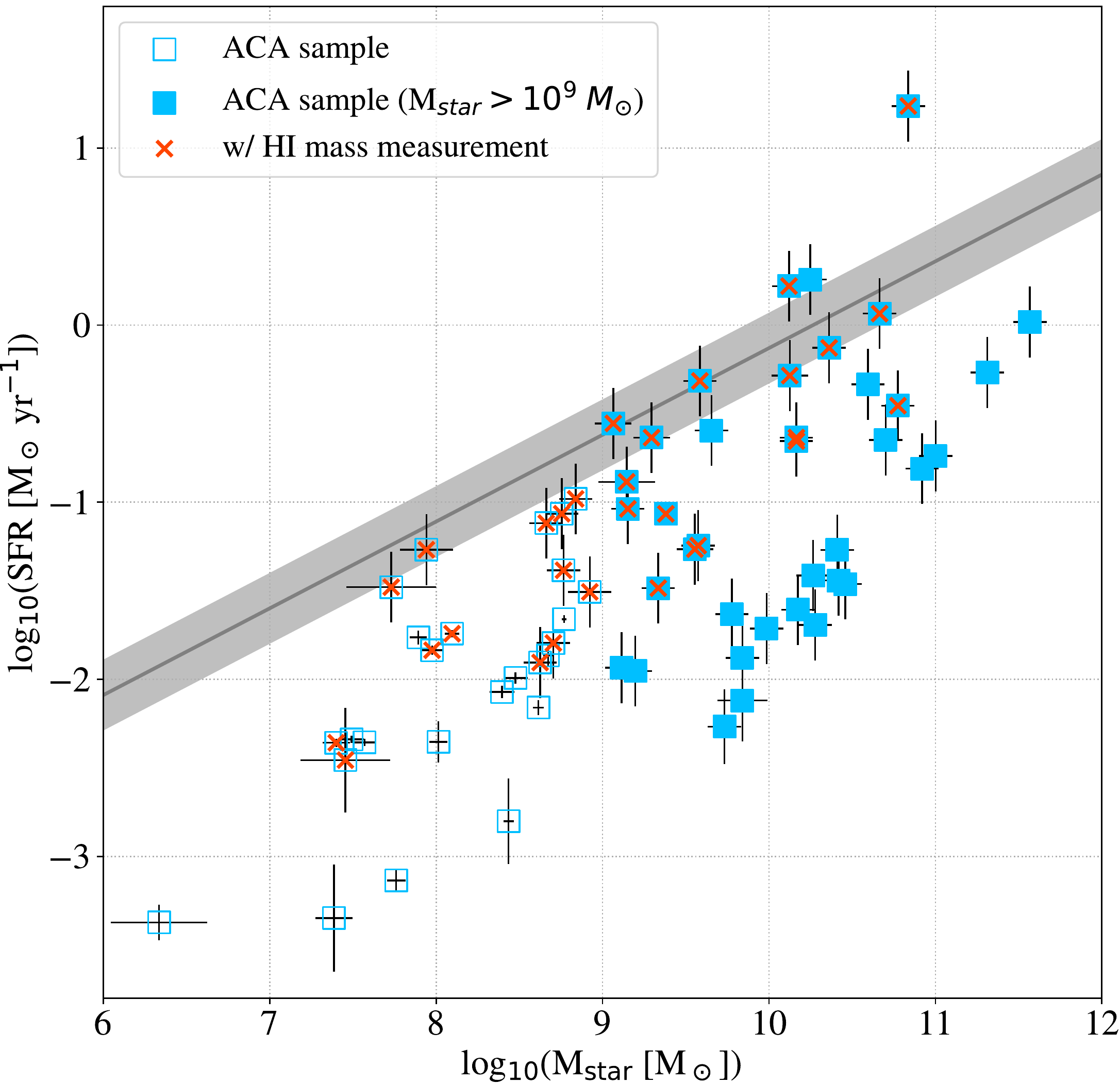}
\end{center}
\caption{
[Left] Sky distributions of Fornax cluster galaxies.
Our entire ACA sample and those with \HI~mass measurements (including upper limits) are indicated with blue open square or orange X marks, respectively.
Those with $M_{\rm star}>10^9~M_\odot$ are indicated with blue filled square.
The grey open circle indicates the FCC galaxies \citep{Ferguson:1989eo}.
The purple contours indicate the X-ray distribution obtained with the Suzaku and XMM-Newton data \citep{Murakami:2011ge}.
Our ACA sample are distributed over the region where the FCC galaxies are distributed.
[Right] Stellar mass versus SFR relation of our sample galaxies.
Grey solid line and grey shaded region indicate the main sequence of star-forming galaxies defined in \cite{Speagle:2014by} and its standard deviation, respectively.
}
\label{fig:radec}
\end{figure*}

Our 64 sample galaxies meet either of the following conditions:
(1) galaxies observed in the AlFoCS project \citep{Zabel:2019ne,Zabel:2020az};
(2) galaxies in the vicinity of the Fornax cluster with archival \HI~data in the HyperLEDA \citep{Makarov:2014wh}.
The galaxies in the condition (1) are selected based on the FCC \citep{Ferguson:1989xz} with stellar masses of $>3\times10^8$~M$_\odot$, and Herschel detection \citep{Fuller:2014ec} or \HI~detection down to $\sim3\times10^7$~M$_\odot$ \citep{Waugh:2002rr}.
The FCC number, coordinates, recessional kinematic LSR velocity, and morphology of the sample galaxies are presented in Table~\ref{tab:basic}.
The $3.4~\mu$m images of our sample galaxies obtained with the Wide-field Infrared Survey Explorer \citep[WISE,][]{Wright:2010oi} are shown in Figure~\ref{fig:wise}.
The sky distribution of our sample galaxies in addition to the entire FCC galaxies is presented in the left panel of Figure~\ref{fig:radec}.
We can see that our sample galaxies are evenly distributed over the region where the FCC galaxies are distributed.
The relationship between stellar mass and SFR is presented in the right panel of Figure~\ref{fig:radec} (see section~\ref{sec:mstarsfr} for the derivation of stellar mass and SFR).
Most sample galaxies are on or below the main-sequence of star-forming galaxies.

\subsection{ALMA 7m and total power observations}

\begin{table*}
\begin{center}
\caption{Observations for the 7M data \label{tab:sbs}}
\begin{tabular}{lccccc}
\tableline\tableline
ID & SB name & Obs. date in 2017 & \# of EBs & Flux/Bandpass cal. & Complex gain cal.\\
\tableline
1 & NGC1316\_a\_03\_7M & 16 Oct.--21 Dec. & $33$ & J0006-0623, J0522-3627 & J0334-4008, J0403-3605, J0424-3756\\
2 & ESO358-G\_a\_03\_7M & 30 Nov.--28 Dec. & $33$ & J0006-0623, J0522-3627 & J0334-4008, J0403-3605, J0406-3826\\
\tableline
\end{tabular}
\end{center}
\end{table*}

ALMA CO observations with the 7m (7M) and 12m total-power (TP) antennas arrays of the 64 Fornax galaxies were performed during the cycle 5 (project code of 2017.1.00129.S and P.I. of Kana Morokuma).
The mapping area for the galaxies is basically determined so that the optical disk of galaxies is covered.

The 7M data were obtained from October 16 to December 21 2017 for the scheduling block (SB) of NGC1316\_a\_03\_7M and from 30 November to 28 December 2017 for the SB of ESO358-G\_a\_03\_7M with 9-11 antennas whose baselines range $8.9-48.9$~m.
The 64 galaxies were divided into the two SBs, and each SB consists of 33 execution blocks (EBs).
The maximum recoverable scales are $40''.7-58''.3$ ($\sim3.9-5.7$~kpc at 20~Mpc) for NGC1316\_a\_03\_7M and $55''.2$ ($\sim5.4$~kpc) for ESO358-G\_a\_03\_7M, respectively.
The flux, bandpass, and complex gain calibrators used in the two SBs are presented in Table~\ref{tab:sbs}.
The systemic velocity for each source was fixed to 1500~km~s$^{-1}$ to cover the entire sample galaxies with a single correlator setup.
The CO line was observed in one of the upper-sideband (USB) spectral windows (SPWs) with bandwidth and resolution of 1875~MHz (4902.5~km~s$^{-1}$) and 1.128~MHz (2.952~km~s$^{-1}$), respectively.

The on-the-fly observations with the TP array were carried out to account for zero spacing information between March to September 2017 with three or four antennas.
There are 64 SBs and each SB consists of one to 13 EBs (13 EBs for NGC~1365).
The mapping area was set to cover the entire area observed with the 7M array.
The correlator and SPW setup was the same as for the 7M observations.

\subsection{ALMA data analysis with CASA}

The obtained ALMA data was calibrated and imaged with the standard ALMA data analysis package, the Common Astronomy Software Applications \citep[CASA,][]{McMullin:2007fu,Petry:2012tx}.
The absolute flux and gain fluctuations of the 7M data were calibrated with the ALMA Science Pipeline (version r40896 of Pipeline-CASA51-P2-B) in the CASA 5.1.1 package.
The resultant fluxes of the phase calibrator at each SPW are mostly consistent within the errors with the values reported in the other ALMA measurements of the same calibrator on the closest date to our observations.
The 7M CO mosaic data cube for each source was generated with the {\tt TCLEAN} task in CASA version 5.5 with standard options as follows: Briggs weighting with a {\tt robust} parameter of 0.5, the {\tt auto-multithresh} mask with standard value for 7M data provided in the CASA guides for the {\tt automasking}, and {\tt niter of} 10000.
The TP data was also reduced with the pipeline version r40896 of Pipeline-CASA51-P2-B in the CASA 5.1.1 package.
The 7M and TP data were combined with the {\tt FEATHER} task in CASA.
The synthesized beam is roughly $15''\times8''$ and the values for each galaxy are presented in Table~\ref{tab:alma}.
The velocity width of the final fits cube is set to $\sim10$~km~s$^{-1}$.
The achieved sensitivity is 12~mJy~beam$^{-1}$.

\subsection{Molecular gas mass}

\begin{figure*}[]
\begin{center}
\includegraphics[width=0.49\textwidth, bb=0 0 515 489]{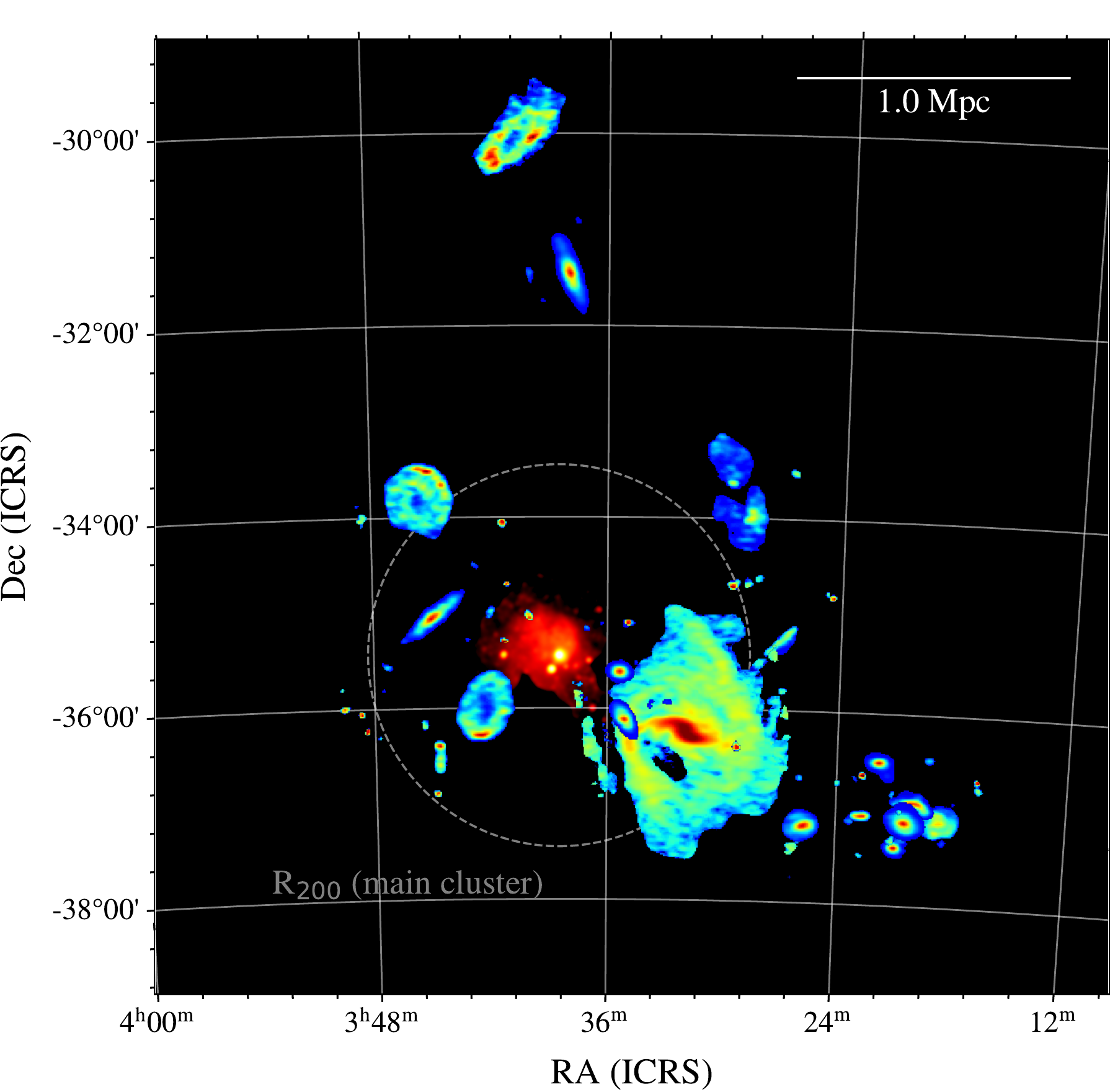}
\includegraphics[width=0.49\textwidth, bb=0 0 648 609]{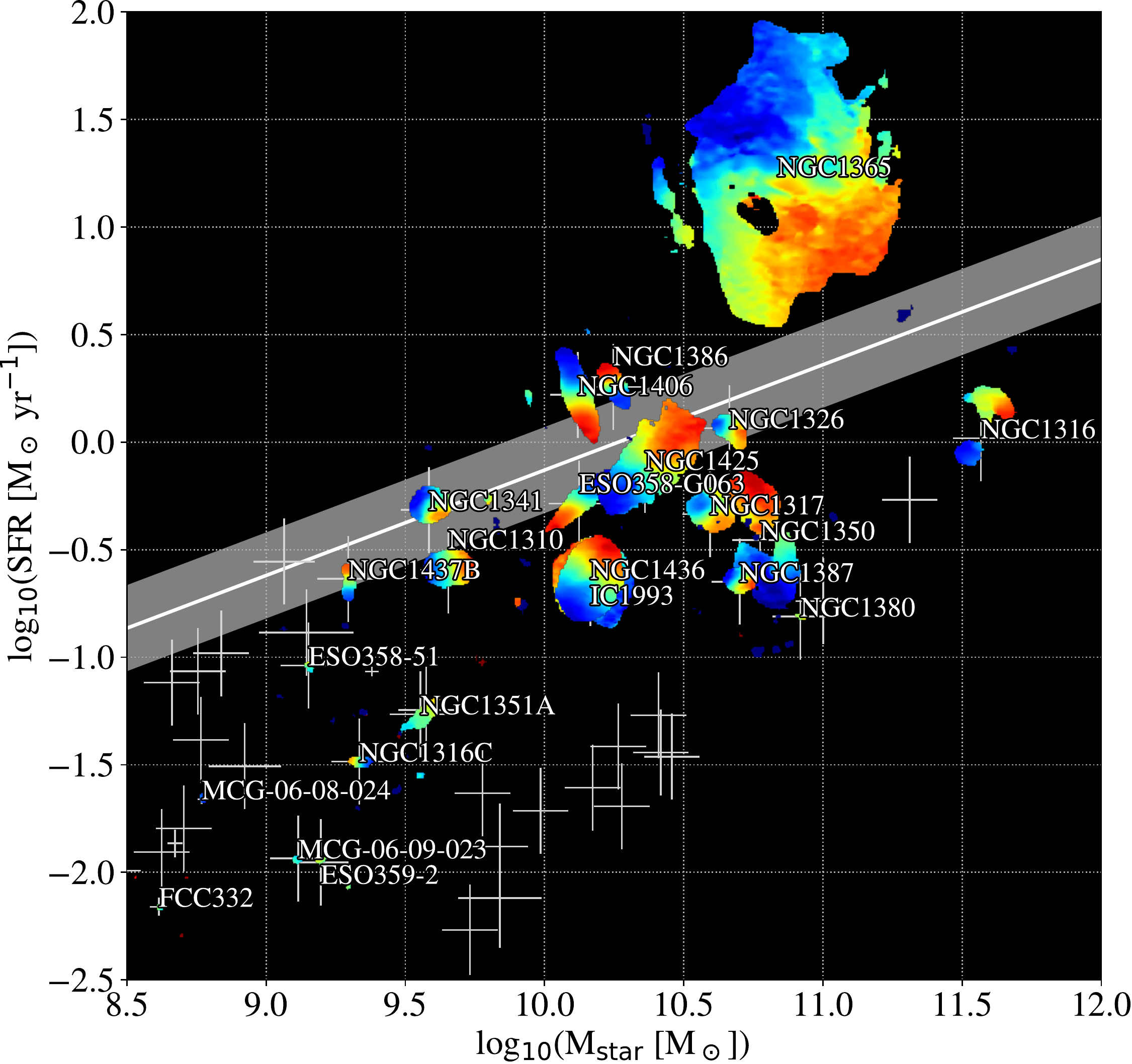}
\end{center}
\caption{
[Left] Sky distributions of 23 Fornax cluster galaxies with CO detections.
The spatial size of CO data (moment 0) is enlarged by a factor of 30.
The linear scaling is adopted for plotting the moment-0 map of the CO-detected galaxies except for NGC~1365 for which the logarithmic scaling is adopted.
Background is the X-ray image obtained by combining the Suzaku and XMM-Newton data for the Fornax cluster \citep{Murakami:2011ge}.
[Right] Stellar mass versus SFR relation of our sample galaxies.
Moment-1 maps of the CO detected galaxies are shown.
White solid line and shaded region indicate the SFMS of \cite{Speagle:2014by} and its standard deviation, respectively.
}
\label{fig:radec_co}
\end{figure*}

CO emission is detected in 23 galaxies out of the 64 sample galaxies.
Through inspecting the output masks, we adjust and decide an optimized set of configuration parameters of the Source Finding Application \citep[SoFiA,][]{Serra:2015mx,Westmeier:2021az}. 
We use the smooth and clip algorithm, which smoothes the cube with gaussian kernels of different sizes, select voxels with a 5-$\sigma$ threshold in each smoothed cube as well as in the original cube, and then use the subset of voxels selected from any of the smoothed cubes or from the original cube. 
The kernels have full width half maximums of 8 and 16 pixels (1 and 2$\times$ the major axis of the synthesis beam) in the R.A. and Dec. directions, and 3 pixels (30 km s$^{-1}$) in the velocity direction. 
We used the reliability module, having verified that we have enough statistics to apply that method.
The SoFiA reliability of 0.95 is required to rule out false detections. 
Then the detection mask is dilated with a maximum growing size of 10  pixels and 3 channels. 
By this stage, the targets have been separated into detections and non-detections. 
For non-detections, we further run the {\it optical finder} of SoFiA, which searches fluxes around input optical positions. 
We combine the final signal to noise ratio and moment maps to decide whether the detections are reliable. 
The final products for each source from SoFiA include a detection mask, moment maps and integral measurements of the fluxes.

For CO detected galaxies, their CO intensities are calculated with galactic total CO spectra which are generated by averaging the spectra of the pixels enclosed by rectangles.
Here, we use the original data to generate the total CO spectra since the spectra generated using the SoFiA masks do not show the level of rms.
The rectangle area to generate the total CO spectrum of each galaxy and the velocity range to calculate integrated intensities are determined by eye to cover the entire area where CO emission is detected.
In case of non-detections, we calculate the upper limit of their CO integrated intensities with galactic total CO spectra. The galactic total CO spectrum of a CO non-detected galaxy is generated with an ellipse whose major and minor axes are two times smaller than the ones used to calculate the total fluxes in the infrared and ultraviolet wavelengths (see section~\ref{sec:mstarsfr}).
For the CO line profile, we assume boxcar function with the 30 velocity channels ($\sim300$~km~s$^{-1}$).
In Figure~\ref{fig:radec_co}, we present the CO moment maps of the CO detected galaxies on the X-ray image \citep{Murakami:2011ge} and the stellar mass-SFR plane.
We can see that CO-detected galaxies tend to avoid being distributed near the core of the Fornax cluster and are biased to relatively massive galaxies with stellar mass of $M_{\rm star}>10^9$~M$_\odot$.
Moment-1 maps of most samples indicate rotation of molecular gas, whereas some also show non-circular motion (NGC~1365 in \citealp{Gao:2021vi}; NGC~1316 in \citealp{Morokuma-Matsui:2019hu}).
The CO moment maps of the 23 CO-detected galaxies are presented in Figures~\ref{fig:mommaps_ESO358-51}-\ref{fig:mommaps_NGC1437B} in the Appendix section.
Their individual CO distribution and kinematics will be discussed in a separated paper.

With the CO integrated intensities obtained above, we calculate molecular gas masses ($M_{\rm mol}$) in the Fornax galaxies using a constant CO-to-H$_2$ conversion factor of the Milky Way of $\alpha_{\rm CO, MW}=4.36$~M$_\odot$ (K~km~s$^{-1}$~pc$^2$)$^{-1}$ \citep{Bolatto:2013vn} as we did in our previous study on the Virgo cluster galaxies \citep{Morokuma-Matsui:2021to}.
This is because we do not have homogeneous estimation of the metallicity of our sample.
Furthermore, we focus on the relatively massive systems with $M_{\rm star}>10^9$~M$_\odot$ in the analysis in the following sections.
Note that $M_{\rm star}\sim10^9$~M$_\odot$ is near the turnover mass of the local mass-metallicity relation of galaxies \citep{Andrews:2013nx} and the uncertainty of the metallicity dependence of the CO-to-H$_2$ conversion factor is expected to be small for the massive galaxies.
The Helium contribution to mass is already included in $\alpha_{\rm CO,MW}$ by a factor of 1.36.

\begin{figure}[]
\begin{center}
\includegraphics[width=.5\textwidth, bb=0 0 512 493]{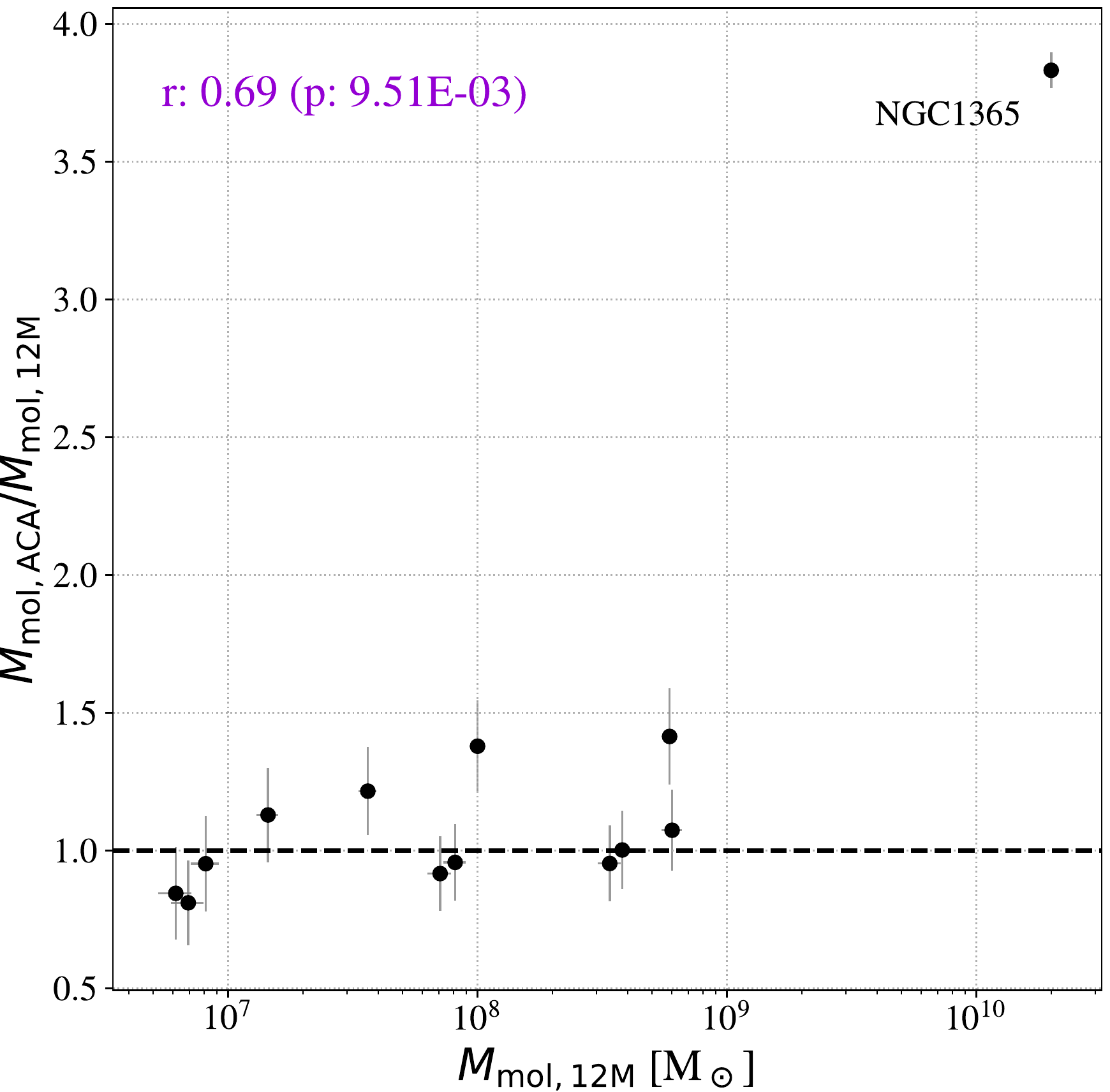}
\end{center}
\caption{
$M_{\rm mol}$ comparison between \cite{Zabel:2019ne} ($M_{\rm mol, 12M}$) and this study ($M_{\rm mol, ACA}$).
The Spearman's correlation coefficient and its $p$-value is presented at the upper left corner in purple.
The galaxies with a large $M_{\rm mol}$ value tend to have a high $M_{\rm mol, ACA}/M_{\rm mol, 12M}$ ratio, suggesting the missing flux in the 12-m data.
}
\label{fig:h2_comp}
\end{figure}

For a comparison with the AlFoCS project \citep{Zabel:2019ne,Zabel:2020az}, we find that molecular gas masses for the CO-detected objects in both the AlFoCS and this studies (13 objects) are mostly consistent within a factor of $\sim1.4$ except for NGC~1365 whose $M_{\rm mol}$ estimation in this study is $3.7$ times larger than the one in the AlFoCS.
Note that the molecular gas mass in the AlFoCS data is recalculated with the constant CO-to-H$_2$ conversion factor of $\alpha_{\rm CO, MW}=4.36$~M$_\odot$ (K~km~s$^{-1}$~pc$^2$)$^{-1}$ \citep{Bolatto:2013vn} to match our data.
If we compare the moment-0 maps of NGC~1365 with a large difference in the $M_{\rm mol}$ estimation between this study and the AlFoCS project, there would be missing flux in the 12-m array data, confirming the importance of the ACA data for calculating total fluxes of the extended nearby objects.
Note that the maximum recoverable scales of the AlFoCS project are $20''.8-23''.3$ ($\sim2.0-2.3$~kpc at 20~Mpc).
The AlFoCS detected CO emission from FCC~207 and FCC~261, whereas we could not since $M_{\rm mol}$ values of these galaxies are below the sensitivity of our observation.
There is a weak $M_{\rm mol}$ dependence in the $M_{\rm mol}$-estimation difference of the CO-detected objects where our estimation tends to be lower/higher for less-massive/massive objects compared to the AlFoCS estimation (Figure~\ref{fig:h2_comp}), although the dependence becomes insignificant if NGC~1365 is excluded ($r=0.55$ and $p=0.07$).
For all the non-detected sources in this study, our upper limits are larger (mostly five times larger) than those estimated in the AlFoCS project, i.e., a more conservative estimation in this study.
This is partially due to the difference in the sensitivity and the assumed line width to calculate the upper limit of the CO intensity.
We assumed $300$~km~s$^{-1}$ for all the CO-non-detections including dwarf and early-type galaxies whereas the AlFoCS assumed $50$~km~s$^{-1}$ since their CO-non-detections are all dwarf galaxies.

\section{Ancillary data}\label{sec:ancillary}

\subsection{Stellar mass and star-formation rate}\label{sec:mstarsfr}

\begin{figure*}[]
\begin{center}
\includegraphics[width=\textwidth, bb=0 0 955 493]{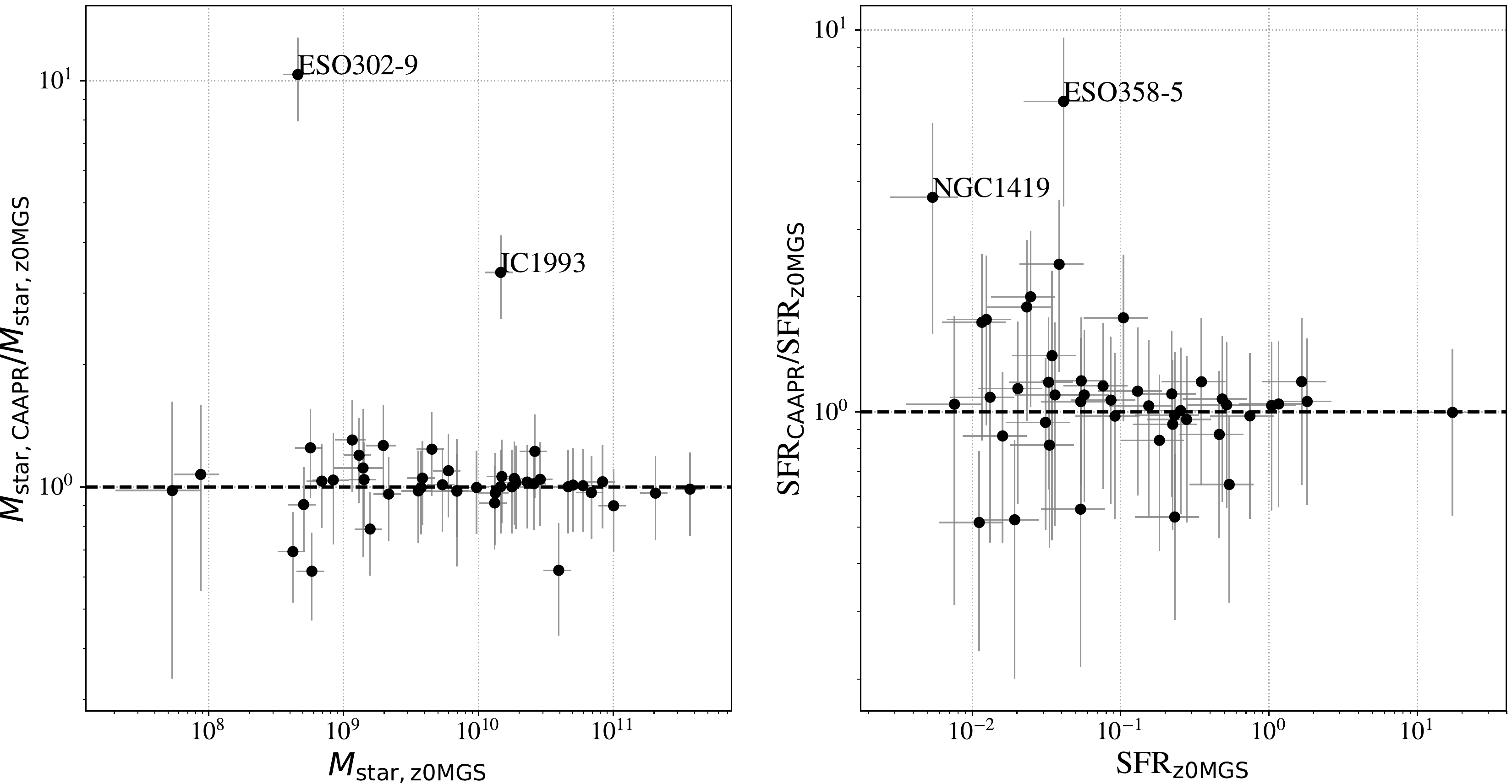}
\end{center}
\caption{
Comparison of $M_{\rm star}$ and SFR estimations in \cite{Leroy:2019cu} ($M_{\rm star, z0MGS}$ and SFR$_{\rm z0MGS}$) and the ones in this study ($M_{\rm star, CAAPR}$ and SFR$_{\rm CAAPR}$).
}
\label{fig:caapr}
\end{figure*}

We adopted $M_{\rm star}$ and SFR values derived in the project, $z=0$ Multiwavelength Galaxy Synthesis \citep[z0MGS][]{Leroy:2019cu}.
\cite{Leroy:2019cu} estimated $M_{\rm star}$ using the WISE $3.4~\mu$m data and SFR using the WISE $22~\mu$m data and far-ultraviolet data obtained with the Galaxy Evolution Explorer \citep[GALEX,][]{Martin:2005wd} to account for both obscured and unobscured components.
$M_{\rm star}$ and SFR values of 46 out of 64 galaxies are found in the z0MGS catalogue.
For the rest of the 18 galaxies, we adopt the $M_{\rm star}$ and SFR values estimated by ourselves as follows.

We first measure the total fluxes in the WISE and GALEX bands of the 64 Fornax galaxies using the Comprehensive \& Adaptable Aperture Photometry Routine \citep[CAAPR\footnote{\url{https://github.com/Stargrazer82301/CAAPR}},][]{Clark:2015zu,Clark:2018tc}.
The CAAPR can remove the foreground and background objects with the Python Toolkit for SKIRT \citep[PTS,][]{Camps:2015yd} and also remove the any large-scale background structure if desired.
The aperture ellipse in each wavelength is determined to cover the pixels where the signal-to-noise ratio $S/N>2$.
The $S/N=2$ ellipse is then enlarged by a factor of 1.25, which is a default expansion factor in the CAAPR.
The final aperture area is the smallest ellipse which contain all the ellipses in different wavelengths.
The detailed description on the CAAPR is found in \cite{Clark:2018tc}.
For MCG-06-08-024 and PGC~012625, we manually determined the photometry aperture to avoid including nearby stars.
The parameters for the final ellipses obtained with the CAAPR for our sample are presented in Table~\ref{tab:caapr}.
With the total fluxes of galaxies in the WISE and GALEX bands, the $M_{\rm star}$ and SFR values are calculated in the same manner adopted in the z0MGS.
The relationship between $M_{\rm star}$ and SFR of our sample galaxies is presented in the right panel of Figure~\ref{fig:radec}.

To assess our estimations of $M_{\rm star}$ and SFR, Figure~\ref{fig:caapr} compares $M_{\rm star}$ and SFR values estimated in this study and in the z0MGS for the 46 Fornax galaxies with the z0MGS data.
Overall, our estimations of $M_{\rm star}$ and SFR are consistent with those of the z0MGS project within the errors.
We find that the $M_{\rm star}$ estimations for two galaxies (ESO~302-9 and IC~1993) and the SFR estimations for two galaxies (NGC~1419 and ESO~358-5) are different from those in the z0MGS.
Given that we adopted the same calibration methods of the z0MGS, the difference in the $M_{\rm star}$ and SFR values is mainly caused by the differences in the photometry processes such as the determination of the photometry/background areas and the subtraction of the foreground and background objects.
Note that the $M_{\rm star}$ and SFR values of the massive galaxies analyzed in section~\ref{sec:coldgassfproperties} are mostly from the z0MGS (37 out of 38 galaxies), thus the effects to obtained results due to the difference in the photometry processes are expected to be negligible.

\subsection{Atomic gas mass}\label{sec:atom}

We make use of the atomic-gas mass ($M_{\rm atom}$) of the Fornax galaxies from \cite{Loni:2021qe} and \cite{Kleiner:2021bx}, which conducted \HI~observations with the Australia Telescope Compact Array (ATCA) and the Meer Karoo Array Telescope (MeerKAT).
For the galaxies whose \HI~mass are not listed in these studies, we utilize the \HI~fluxes that are mostly measured in the \HI~Parkes All Sky Survey (HIPASS) in the HyperLEDA catalogue.
\HI~data were found for 42 out of the 64 Fornax galaxies: 20 main-cluster galaxies from \cite{Loni:2021qe}; seven Fornax-A-group galaxies from \cite{Kleiner:2021bx}; 15 galaxies from the Hyper-LEDA.
$M_{\rm atom}$ is calculated based on the optically thin assumption as
\begin{equation}
M_{\rm atom} = 1.36 \times 2.36 \times 10^5 D^2 \times \sum_{i} S_i \Delta v_i,
\end{equation}
where the factor of 1.36 accounts for the Helium contribution to mass, $D$ is the luminosity distance to the galaxies in Mpc (20~Mpc) and $\Sigma_i S_i \Delta v_i$ is the total integrated flux in Jy~km~s$^{-1}$.

\section{Cold-gas and star-formation properties}\label{sec:coldgassfproperties}

\begin{table}
\begin{center}
\caption{Numbers of the 64 Fornax galaxies for the subsamples.\label{tab:subsample}}
\begin{tabular}{lcc}
\tableline
\tableline
Subsample & w/ data & w/ detection\\
\tableline
H$_2$ & 64 & 23\\
\HI  & 42 & 36 \\
H$_2$ \& \HI & 42 & 16 \\
\tableline
H$_{2, M_{\rm star9}}$$^{\rm a}$  & 38 & 21\\
\HI$_{,M_{\rm star9}}$ & 23 & 20 \\
H$_{2, M_{\rm star9}}$ \& \HI$_{,M_{\rm star9}}$ & 23$^{\rm b}$ & 16$^{\rm c}$\\
\tableline
\end{tabular}
\end{center}
\vspace{-4mm}
\tablecomments{
$^{\rm a}$ Galaxies with $M_{\rm star}>10^9$~M$_\odot$, i.e., with $\Delta$(Field) measurements,\\
$^{\rm b}$ {\it CO+\HI-obs. sample},\\
$^{\rm c}$ {\it CO+\HI-det. sample}.
}
\end{table}

In the following sections, we basically follow the analysis procedure adopted in \cite{Morokuma-Matsui:2021to} for the Virgo cluster galaxies.
The comparison among different local clusters including the Fornax and Virgo will be presented in a separated paper (Morokuma-Matsui et al.~in prep.).
We present cold-gas and star-formation properties of the Fornax galaxies focusing on the following key quantities (Table~\ref{tab:key}):
(1) SFR,
(2) specific SFR (${\rm sSFR}={\rm SFR}/M_{\rm star}$),
(3) molecular-gas mass to atomic-gas mass ratio ($R_{\rm H_2}=M_{\rm mol}/M_{\rm atom}$),
(4) total-gas mass to stellar mass ratio ($\mu_{\rm gas}=M_{\rm gas}/M_{\rm star}=[M_{\rm mol}+M_{\rm atom}]/M_{\rm star}$),
(5) atomic-gas mass to stellar mass ratio ($\mu_{\rm HI}=M_{\rm atom}/M_{\rm star}$),
(6) molecular-gas mass to stellar mass ratio ($\mu_{\rm H_2}=M_{\rm mol}/M_{\rm star}$),
(7) star-formation efficiency from total gas (${\rm SFE_{gas}}={\rm SFR}/M_{\rm gas}$),
(8) star-formation efficiency from atomic gas (${\rm SFE_{HI}}={\rm SFR}/M_{\rm HI}$), and
(9) star-formation efficiency from molecular gas (${\rm SFE_{H_2}}={\rm SFR}/M_{\rm mol}$).

\begin{table*}
\begin{center}
\caption{Key quantities.\label{tab:key}}
\begin{tabular}{lll}
\tableline
\tableline
Quantity & Unit & Description\\
\tableline
SFR & [M$_\odot$~yr$^{-1}$] & stat-formation rate\\
sSFR & [yr$^{-1}$] & ${\rm SFR}/M_{\rm star}$, specific SFR\\
$R_{\rm H_2}$ &  & $M_{\rm mol}/M_{\rm atom}$, molecular-gas mass to atomic-gas mass ratio\\
$\mu_{\rm gas}$ & & $M_{\rm gas}/M_{\rm star}=[M_{\rm mol}+M_{\rm atom}]/M_{\rm star}$, total-gas mass to stellar mass ratio\\
$\mu_{\rm HI}$ & & $M_{\rm atom}/M_{\rm star}$, atomic-gas mass to stellar mass ratio\\
$\mu_{\rm H_2}$ & & $M_{\rm mol}/M_{\rm star}$, molecular-gas mass to stellar mass ratio\\
${\rm SFE_{gas}}$ & [yr$^{-1}$] & ${\rm SFR}/M_{\rm gas}$, star-formation efficiency from total gas\\
${\rm SFE_{HI}}$ & [yr$^{-1}$] & ${\rm SFR}/M_{\rm atom}$, star-formation efficiency from atomic gas\\
${\rm SFE_{H_2}}$ & [yr$^{-1}$] & ${\rm SFR}/M_{\rm mol}$, star-formation efficiency from molecular gas\\
\tableline
\end{tabular}
\end{center}
\end{table*}

We consider the property distributions of two samples to be different if the $p$-value of the Kolmogorov-Smirnov (KS) test is smaller than 0.05.
We also assess the correlation between two quantities by Spearman's rank-order correlation method.
We consider that the correlation coefficient, $r$, is reliable if the $p$-value is smaller than 0.05
\footnote{
We use the term ``significant correlation'' if its $p$-value is smaller than 0.05 and the term ``stronger/weaker'' even for a moderate correlation with $r\sim0.5$ just in a relative sense, i.e., in a comparison of two $r$ values.}.
Here, we make use of {\tt ks\_2samp} and {\tt spearmanr} in {\tt Scipy} for the KS test and the estimation of the correlation coefficient and $p$-value, respectively.
The key quantities of the Fornax galaxies are compared on the best effort basis, i.e., all the galaxies with measurements (including upper limits) of the numerators and denominators of the key quantities (``{\it best-effort sample}'').
However, for $R_{\rm H_2}$, we limit the sample to the galaxies with detection in both CO and \HI.
Depending on the quantities to be compared, the number of galaxies in the following plots is different.
The galaxies with the \HI~measurement are mostly \HI~detected and biased to star-forming main sequence (SFMS) galaxies (Figure~\ref{fig:radec} right).
Therefore, the quantities with \HI~data, such as $R_{\rm H_2}$, $\mu_{\rm HI}$, $\mu_{\rm gas}$, ${\rm SFE_{gas}}$, and ${\rm SFE_{HI}}$, may not represent the majority of our sample galaxies.
To assess the sample bias caused by this treatment, the key quantities are also compared for galaxies with all the measurements (``{\it CO+\HI-obs. sample}'') or detections of \HI~and CO (``{\it CO+\HI-det. sample}'') in Section~\ref{sec:results_subsamples} although the sample size becomes small.
The summary of the subsamples are presented in Table~\ref{tab:subsample}.

We first compare the Fornax galaxies and the field galaxies in Section~\ref{sec:field}, and then investigate the dependence of the key quantities on three environmental parameters: clustocentric distance in Section~\ref{sec:distance}, the distance to the 5th nearest galaxy in Section~\ref{sec:density}, and the accretion phase in the cluster in Section~\ref{sec:accretion}.

\subsection{Comparison with field galaxies}\label{sec:field}

\begin{figure*}[]
\begin{center}
\includegraphics[width=\textwidth, bb=0 0 1412 1384]{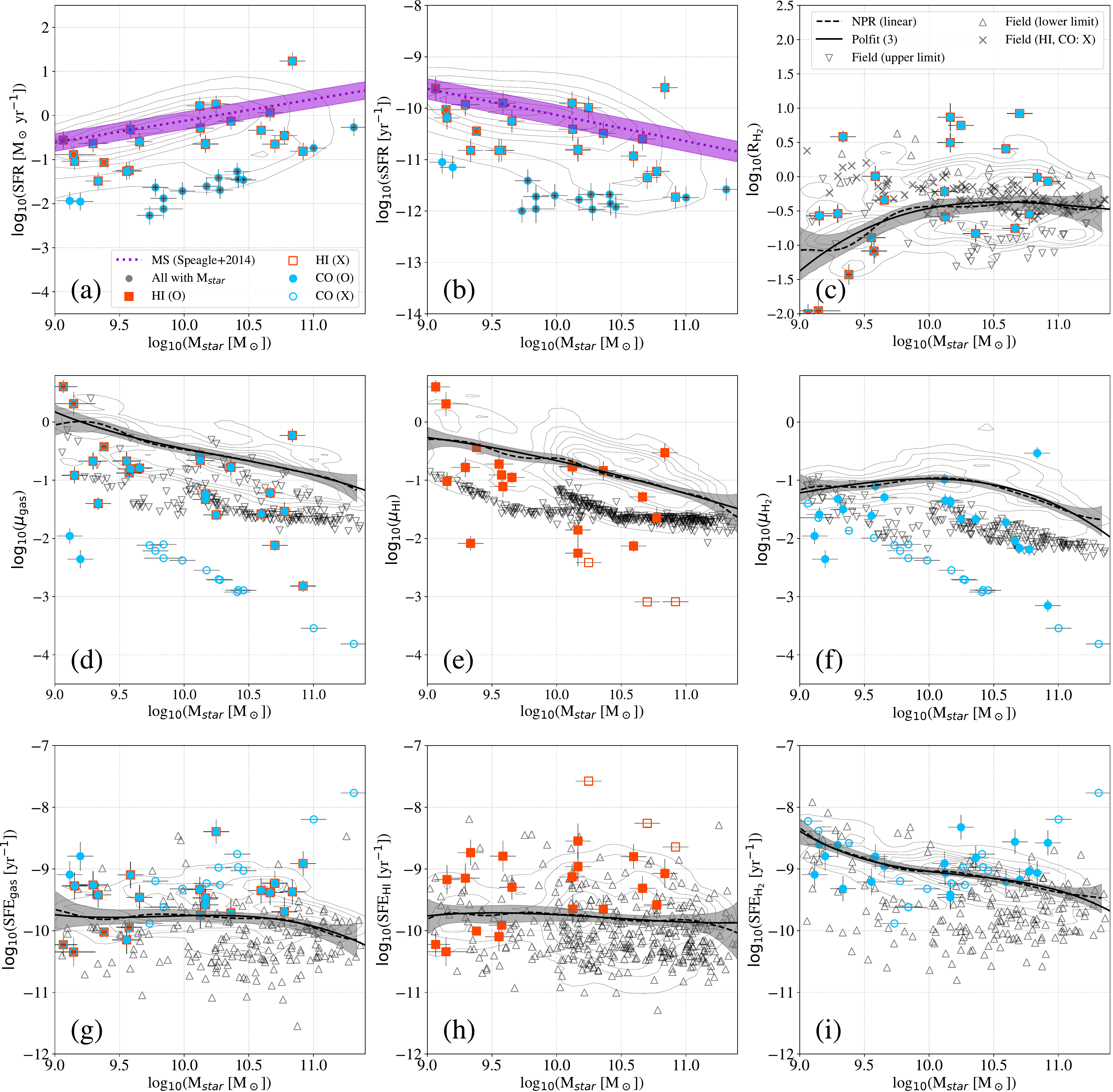}
\end{center}
\caption{
Comparison between the Fornax galaxies and field galaxies from the z0MGS \citep{Leroy:2019cu}, xGASS \citep{Catinella:2018ib} and xCOLDGASS \citep{Saintonge:2017ve} projects: $M_{\rm star}$ dependences of
(a) SFR,
(b) sSFR,
(c) $R_{\rm H_2}=M_{\rm H_2}/M_{\rm HI}$,
(d) $\mu_{\rm gas}=M_{\rm gas}/M_{\rm star}$,
(e) $\mu_{\rm HI}=M_{\rm HI}/M_{\rm star}$,
(f) $\mu_{\rm H_2}=M_{\rm H_2}/M_{\rm star}$,
(g) SFE$_{\rm gas}=$SFR/$M_{\rm gas}$,
(h) SFE$_{\rm HI}=$SFR/$M_{\rm HI}$,
(i) SFE$_{\rm H_2}=$SFR/$M_{\rm H_2}$.
The main-sequence of star-forming galaxies defined in \cite{Speagle:2014by} is indicated as purple dotted lines and the 0.2 dex scatter range is indicated as purple shading in the panels (a) and (b).
The field relations are estimated with CO- or \HI-detected galaxies, i.e., star-forming galaxies, and indicated as dashed (with non-parametric fitting) and (with 3rd-order polynomial fitting) solid lines in the panels (c)-(i).
Grey-shaded regions are the errors of the non-parametric fitting and estimated by the bootstrap method with a 95~\% confidence interval.
The Fornax galaxies with CO/\HI~detection and non-detection are indicated as blue filled circles/orange filled squares and blue open circles/orange open squares, respectively.
The contours indicate field star-forming galaxies used to derive the field relations.
The lower/upper limits of the field galaxies are indicated as open triangles/inverted triangles.
In the panels (d) and (g), $M_{\rm H_2}/M_{\rm star}$ and SFR/$M_{\rm H_2}$ are plotted for the galaxies without \HI~data, respectively.
For the $R_{\rm H_2}$ panel, the field galaxies with \HI~and CO upper limits are indicated as X marks.
}
\label{fig:comp_field}
\end{figure*}

\begin{figure*}[]
\begin{center}
\includegraphics[width=.8\textwidth, bb=0 0 1398 1384]{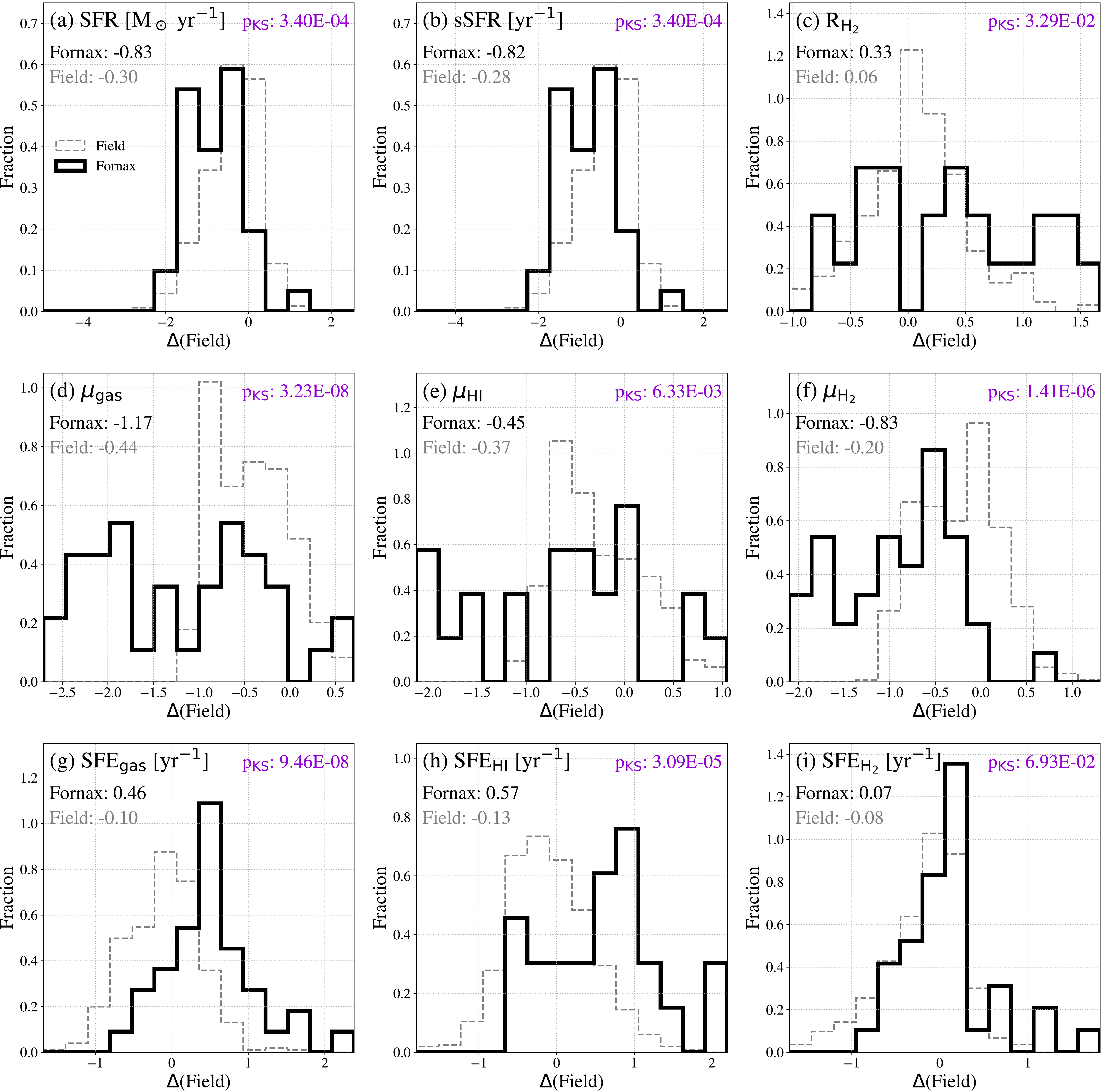}
\end{center}
\caption{
Comparison between the Fornax galaxies (solid line) and field galaxies (dashed line) with $M_{\rm star}>10^9$~M$_\odot$: histograms for the offsets from the field relations, $\Delta({\rm Field})$.
Note that $\Delta({\rm Field})$s of SFR and sSFR are calculated based on the SFMS definition in \cite{Speagle:2014by}.
The medians for the Fornax and field galaxies are indicated at the upper left corner.
The $p$-value for the KS test comparing the Fornax and field galaxies is indicated at the upper right corner.
Our Fornax sample has lower SFR, sSFR, gas fractions and higher $R_{\rm H_2}$, SFE$_{\rm gas}$, SFE$_{\rm HI}$ for their masses.
SFE$_{\rm H_2}$ of our Fornax sample is not significantly different from the one of the field galaxies. 
}
\label{fig:comp_field2}
\end{figure*}

To clarify the environmental effects on cluster galaxies independently of the mass dependent galaxy evolution, we first determine the field relations, i.e., stellar mass dependence of the key quantities of the field {\it star-forming} galaxies (hereafter ``field SFG relations''), by making use of the data of the extended GALEX Arecibo SDSS Survey \citep[xGASS, 1,179 galaxies,][]{Catinella:2018ib} for \HI-related relations, and the extended CO Legacy Database for GASS \citep[xCOLD GASS, 532 galaxies,][]{Saintonge:2017ve} for H$_2$-related relations.
The data of xGASS and xCOLD GASS projects are used because their samples are selected only by stellar mass and redshift.
For consistency with our Fornax data, the molecular gas mass of the xCOLD-GASS galaxies is recalculated with the constant $\alpha_{\rm CO, MW}$ of 4.36~M$_\odot$ (K~km~s$^{-1}$ pc$^{2}$)$^{-1}$, and the atomic gas mass of the xGASS galaxies is scaled with $1.36$ to account for the helium.
The field SFG relations are derived non-parametrically and by third-order polynomial fitting using the data of CO-detected xCOLD GASS galaxies and/or \HI-detected xGASS galaxies.
For the SFR-related relations, we adopted the SFMS derived based on the orthogonal distance regression using data of $>10^6$ galaxies at $0<z<6$ in \cite{Speagle:2014by}.

The comparison of the key quantities among the Fornax sample, the field (xGASS and xCOLD GASS) sample including both SF and passive galaxies, and the field SFG relations is shown in Figure~\ref{fig:comp_field}.
The contours in the panels (a) and (b) indicate the entire field galaxies, and the contours in the rest of the panels indicate the field galaxies with \HI~or CO detection, i.e., star-forming galaxies.
The open triangles/inverse-triangles indicate the field galaxies with lower/upper limits of the quantities.
We can see that most of our Fornax galaxies (orange and blue symbols) are on or below the SFMS.
There seems to be a tendency where all the gas fractions of the Fornax galaxies are lower than those of the field galaxies even when we limit to the galaxies with \HI~or CO detections.
Although the $R_{\rm H_2}$ values of the Fornax galaxies seem to be similar to those of the field galaxies, some Fornax galaxies have higher $R_{\rm H_2}$.
In terms of the SFEs, most Fornax galaxies seem to have higher SFE$_{\rm gas}$ and SFE$_{\rm HI}$ than the field galaxies, especially for the massive galaxies with $>10^{10}$~M$_\odot$.
Although a similar mass dependence is observed for SFE$_{\rm H_2}$, the difference between the Fornax and field galaxies seems insignificant compared to SFE$_{\rm gas}$ and SFE$_{\rm HI}$.

For both the Fornax and field samples, the deviation from the field SFG relations is calculated by 
\begin{equation}
\Delta({\rm Field})=\log_{10}{\rm (X_{\rm Fornax/Field}(M_{\rm star})/X_{\rm Field, SFG}(M_{\rm star}))}.
\end{equation}
$X_{\rm Fornax/Field}(M_{\rm star})$ is a key quantity of the Fornax or field galaxies with a stellar mass of $M_{\rm star}$.
$X_{\rm Field, SFG}(M_{\rm star})$ is the field SFG relation that is derived based on the third-order polynomial regression (except for SFR and sSFR where Speagle's SFMS is used).
Figure~\ref{fig:comp_field2} shows the histograms of key quantities in the form of $\Delta({\rm Field})$ of the Fornax and field galaxies.
According to the KS test, the tendency seen in Figure~\ref{fig:comp_field} is confirmed quantitatively.
In the Fornax cluster a larger fraction of galaxies tend to have lower SF activity, a smaller amount of atomic and molecular gas, and a higher SFE from the atomic or the total cold (atomic plus molecular) gas, compared to the field galaxies with the same stellar mass.
Furthermore, the Fornax galaxies have a wider range of $R_{\rm H_2}$ and have a higher median value of $R_{\rm H_2}$ than the field galaxies.
Although the upper/lower limits of the key quantities for the field galaxies are looser than those of the Fornax galaxies (Figure~\ref{fig:comp_field}), we confirmed that the obtained result does not change even when limiting to the galaxies with \HI~or CO detection.

\begin{figure*}[]
\begin{center}
\includegraphics[width=\textwidth, bb=0 0 1877 943]{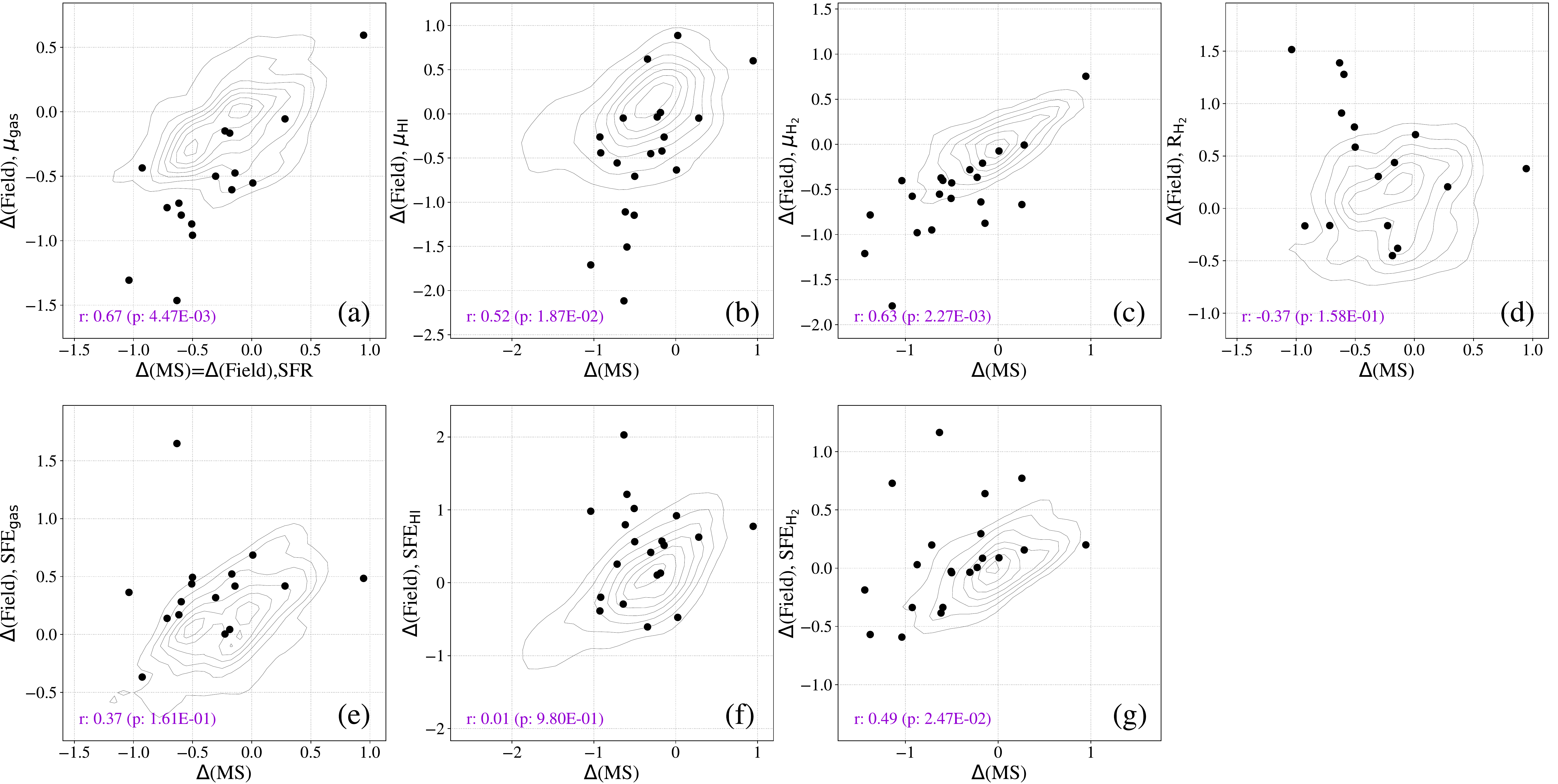}
\end{center}
\caption{
Comparison between the Fornax galaxies and field galaxies with $M_{\rm star}>10^9$~M$_\odot$: relationships between the offsets from the main sequence of star-forming galaxies [$\Delta({\rm MS})=\Delta({\rm Field})_{\rm SFR}$] with $\Delta({\rm Field})$ values of gas fractions, $R_{\rm H_2}$, and SFEs.
The contours indicate field galaxies with CO/\HI~detection and filled black circles indicate the Fornax galaxies with CO/\HI~detection.
The Spearman's rank-order correlation coefficient $r$ and the $p$-value are shown on the lower left corner of each panel in purple.
Our Fornax sample has lower gas fractions and higher $R_{\rm H_2}$ and SFEs than the field galaxies at $\Delta({\rm MS})\sim-1$.
}
\label{fig:comp_field3}
\end{figure*}

Both the $\mu_{\rm H_2}$ and SFE$_{\rm H_2}$ values of galaxies have been claimed to positively correlate with the offset from the SFMS of the field galaxies,
\begin{equation}
\Delta({\rm MS})=\log_{10}{\rm (SFR/SFR_{MS})}
\end{equation}
\citep[e.g.,][]{Saintonge:2012nj,Genzel:2015gn,Scoville:2017jw,Ellison:2020kf}, where ${\rm SFR_{MS}}$ is SFR of the SFMS of the field galaxies for a fixed stellar mass and redshift, i.e., $\Delta({\rm MS})=\Delta({\rm Field})_{\rm SFR}$.
These relations are claimed to be independent of the galaxy environment at least for field and group galaxies \citep{Koyama:2017lf}.
For denser environments, \cite{Morokuma-Matsui:2021to} found that the Virgo galaxies with low $\Delta$(MS) values offset to lower $\mu_{\rm H_2}$ and higher SFE$_{\rm H_2}$ regimes with respect to the field and group relations.

Figure~\ref{fig:comp_field3} shows the distribution of the Fornax and field galaxies with $M_{\rm star}>10^9$~M$_\odot$ on the plane of $\Delta$(MS) versus $\Delta({\rm Field})$ values of the cold-gas mass fractions, $R_{\rm H_2}$ or SFEs.
The comparison to the original values (before the mass dependence is subtracted) is presented in Figure~\ref{fig:comp_field4} in the Appendix~\ref{sec:supplementaryplots}.
Here we limit the galaxies to those with \HI~or CO detections, this is because the observation sensitivity, i.e., the strength of the upper/lower limits is different between our ALMA observation and the ones for the field galaxies.
The Fornax galaxies below the main sequence [$-1.5<\Delta$(MS)$<-0.5$] tend to have smaller gas fractions and larger $R_{\rm H_2}$ ratio, SFE$_{\rm gas}$, and SFE$_{\rm HI}$ compared to field galaxies according to the KS test.
$\mu_{\rm H_2}$ and $\mu_{\rm gas}$ of the Fornax galaxies is lower than those of the field galaxies even on the main sequence [$-0.5<\Delta$(MS)$<0.5$].

\begin{figure*}[]
\begin{center}
\includegraphics[width=.8\textwidth, bb=0 0 958 495]{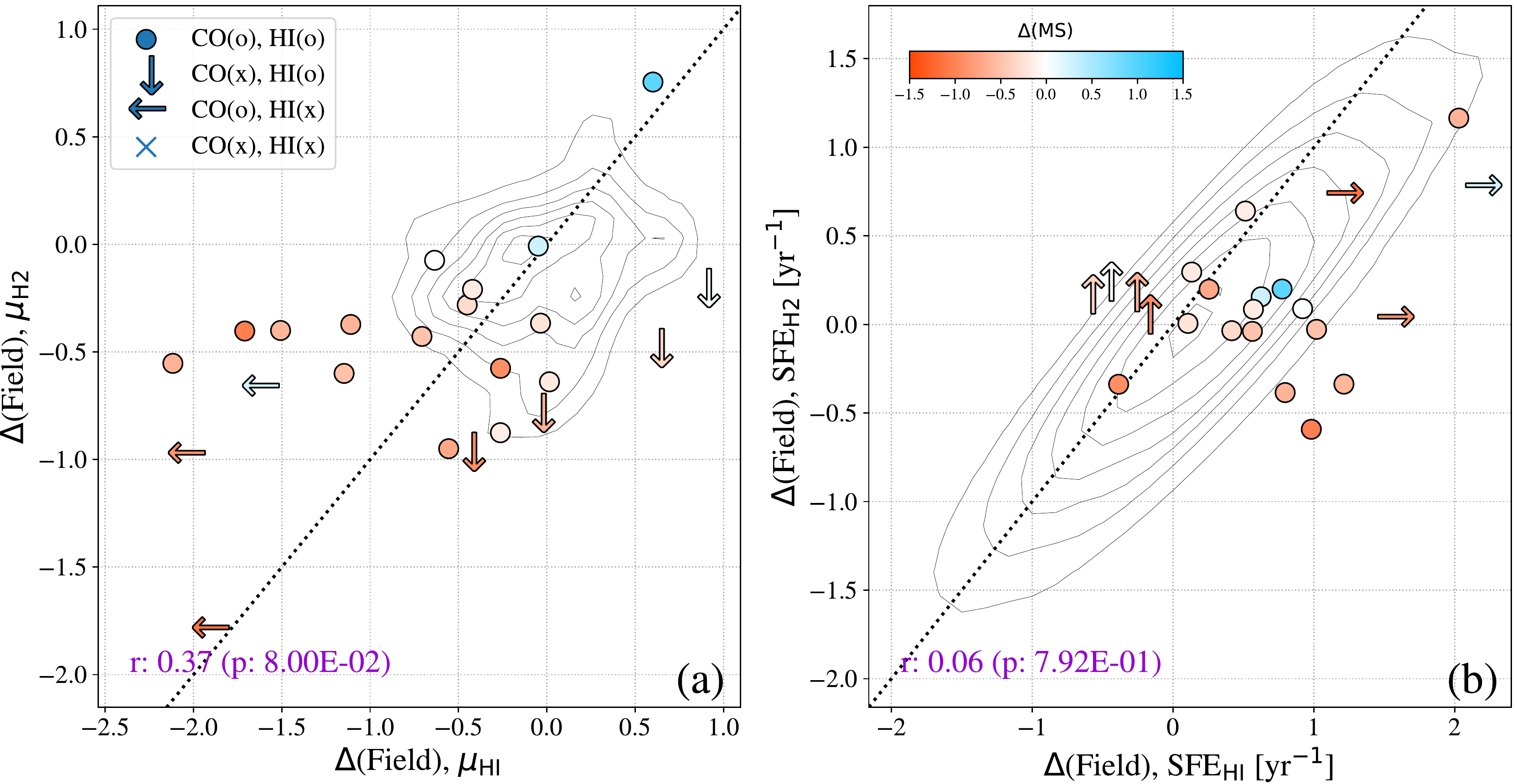}
\end{center}
\caption{
\HI-H$_2$ comparison of the offsets from the field SF galaxies at fixed stellar masses of gas fraction (a) and SFE (b).
The galaxies with both CO and \HI~detection are indicated as circles and those with upper limits are indicated as arrows.
The color of the symbol indicates $\Delta({\rm MS})$.
The contours indicate the field galaxies with CO and \HI~detection.
The Spearman's rank-order correlation coefficient $r$ and the $p$-value are shown on the lower left corner of each panel in purple.
Although there is no clear correlation between $\Delta({\rm Field),\mu_{HI}}$ and $\Delta({\rm Field),\mu_{H_2}}$ nor $\Delta({\rm Field),SFE_{HI}}$ and $\Delta({\rm Field),SFE_{H_2}}$, our Fornax sample with \HI~and H$_2$ measurements tend to be more depleted in \HI~than H$_2$ and tend to have higher SFE$_{\rm HI}$ than field galaxies whereas their SFE$_{\rm H_2}$ is comparable to that of the field galaxies.
}
\label{fig:hih2}
\end{figure*}

To understand the properties of galaxies whose molecular- and/or atomic gas is depleted, Figure~\ref{fig:hih2} compares the offsets of the \HI~and H$_2$ gas fractions and SFEs from the values of the field galaxies for fixed stellar masses.
The color of the symbols indicates $\Delta({\rm MS})$ of galaxies.
Note that the sample is limited to the galaxies with both \HI~and H$_2$ measurements and the galaxies with large \HI~depletion are largely missed in this analysis.
For a reference, the field galaxies with both CO and \HI~detections are shown as contours.
There is no significant correlation between $\mu_{\rm HI}$ and $\mu_{\rm H_2}$ nor SFE$_{\rm HI}$ and SFE$_{\rm H_2}$ of the Fornax galaxies.
The Fornax galaxies with lower $\Delta({\rm MS})$ tend to have lower $\mu_{\rm HI}$ and $\mu_{\rm H_2}$ while the location of galaxies on the SFE$_{\rm HI}$-SFE$_{\rm H_2}$ plot does not strongly depend on $\Delta({\rm MS})$.

For the comparison of the gas fractions, we can see that the Fornax galaxies tend to distribute at the third quadrant in the plot, i.e., both the molecular and atomic gas is depleted in the Fornax galaxies compared to the field galaxies for their stellar masses.
The dotted line indicates $\Delta({\rm Field), \mu_{H_2}}=\Delta({\rm Field), \mu_{HI}}$, i.e., the galaxies on this line have the same degree of depletion in molecular and atomic gas.
In the galaxies that distribute above the dotted line in the third quadrant, \HI~gas is more deficient than the H$_2$ gas compared to the field galaxies.
For some galaxies with the smallest $\mu_{\rm HI}$, molecular gas is not depleted as much as atomic gas.
Next, for the comparison of the SFE, most of the galaxies have comparable SFE$_{\rm H_2}$ to the field galaxies while many galaxies have enhanced SFE$_{\rm HI}$ compared to the field galaxies.

Hereafter, we use the $\Delta({\rm Field})$ values, i.e., the values from which $M_{\rm star}$-dependences of field galaxies have been subtracted, in order to investigate the environmental effect without the $M_{\rm star}$ effect.

\subsection{Dependence on the projected clustocentric distance}\label{sec:distance}

\begin{figure*}[]
\begin{center}
\includegraphics[width=150mm, bb=0 0 1442 1833]{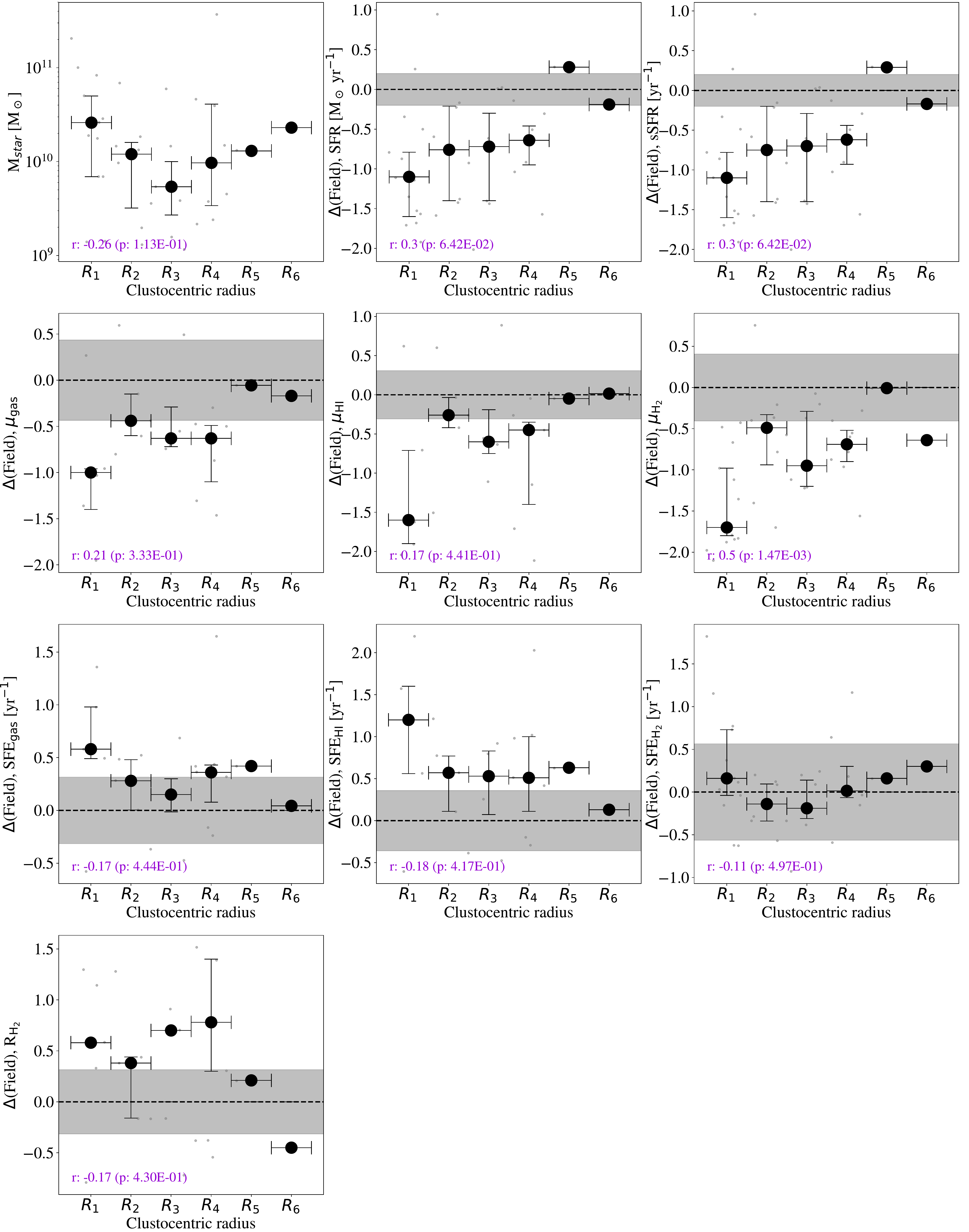}
\end{center}
\caption{
Radial variation of key quantities whose $M_{\rm star}$ dependence have been subtracted, i.e., $\Delta({\rm Field})$ values.
The black filled circle and its upper/lower limits indicate the median and 1st/3rd quartiles, respectively.
The individual data is indicated as grey filled circles.
Zero of each key quantity indicates the values of the field galaxies (dashed line) and the grey-shaded regions indicate the 1-$\sigma$ of the $\Delta({\rm Field})$ values for the field galaxies.
Note that the numbers of galaxies used to plot in each panel are different.
The Spearman's rank-order correlation coefficient $r$ and the $p$-value are shown on the lower left corner of each panel in purple.
Only $\Delta({\rm Field),\mu_{H_2}}$ significantly correlates to the clustocentric distance.
}
\label{fig:radial_wul}
\end{figure*}

The dependence of the galaxy properties on the projected clustocentric distance ($R_{\rm proj}$) has been explored in various studies \citep[e.g.][]{Giovanelli:1985kt,Solanes:2001nq}.
$R_{\rm proj}$ is an easy indicator of the elapsed time since the galaxy is trapped by the cluster potential, i.e., accretion phase, as well as an indicator of the ICM density since the ICM density is higher at the cluster core than the outskirt to a first approximation.
The clustocentric radius of our sample is divided into six radial bins with each bin spanning $0.5~R_{200}$, that is
$R_{\rm proj}/R_{200}<0.5$ ($R_1$),
$0.5 \leq R_{\rm proj}/R_{200}<1.0$ ($R_2$),
$1.0 \leq R_{\rm proj}/R_{200}<1.5$ ($R_3$),
$1.5 \leq R_{\rm proj}/R_{200}<2.0$ ($R_4$),
$2.0 \leq R_{\rm proj}/R_{200}<2.5$ ($R_5$), and
$R_{\rm proj}/R_{200} \geq 2.5$ ($R_6$),
where $R_{200}$ is the radius where the mean interior density is 200 times the critical density of the universe and we adopted the virial radius of $2.0$~degree for $R_{200}$ \citep{Drinkwater:2001so}.
The most outskirt galaxy in our ``{\it best-effort sample}'' locates at $\sim2.5 R_{200}$.

Figure~\ref{fig:radial_wul} shows the medians, 1st and 3rd quartiles of the key quantities for galaxies in each category.
There seems a radial trend in SFR, sSFR and $\mu_{\rm H_2}$ where galaxies that are nearer to the cluster center have a lower value compared to the galaxies at the cluster outskirt.
However, only the trend of $\mu_{\rm H_2}$ is further confirmed by the Spearman's rank-order correlation coefficient ($0.5$) and the $p$-value ($<0.05$).
A significant radial dependence is not seen in the other quantities.
When compared to the field SF galaxies represented by zero value in Figure~\ref{fig:radial_wul}, the key quantities of the Fornax galaxies are not significantly different from those of the field galaxies except for SFR, sSFR and $\mu_{\rm H_2}$.


\subsection{Dependence on the galaxy number density}\label{sec:density}

\begin{figure*}[]
\begin{center}
\includegraphics[width=150mm, bb=0 0 1422 1390]{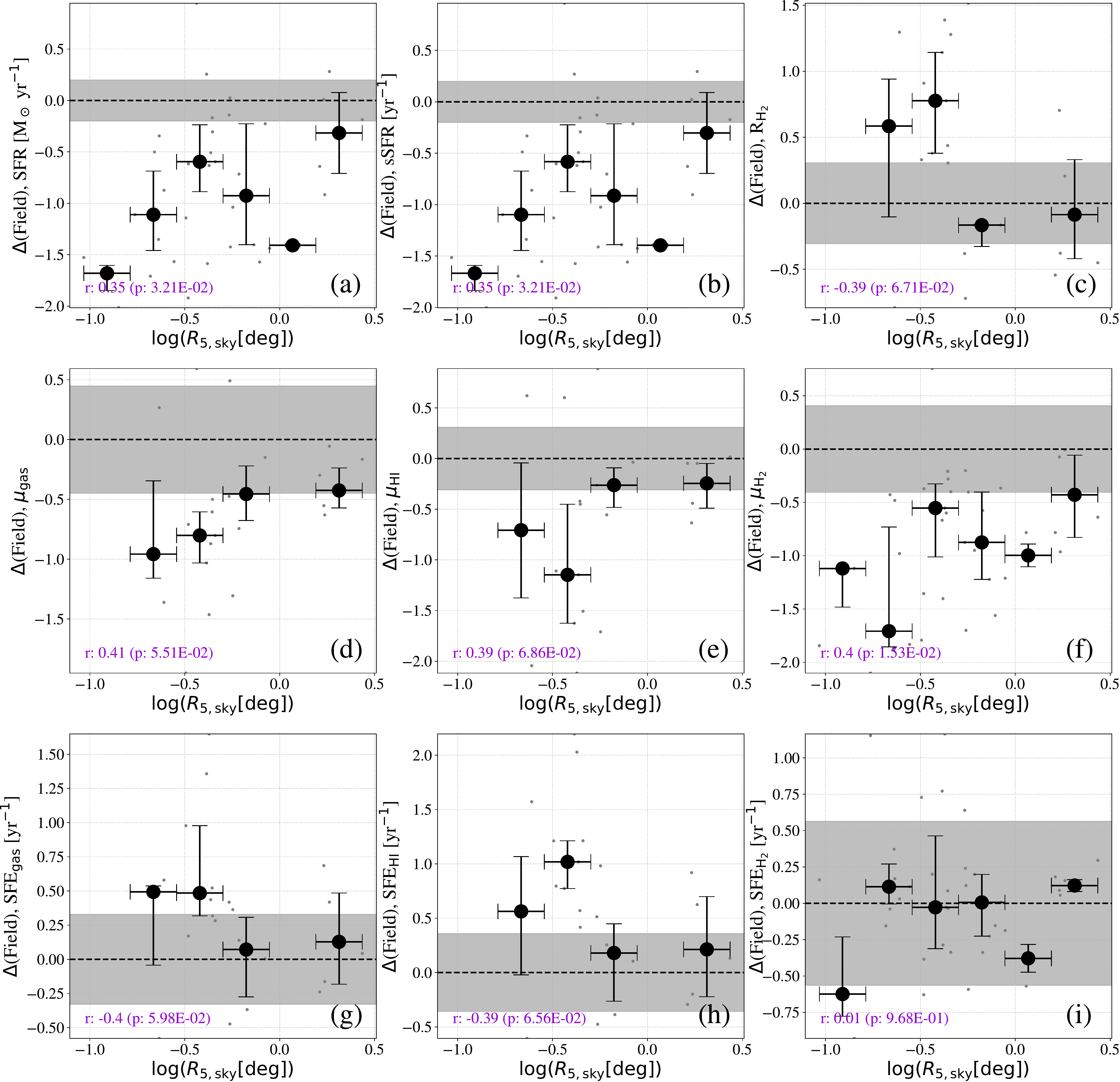}
\end{center}
\caption{
The relationships between $R_{\rm 5, sky}$ with the key quantities [$\Delta({\rm Field})$ values].
Symbols are same as in Figure~\ref{fig:radial_wul}.
$R_{\rm 5,sky}$ positively correlate to SFR, sSFR, $\mu_{\rm gas}$, and $\mu_{\rm H_2}$ and negatively correlates to $R_{\rm H_2}$.
}
\label{fig:numberdensity_radec}
\end{figure*}

Galaxy number density is also one of the important parameters to understand the dominant quenching mechanism in cluster galaxies.
If the star-formation and cold-gas properties of galaxies more strongly depend on the local galaxy number density than the other parameters of the parent clusters, galaxy-galaxy interactions are expected to be relatively important.

We adopt two distances to characterize the local galaxy number density:
the distances to the 5-th nearest galaxy (1) on the sky ($R_{\rm 5, sky}$) and (2) on the PSD ($R_{\rm 5, PSD}$).
The vertical and horizontal axes of the PSD are respectively $R_{\rm proj}/R_{200}$ and $\Delta v_{\rm gal}/\sigma_{\rm Fornax}$, where $\Delta v_{\rm gal}$ and $\sigma_{\rm Fornax}$ are the velocity offset of a galaxy to the cluster center and the velocity dispersion of the Fornax cluster, respectively.
Both values are dimensionless quantities.
We assume a unity weight for both the x-axis and y-axis of the PSD for calculating $R_{\rm 5, PSD}$.
To calculate the distances to the 5-th nearest galaxy, we utilize all the FCC galaxies with the information of their recession velocity in the NED database, resulting in 148 galaxies.
$R_{\rm 5, sky}$ positively correlates to $R_{\rm 5, PSD}$ with a correlation coefficient of 0.32 ($p$-value of $8\times10^{-5}$).
Although there are several ways to estimate the galaxy local density, the $N$-th nearest-neighbor-based measures are considered to be the best probe of the internal density of massive halo \citep{Cooper:2005pj,Muldrew:2012fq} and $N$ generally ranges from three to ten \citep[e.g.,][]{Dressler:1980yl,Gomez:2003hb,Cooper:2005pj,Muldrew:2012fq}. 

The two distances work complementarily to evaluate the different galaxy-galaxy interactions.
The tidal force between galaxies is inversely proportional to the cube of their physical distance.
$R_{\rm 5, sky}$ is simply the projected distance between the galaxies on the sky and it can be diluted by apparently close galaxy pairs whose distance along line-of-sight is large.
On the other hand, $R_{\rm 5, PSD}$ takes the relative line-of-sight velocity of the galaxy pairs into consideration in addition to the projected distance.
However, $R_{\rm 5, PSD}$ would miss the effect of the galaxy harassment since galaxies with a large relative velocity could be neglected due to their large distance on the PSD.

Before comparing the two distances and the key quantities, $R_{\rm 5, sky}$ or $R_{\rm 5, PSD}$ needs to be compared to the projected clustocenteric distance since there does exist radial tendency for some quantities (see Section~\ref{sec:distance}).
The results are shown in Appendix~\ref{sec:supplementaryplots}.
$R_{\rm 5, sky}$ and $R_{\rm 5, PSD}$ both positively correlate to the clustocentric distance with a correlation coefficient of $0.85$ and $0.35$, respectively, i.e., it is denser near the center than the outskirt of the cluster (Figure~\ref{fig:dist_r5}).
These positive correlations suggest that the dependence of the key quantities on the clustocentirc distance seen in Figure~\ref{fig:radial_wul} might just reflect the dependence on the local galaxy number densities, and vice versa.

The comparison between the key quantities and $R_{\rm 5, sky}$ or $R_{\rm 5, PSD}$ of our sample is shown in Figures~\ref{fig:numberdensity_radec} and \ref{fig:numberdensity}, respectively.
$R_{\rm 5, sky}$ positively correlates to SFR, sSFR, and $\mu_{\rm H_2}$ with correlation coefficients of $\sim0.4$.
$R_{\rm 5, PSD}$ positively correlates to SFR, sSFR, and $\mu_{\rm H_2}$ with correlation coefficients of $\sim0.4$, and negatively correlates to SFE$_{\rm gas}$ with correlation coefficients of $\sim-0.4$.
Considering the stronger correlation between $\mu_{\rm H_2}$ and the clustocentric distance than $R_{\rm 5, sky}$ or $R_{\rm 5, PSD}$ and the positive correlation between the clustocentric distance and $R_{\rm 5, sky}$ or $R_{\rm 5, PSD}$, the observed $\mu_{\rm H_2}$ dependence on $R_{\rm 5, sky}$ or $R_{\rm 5, PSD}$ could just reflect the more intrinsic dependence of $\mu_{\rm H_2}$ on the clustocentric distance.
Following the same logic, the unique correlations between $R_{\rm 5, PSD}$ with SFE$_{\rm gas}$ indicates that SFE$_{\rm gas}$ intrinsically depends on the distances to the 5-th nearest galaxy on the PSD.
On the other hand, SFR and sSFR correlate to $R_{\rm 5, sky}$ and $R_{\rm 5, PSD}$ but do not correlate to the projected clustocentric distance.

\subsection{Dependence on the accretion phase to the cluster}\label{sec:accretion}

\begin{figure*}[]
\begin{center}
\includegraphics[width=\textwidth, bb=0 0 1572 498]{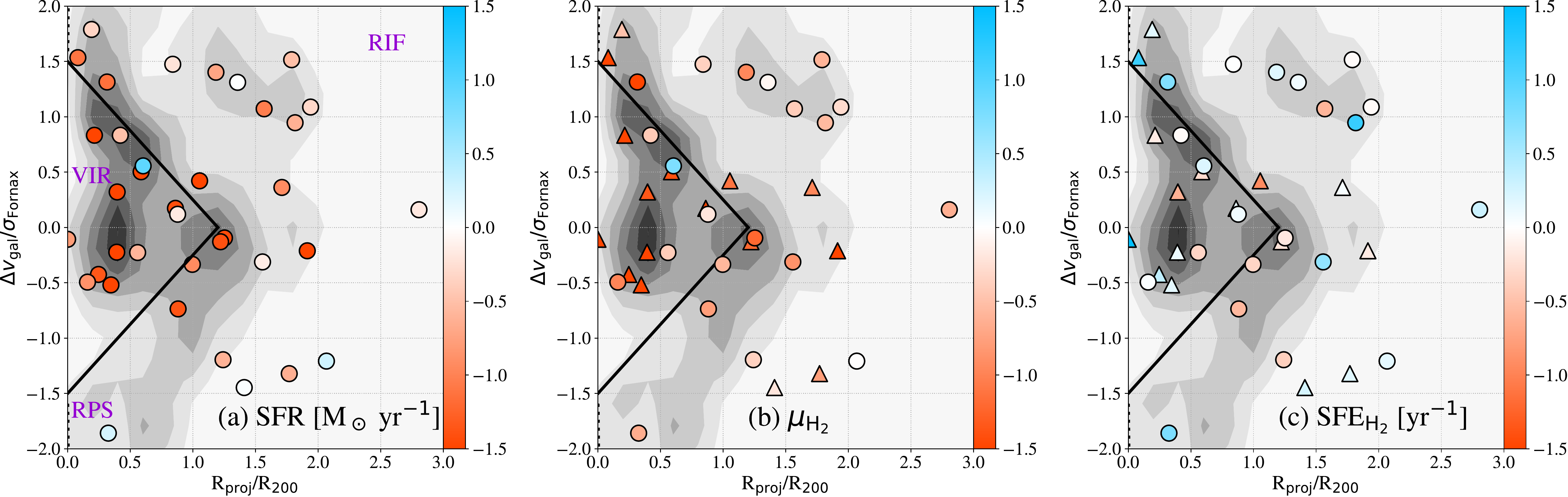}
\end{center}
\caption{
(a) $\Delta({\rm Field})_{\rm SFR}$, 
(b) $\Delta({\rm Field})_{\mu_{\rm H_2}}$, and
(c) $\Delta({\rm Field})_{\rm SFE_{\rm H_2}}$ on the PSD.
The grey contours indicate the FCC galaxies with the NED velocity.
The triangle region enclosed by the solid black line is the VIR region.
The boundary for the RPS region is indicated with the dotted line and is too close to the Y-axis of the plot to be recognized in this plot.
The rest of the region is the RIF region.
The CO non-detections are indicated with triangles.
}
\label{fig:psd_co}
\end{figure*}

\begin{figure*}[]
\begin{center}
\includegraphics[width=150mm, bb=0 0 1422 1390]{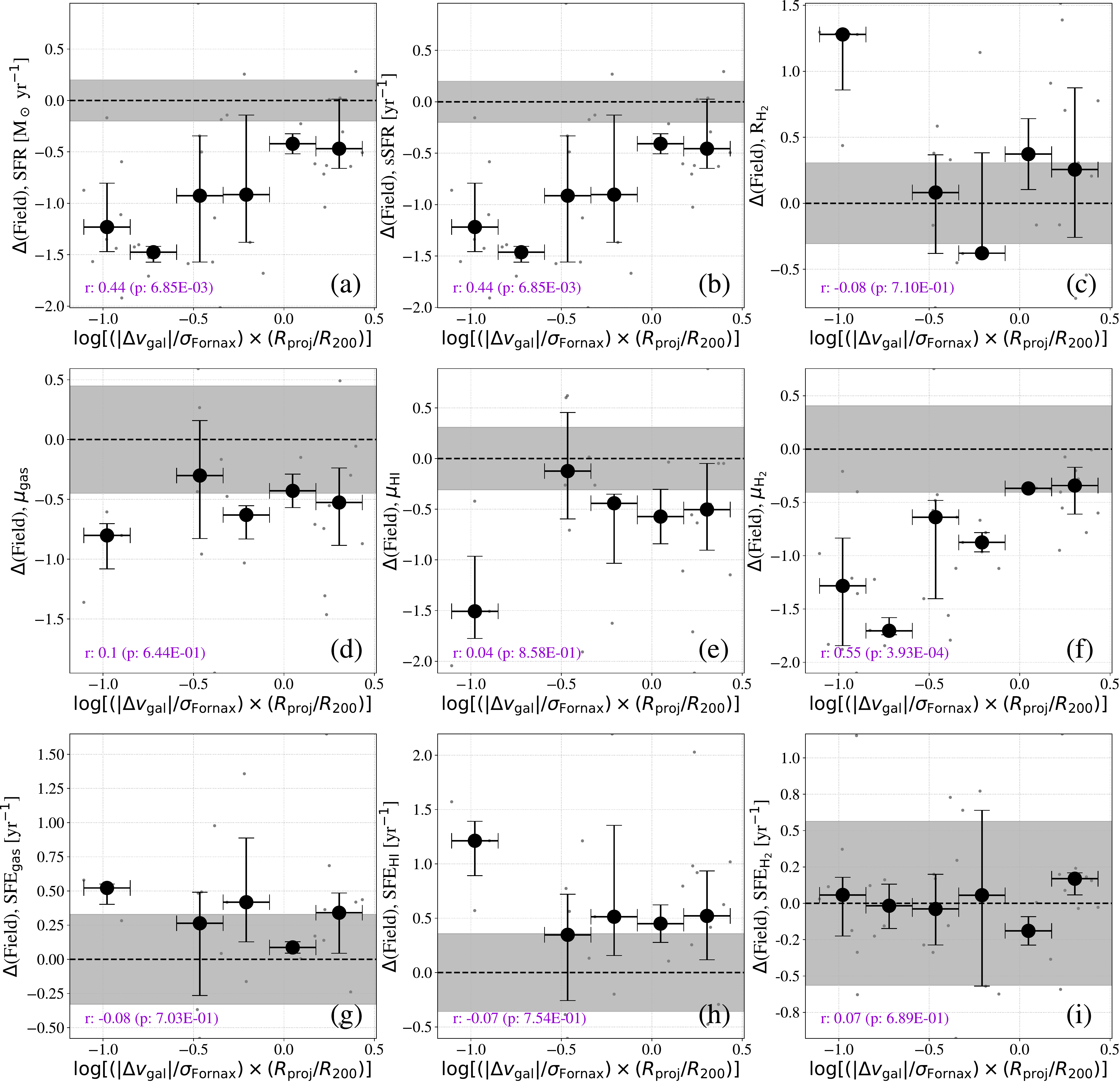}
\end{center}
\caption{
Relationship between the accretion phase, $(|\Delta v_{\rm gal}|/\sigma_{\rm Fornax})\times(R_{\rm proj}/R_{200})$, with the key quantities.
Symbols are same as in Figure~\ref{fig:radial_wul}.
Among all the environmental parameters, SFR, sSFR, and $\mu_{\rm H_2}$ most strongly depends on the accretion phase, where ancient infallers tend to have lower SFR, sSFR, and $\mu_{\rm H_2}$ values than the recent infallers.
}
\label{fig:accretionphase}
\end{figure*}

\begin{table}
\begin{center}
\caption{ICM and galaxy parameters for the RPS boundary line\label{tab:str_model}}
\begin{tabular}{lcc}
\tableline\tableline
Parameters & Values & References$^{a}$\\
\tableline
&--- ICM model$^{b}$ ---&\\
$\rho_0$ & $0.018$ cm$^{-3}$ & 1\\
$R_{\rm C}$ & 4.36 kpc & 1\\
$\beta$ & $0.35$ & 1\\
\tableline
&--- Galaxy model$^{c}$ ---&\\
$M_{\rm d,star}$ & $10^{9}$ M$_\odot$ & 2\\
$M_{\rm d,gas}$ & $1.0 \times M_{\rm d,star}$ & 3\\
$R_{\rm d,star}$ & $0.9$ kpc & 4\\
$R_{\rm d,gas}$ & $1.7 \times R_{\rm d,star}$ & 5\\
\tableline
\end{tabular}
\end{center}
\tablecomments{
$^{\rm a}$ 1: \cite{Jones:1997aa}; 2: this study; 3: \cite{Hunt:2015yd}; 4: \cite{Morokuma-Matsui:2021to}; 5: \cite{Cayatte:1994hr}.
$^{\rm b}$ Standard $\beta$-model of ICM gas density distribution \citep{Cavaliere:1976mb}:
$\rho_{\rm ICM}(r_{\rm 3D})=\rho_0\left[1+\left(\frac{r_{\rm 3D}}{R_{\rm C}}\right)^2\right]^{-3\beta/2}$, where $r_{\rm 3D}=(\pi/2)R_{\rm proj}$, $\rho_0$, $R_{\rm C}$, and $\beta$ are 3D distance from the cluster center, the number density at the center of the cluster, core radius, and the power of the distribution.\\
$^{\rm c}$ Exponential surface density distribution:
$\Sigma_{\rm d, star~or~gas}(r) = \frac{M_{\rm d, star~or~gas}}{2 \pi R_{\rm d, star~or~gas}^2} \exp{\left(\frac{-r}{R_{\rm d, star~or~gas}}\right)}$, where $M_{\rm d,star}$ and $M_{\rm d,gas}$ are the stellar and gas masses, and $R_{\rm d,star}$ and $R_{\rm d,gas}$ are the scale lengths of stellar and gas disk.
We adopted a RPS boundary for the $M_{\rm star}=10^9$~M$_\odot$ galaxies where their entire gas (both atomic and molecular gas) is removed.
The $R50$ for the $M_{\rm star}\sim10^9$~M$_\odot$ galaxies listed in EVCC is converted to the scale length with an empirical relation between $R50$ and scale length of $R50=1.69 \times R_{\rm d,star}$.
}
\end{table}

The galaxy distribution on the PSD is useful to trace the accretion phase of galaxies in a cluster as suggested by previous studies \citep[e.g.,][]{Jaffe:2015pq,Yoon:2017jl,Wang:2021yp}, where they show that \HI~gas removal in cluster galaxies can be explained by the ram pressure stripping effects.
The galaxies trapped by a cluster potential generally follow a ``wedge'' trajectory on the PSD, i.e., oscillating in position and velocity, and eventually fall into the cluster core \citep[see Figure~4 of][]{Jaffe:2015pq}.
In order to clarify the effects of the ram pressure and the other interactions with ICM or the cluster potential, the galaxies in our sample are divided into two groups according to their locations on the PSD of the cluster:
(1) ``virialized (VIR)'' and ``ram-pressure stripping (RPS)'' galaxies, and
(2) ``recent infall (RIF)'' galaxies.
The classification is done so that the RPS and VIR galaxies are currently affected and have been affected by the ram pressure, respectively, while the RIF galaxies are unlikely to have been strongly affected by the ram pressure so far.
It should be noted that the position on the PSD provides just a statistical probability that each galaxy is in a different accretion phase and is not univocally associated to the quenching mechanisms.

The VIR region is enclosed by the lines connecting the points of $(R_{\rm proj}/R_{200}$, $|\Delta v_{\rm gal}|/
\sigma_{\rm Fornax})=(1.2, 0.0)$ and $(0.0,1.5)$ \citep{Jaffe:2015pq,Yoon:2017jl}.
The boundary of the RPS region is defined so that the gas in a typical $M_{\rm star}\sim10^9$~M$_\odot$ galaxy can be totally stripped, i.e., even the gas at the galaxy center, where the anchoring force by the galaxy potential is the strongest, can be removed.
Specifically, the boundary line is determined by the balance between the ram pressure, 
\begin{equation}
P_{\rm ram}=\rho_{\rm ICM} v_{\rm 3D,gal}^2,
\end{equation}
and the galaxy's anchoring force per area, 
\begin{equation}
\Pi_{\rm gal}=2\pi G \Sigma_{\rm star} \Sigma_{\rm gas},
\end{equation}
where $\rho_{\rm ICM}$, $v_{\rm 3D,gal}$, $G$, $\Sigma_{\rm d,star}$, and $\Sigma_{\rm d,gas}$ are the ICM density distribution, 3D velocity of the galaxy ($v_{\rm 3D,gal}=\sqrt{3} v_{\rm obs,gal}$), gravitational constant, surface mass density of stellar and gas disks, respectively \citep{Gunn:1972kc}.
The model parameters used to calculate the RPS boundary line are summarized in Table~\ref{tab:str_model}.
The galaxies that are not in the VIR or RPS regions are classified as the RIF galaxies.

\begin{figure*}[]
\begin{center}
\includegraphics[width=150mm, bb=0 0 1402 1824]{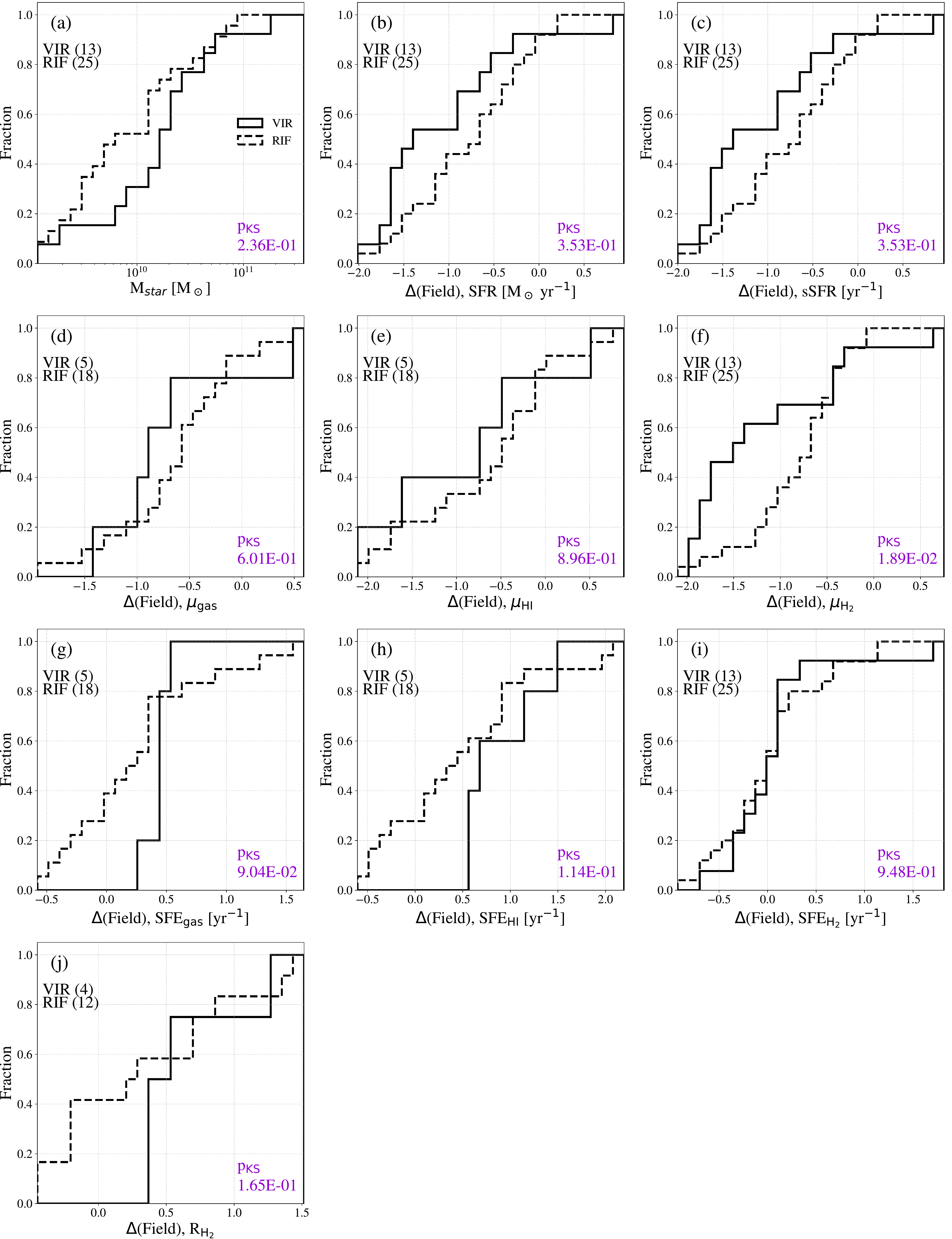}
\end{center}
\caption{
Cumulative histograms of stellar mass and the key quantities of galaxies in the VIR (solid line) and RIF (dashed line).
The numbers of galaxies in the VIR and RIF regions are indicated at the upper left corner in each panel.
The $p$-values of the KS test is indicated in purple at the lower right corner.
}
\label{fig:hist_c2_inclp}
\end{figure*}

The galaxy distributions on the PSD that are color-coded for SFR, $\mu_{\rm H_2}$, and SFE$_{\rm H_2}$ are presented in Figure~\ref{fig:psd_co} (the one with color-coded for all the key quantities is presented in Figure~\ref{fig:psd_all} in Appendix~\ref{sec:supplementaryplots}).
The black solid and dotted lines indicate the boundary lines for the VIR and RPS regions, respectively.
As seen in this figure, there is no RPS galaxy in our sample since the RPS region defined in this study is quite small.
Note that the RPS region is wider for the less massive galaxies.
We compare the cumulative histogram of the key quantities for the VIR and RIF galaxies in Figure~\ref{fig:hist_c2_inclp}.
There seem to be differences between the two samples: compared to the RIF galaxies, the VIR galaxies have smaller median values of SFR, sSFR, and gas fractions and higher median values of SFE$_{\rm gas}$ and SFE$_{\rm HI}$.
The KS test confirms that only the difference in $\mu_{\rm H_2}$ between VIR and RIF galaxies is significant.

We further study the variation of the key quantities with the product of $(|\Delta v_{\rm gal}|/\sigma_{\rm Fornax})\times(R_{\rm proj}/R_{200})$ which is claimed to be a measure of the accretion epoch.
The galaxies with lower value of this quantity, i.e., smaller clustocentric distance and/or smaller relative velocity with respect to the cluster, are considered to be accreted earlier.
The result is shown in Figure~\ref{fig:accretionphase}.
From the figure, $(|\Delta v_{\rm gal}|/\sigma_{\rm Fornax})\times(R_{\rm proj}/R_{200})$ positively correlates to SFR, sSFR and $\mu_{\rm H_2}$ whose correlation coefficients are among $\sim0.4-0.6$, i.e., the ancient infallers tend to have lower SF activity and smaller molecular-gas contents for their masses. 

We find that $(|\Delta v_{\rm gal}|/\sigma_{\rm Fornax})\times(R_{\rm proj}/R_{200})$ positively correlates to the $R_{\rm 5,sky}$ or $R_{\rm 5,PSD}$ with correlation coefficients of 0.42 and 0.54, respectively, i.e., the ancient infallers tend to reside in denser region than the recent infallers (Figure~\ref{fig:dist_r5} in the Appendix).
The relationships between $R_{\rm 5,sky}$ and $R_{\rm 5,PSD}$ with SFR, sSFR and $\mu_{\rm H_2}$ would just reflect the more intrinsic dependence of SFR, sSFR and $\mu_{\rm H_2}$ on the accretion phase, comparing the correlation coefficients.
The null correlation between the accretion phase and $\mu_{\rm gas}$ further support the scenario that $\mu_{\rm gas}$ intrinsically depends on the local galaxy number density.


\section{Discussions\label{sec:discussion}}

\subsection{Sample biases for the \HI~related properties}\label{sec:results_subsamples}

As suggested at the beginning of Section~\ref{sec:coldgassfproperties}, there may exist sample bias when we use the best-effort sample.
Especially, the galaxies with \HI~measurements are biased to the recent infallers that are mostly SFMS galaxies with $\Delta({\rm MS})>-1$ compared to the entire ALMA sample (Figures~\ref{fig:radec}, \ref{fig:comp_field}, \ref{fig:accretionphase}, and \ref{fig:psd_all}).
Thus, we also perform all the analysis done in the previous sections for the galaxies with the measurements or detections of CO~and \HI~(the CO+\HI-obs. or CO+\HI-det. samples) to see if any of the obtained results changed.
As a result, all the correlations with respect to SFR, sSFR, and $\mu_{\rm H_2}$ seen in previous sections disappear in the case of CO+\HI-obs. samples.
On the other hand, SFE$_{\rm gas}$ correlates to $R_{\rm 5,PSD}$ ($r\sim-0.4$ and $p=0.03$) with the CO+\HI-obs. samples.
When based the CO+\HI-det. samples, all the correlation disappears except for the relation between $R_{\rm H_2}$ and $R_{\rm 5,sky}$ ($r\sim-0.5$ and $p=0.04$).

Additionally, we investigate how the results change when assuming the upper limits of the atomic gas mass for the galaxies without the M$_{\rm atom}$ estimation.
The \HI~data of our sample galaxies is retrieved from the unbiased wide field surveys.
Thus, it is a reasonable approximation to apply the upper limits of the Fornax galaxies obtained in the surveys to other galaxies in the survey areas but not listed in the previous studies.

\cite{Loni:2021qe} carried out the \HI~survey with the ATCA covering an area of $15$~deg$^2$ roughly centered on NGC~1399 and the spatial/velocity resolutions and the \HI~mass sensitivity are $67''\times95''$ ($\sim6\times9$~kpc)/$6.6$~km~s$^{-1}$ and $\sim4.7\times10^7$~M$_\odot$ ($3\sigma$ over $300$~km~s$^{-1}$), respectively.
\cite{Kleiner:2021bx} observed the Fornax-A group using the MeerKAT.
The spatial/velocity resolutions and the \HI~mass sensitivity of the MeerKAT observations are $33''.0\times29''.2$ ($\sim3.2\times2.8$~kpc)/$44.1$~km~s$^{-1}$ and $2.9\times10^6$~M$_\odot$ ($3\sigma$ over $300$~km~s$^{-1}$ at the most sensitive area in case of the point source), respectively.
The HIPASS covers the area where Dec. $<2^\circ$, i.e., the Fornax cluster area is covered in this survey.
Its spatial/velocity resolution and the \HI~mass sensitivity are $14'/18.0$~km~s$^{-1}$ and $4.7\times10^8$~M$_\odot$ ($3\sigma$ over $300$~km~s$^{-1}$ at the distance of $20$~Mpc), respectively.
All the \HI~mass sensitivities have been multiplied by 1.36 to account for Helium.

As a simple test, we assume $4.7\times10^7$~M$_\odot$ as an upper limit of $M_{\rm atom}$ for the galaxies that are within the survey areas of \cite{Loni:2021qe} and \cite{Kleiner:2021bx} but without M$_{\rm atom}$ estimations in those studies, and $4.7\times10^8$~M$_\odot$ as an upper limit of $M_{\rm atom}$ for the galaxies that are outside of the survey areas of \cite{Loni:2021qe} and \cite{Kleiner:2021bx}.
We obtain the qualitatively same results for the comparison of the key quantities between the field and the Fornax galaxies.
However, the $\Delta({\rm Field})_{\mu_{\rm HI}}$ [$\Delta({\rm Field})_{\rm SFE_{HI}}$] becomes to correlate to $\Delta({\rm Field})_{\mu_{\rm H_2}}$ [$\Delta({\rm Field})_{\rm SFE_{H_2}}$] with a correlation coefficient of 0.64 [0.41].
For a comparison between the key quantities and the environment parameters, we find that the $\mu_{\rm gas}$ and $\mu_{\rm HI}$ become to correlate to all the environment parameters (clustocentric distance, $R_{\rm 5,sky}$, $R_{\rm 5,PSD}$, and the accretion phase) while SFE$_{\rm gas}$ and SFE$_{\rm HI}$ remain independent on the environment parameters except for the clustocentric distance.
$\mu_{\rm HI}$ most strongly correlates to the clustocentric distance ($r$ of 0.55) unlike $\mu_{\rm H_2}$ that most strongly depends on the accretion phase.
Furthermore, $\mu_{\rm HI}$ and $\mu_{\rm gas}$ (SFE$_{\rm HI}$) of the VIR galaxies become to be significantly lower (higher) than those of the RIF galaxies.

These results suggest there exists the observational bias and that the weaker dependence of \HI~related quantities than the H$_2$ ones on the environment properties when we use the best-effort sample would {\it not} simply indicate that the molecular gas in the Fornax galaxies is more affected by the cluster environment than their atomic gas.
We speculate that the tendencies seen in the H$_2$ properties would be also observed in the \HI~properties if we could investigate the \HI~properties of all the ALMA sample with $M_{\rm star}$ of $10^9~M_\odot$.

\subsection{Comparison with previous studies on the cold-gas properties in the Fornax galaxies}

We find that our Fornax sample with $M_{\rm star}>10^9$~M$_\odot$ has lower gas fractions and higher SFE$_{\rm gas}$ and SFE$_{\rm HI}$ than the field galaxies for their stellar masses while the difference in SFE$_{\rm H_2}$ between the Fornax and field galaxies is not significant.
The deficiency of the cold-gas reservoirs observed in this study is consistent with previous studies on the atomic- and molecular-gas properties in the Fornax galaxies \citep{Horellou:1995iq,Zabel:2019ne,Loni:2021qe,Kleiner:2021bx}.

\cite{Zabel:2020az} have already reported that depletion time (the inverse number of SFE) of the Fornax galaxies is comparable to those of the field galaxies using ALMA CO and H$\alpha$ data obtained with the Multi-Unit Spectroscopic Explorer (MUSE) on the ESO Very Large Telescope (VLT).
They also report that some Fornax dwarf galaxies have smaller depletion time (higher SFE) than the field galaxies.

We find a higher $R_{\rm H_2}$ in the Fornax galaxies than the field galaxies and that the Fornax galaxies show a wider range of $R_{\rm H_2}$ than the field galaxies as claimed in \cite{Loni:2021qe} and \cite{Kleiner:2021bx}.
These studies discuss that the wide $R_{\rm H_2}$ range indicates that there are various galaxies in different stages of the gas removal/consumption processes in the Fornax cluster.

\subsection{Star-formation quenching in the Fornax galaxies}\label{sec:quenching}

\paragraph{Amount of gas or conversion efficiency from gas to stars?}
When limiting to the galaxies at the lower $\Delta({\rm MS})$ regime, we find that the CO- or \HI-detected Fornax galaxies with $M_{\rm star}>10^9$~M$_\odot$ have lower gas fractions ($\mu_{\rm gas}$, $\mu_{\rm HI}$ and $\mu_{\rm H_2}$) and higher SFE$_{\rm gas}$, SFE$_{\rm HI}$, and $R_{\rm H_2}$ than the field galaxies (Figure~\ref{fig:comp_field3}).
SFE$_{\rm H_2}$ of the Fornax galaxies is higher than or comparable to that of the field galaxies at $\Delta({\rm MS})\sim-1.0$, although the difference is statistically insignificant.
In addition, even for the SFMS galaxies, $\mu_{\rm H_2}$ in the Fornax galaxies is lower than that in the field galaxies.
This suggests that the decrease in the cold-gas reservoir is more important for the SF quenching in the Fornax galaxies than the decrease of SFE (our Fornax sample has higher SFE than the field galaxies!).
Furthermore, the \HI-gas reservoir is more heavily reduced than the H$_2$-gas reservoir in the low $\Delta({\rm MS})$ Fornax galaxies.

\paragraph{Timescale of the gas removal}
Our results also imply that the cold-gas reservoir in the Fornax galaxies with $\Delta({\rm MS})<0$ is decreased (stripped or consumed) in a timescale shorter than the typical gas-depletion timescale by star formation in the field SF galaxies of $M_{\rm gas}/{\rm SFR}\sim3$~Gyr, i.e., SFR does not have enough time to change according to the change in the amount of the cold gas.
This is incompatible with the strangulation whose typical timescale is claimed to be $\sim4$~Gyr \citep{Peng:2015kx}.
A similar timescale for the gas depletion is estimated in a different way in \cite{Loni:2021qe} based on the blind \HI~survey of the Fornax cluster.
They find that the \HI-deficient galaxies tend to distribute in the virialized region on the PSD and they consequently discussed the \HI~removal occurs on a timescale shorter than the cluster's crossing time of $R_{200}/\sigma_{\rm Fornax}\sim2$~Gyr.
On the other hand, for the galaxies in the cluster outskirt, they do not find a significant offset from the field relation between the $\Delta({\rm Field})_{\rm SFR}$ and $\Delta({\rm Field})_{\mu_{\rm HI}}$ \citep[Figure~11 in ][]{Loni:2021qe}, suggesting that the SFR has had sufficient time to respond to a variation in \HI~mass as the field galaxies.
In the same figure, they also find that some \HI-non-detected Fornax galaxies would have lower $\mu_{\rm HI}$ than the field galaxies for fixed $\Delta({\rm MS})$.
The shortage of the \HI~gas can lead to the reduction of the H$_2$-gas reservoirs in galaxies without a significant enhancement in SFE$_{\rm H_2}$.

\paragraph{Ram-pressure gas stripping}
One of the (most famous) fast processes removing gas from galaxies in a cluster is ISM ram-pressure stripping.
\cite{Schroder:2001rc} discussed that the \HI~gas deficiency in the Fornax galaxies is caused by the ram-pressure stripping based on the \HI~gas survey toward 66 Fornax galaxies with the Parkes 64-m telescope.
They find that the \HI~deficient galaxies tend to have small velocity dispersion compared to the other galaxies suggesting that they have more radial orbits than the other galaxies.
\cite{Loni:2021qe} detected \HI~emission from $16$ Fornax galaxies located outside the virialized region, and they find that eight out of the 16 galaxies show disturbed \HI~morphology with their $\sim 6~{\rm kpc}\times 9~{\rm kpc}$ ATCA beam.
Among the eight \HI~disturbed galaxies, five galaxies (ESO~358-G063, NGC~1351A, NGC~1365, NGC~1427A, and NGC~1437B) have $>10^9$~M$_\odot$ in our study.
They discussed the ram-pressure stripping as the process responsible for the disturbed \HI~morphologies of ESO~358-G063 and NGC~1351A among the five galaxies.
Thus, \HI~gas of some Fornax galaxies with $M_{\rm star}>10^9$~M$_\odot$ could be removed by the ram-pressure from the ICM.

However, \cite{Horellou:1995iq} consider that the ISM ram-pressure stripping is not as effective as in the Virgo cluster, given that the X-ray emission in the Fornax cluster is weaker than that of the Virgo cluster.
In general, molecular gas is expected to be less affected by ram pressure of ICM compared to atomic gas \cite[][and references therein]{Cortese:2021vp} since molecular gas resides within optical disks of galaxies whereas atomic gas often distributes far out of the optical disks.
On the other hand, \cite{Zabel:2019ne} find that the low-mass Fornax galaxies with $<3\times10^9$~M$_\odot$ have disturbed molecular-gas morphology and such galaxies tend to have a molecular-gas deficit.
This may possibly suggest that some low-mass galaxies are affected by the ram pressure of the ICM.
\cite{Zabel:2021um} find that the Fornax galaxies generally have a lower gas-to-dust ratio than the field galaxies due to the depletion of both the \HI~and H$_2$ gas even though the dust mass is also reduced compared to the field galaxies.
They discuss various possible explanations for this findings including the more efficient stripping of H$_2$ compared to dust, more efficient enrichment of dust in the star formation process, and altered interstellar medium physics in the cluster environment.
As one of the possible explanations, they suggest a combination of the dust formation and H$_2$ consumption by on-going star formation but inefficient H$_2$ formation owing to the \HI~gas depletion due to the strangulation or stripping.

Our data cannot completely reject the ISM ram-pressure stripping as a dominant physical process responsible for the reduction of molecular-gas reservoirs in the Fornax galaxies studied here.
The short timescale atomic gas stripping by the ram pressure is rather compatible with our results.
Galaxies with a low gas fractions and a high SFEs are naturally explained by ram pressure stripping, provided that galaxies generally tend to have high SFE near the galactic center \citep[a steeper radial gradient for earlier spiral galaxies,][]{Villanueva:2021wm}, and that ram pressure preferentially strips outskirt components in galaxies \citep{Bekki:2014ip,Cortese:2016qt}.
Interestingly, NGC~1351A has a tentative elongated CO structure to the South-East direction which is consistent with the \HI~data in \cite{Loni:2021qe}, whereas there is no clear disturbance in the WISE $3.4\mu$m image (Figure~\ref{fig:mommaps_NGC1351A}).
This would suggest the ram-pressure strips not only \HI~but also H$_2$ gas in NGC~1351A.
A detailed comparison of morphologies (and kinematics) of the stellar, atomic-gas and molecular-gas components of the galaxy is required to confirm this possibility.

As another clue to the dominant quenching processes in the Fornax cluster, we find that $\Delta({\rm Field})_{\rm SFR}$ and $\Delta({\rm Field})_{\mu_{\rm H_2}}$ more strongly correlate to the accretion phase to the cluster than the local galaxy number density and the projected clustocentric distance.
On the other hand, $\Delta({\rm Field})_{\mu_{\rm HI}}$ most strongly depends on the projected clustocentric distance to which $\Delta({\rm Field})_{\rm SFR}$ does not significantly correlate (section~\ref{sec:results_subsamples}).
These results suggest that the reduction of \HI-gas reservoirs in galaxies does not directly lead to their star-formation quenching and is partially attributed to the ram-pressure stripping.
The effect of ram pressure stripping would blur the dependence of key quantities on the accretion phase since the galaxies that are being affected by ram pressure could have a small $R_{\rm proj}/R_{200}$ and a large $|\Delta v_{\rm gal}|/\sigma_{\rm Fornax}$, i.e., distribute at the upper left corner on the $|\Delta v_{\rm gal}|/\sigma_{\rm Fornax}$ versus $R_{\rm proj}/R_{200}$ plot.
The strongest dependence of SFR and $\mu_{\rm H_2}$ on the accretion phase among all the environmental parameters explored here basically suggests the importance of long-timescale processes such as the interaction with the cluster potential and the strangulation for star-formation quenching in the Fornax galaxies.
However, it is also possible that the reduction of \HI~gas due to the ram-pressure stripping drives the H$_2$ gas depletion and consequently star-formation quenching in the Fornax galaxies.

\paragraph{Tidal interaction}
The tidal interaction is expected to strip gas in the galaxy outskirt as well as enhance star-formation activity in a galaxy by inducing gas inflow to the galaxy center \citep[e.g.,][]{Noguchi:1986uv,Noguchi:1988bm,Byrd:1990pd,Hernquist:1995ls} and possibly accelerate the gas depletion in the galaxy along with the strangulation.
We find that some Fornax galaxies with $\Delta({\rm MS})<0$ tend to have lower $\mu_{\rm H_2}$ and higher SFE$_{\rm H_2}$ than the field galaxies.
The depletion timescale of molecular gas (1/SFE$_{\rm H_2}$) in some samples with $\Delta({\rm MS})\lesssim-1.5$ is as short as $\lesssim1$~Gyr (Figure~\ref{fig:comp_field4}).
Since our sample are currently forming stars less actively, their cold-gas reservoirs would have been already depleted when star-formation is enhanced by the tidal interaction, or they would be in the later stage of the tidal interaction.
Although the morphological analysis will be presented in a separated paper, there are galaxies with a ring-like gas structure (Figures~\ref{fig:radec_co}, \ref{fig:mommaps_IC1993}, and \ref{fig:mommaps_NGC1425}, \ref{fig:mommaps_NGC1436}), which is predicted in the later stage of the tidal interaction based on the numerical simulations \citep{Noguchi:1986uv}\footnote{
A ring-like structure in H$\alpha$ is also predicted in some galaxies that are affected by ram pressure \citep{Bekki:2014ip}.}.

We approximate the tidal radius of each galaxy as $r_{\rm tidal}=R_{\rm proj}\left( \frac{M_{\rm galaxy}}{3M_{\rm cluster}} \right)^{1/3}$ and compare it with the galaxy size.
The tidal radius of a galaxy is defined as the point at which galactic stars/ISM escapes the galaxy's potential well and becomes part of the cluster.
We adopt $M_{\rm galaxy}=10\times M_{\rm star}$, $M_{\rm cluster}=5\times10^{13}$~M$_\odot$ \citep{Drinkwater:2001so} and the semi-major axis of the ellipse aperture for the WISE and GALEX photometry as the galaxy size ($r_{\rm CAAPR}$).
The $r_{\rm tidal}/r_{\rm CAAPR}$ ratio mostly ranges from zero to five and the median value is $2.7$.
Four out of the 38 galaxies have $r_{\rm tidal}/r_{\rm CAAPR}<1.0$.
Given that the galaxies often have from a few to several times more extended \HI~disk than the optical disk \citep[e.g.,][]{Walter:2008fr,Koribalski:2018la}, \HI~gas in the galaxy outskirt would have been tidally stripped by the cluster potential.

The dependence of the key quantities on the $r_{\rm tidal}/r_{\rm CAAPR}$ ratio is also investigated.
We find that $r_{\rm tidal}/r_{\rm CAAPR}$ positively correlates to $\mu_{\rm H_2}$ with a correlation coefficient of $0.35$.
Although galaxies with high SFE$_{\rm H_2}$ tend to have smaller $r_{\rm tidal}/r_{\rm CAAPR}$, SFE$_{\rm H_2}$ does not significantly depend on the $r_{\rm tidal}/r_{\rm CAAPR}$ ($r\sim-0.3$ and $p=0.06$).
Considering $r_{\rm tidal}/r_{\rm CAAPR}\propto R_{\rm proj}$ and the stronger dependence of $\mu_{\rm H_2}$ on $R_{\rm proj}$ ($r=0.5$), the correlation between $r_{\rm tidal}/r_{\rm CAAPR}$ and $\mu_{\rm H_2}$ may just reflect the more intrinsic relation between and $R_{\rm proj}$ and $\mu_{\rm H_2}$.
The $\mu_{\rm HI}$ and SFE$_{\rm HI}$ of galaxies with $M_{\rm atom}$ measurements in the previous studies does not depend on $r_{\rm tidal}/r_{\rm CAAPR}$.
However $\mu_{\rm HI}$ and SFE$_{\rm HI}$ correlate to $r_{\rm tidal}/r_{\rm CAAPR}$ with correlation coefficients of $0.42$ and $-0.32$, respectively in case we assume the $M_{\rm atom}$ upper limit as did in section~\ref{sec:results_subsamples}.
These correlation coefficients are smaller than and same with those for the $R_{\rm proj}-\mu_{\rm HI}$ relation ($0.55$) and the $R_{\rm proj}-{\rm SFE_{HI}}$ relation ($-0.32$), respectively.
Although the impact of the tidal effect from the cluster potential on galaxies is not so large in the current state, it may have boosted the SFEs and help decreasing gas fractions of some galaxies in the process of galaxy accretion to the cluster.

The caveats here are that we use $10\times M_{\rm star}$ in stead of the halo mass and $R_{\rm proj}$ instead of the distance at the pericenter.
To calculate $r_{\rm tidal}$, we should use halo mass that is generally $\sim100$ times larger than the stellar mass of galaxies depending on the halo mass \citep[e.g.,][]{Behroozi:2010yj,Behroozi:2013zj}.
However it is difficult to estimate the dark matter mass of cluster galaxies since it is predicted that substantial quantities of dark matter are tidally stripped before stellar stripping begins \citep[galaxies that lose $\sim80$~\% of their dark matter lose only $\sim10$~\% of their stars,][]{Smith:2016xg}.
Therefore $r_{\rm tidal}$ estimated here should be considered to be a measure of the true tidal radius.

\paragraph{Pre-processes in the group environment}
The pre-processes in the group environment would be also important for Fornax galaxies.
Theoretical studies predict that many cluster galaxies ($20-40$~\%) had been accreted as a group of galaxies \citep{McGee:2009oh,De-Lucia:2012ir,Benavides:2020ap} and the pre-processes have non-negligible impact on their evolution \citep[but see also][]{Berrier:2009he}.
The galaxy groups within cluster would be disassembled and virialized into the host cluster after the first pericentric passage depending on their mass concentrations \citep{Taffoni:2003ug}.
\cite{Kleiner:2021bx} investigated the cold-gas properties of the galaxies in the Fornax A group located at $\sim2~R_{\rm vir}$ from NGC~1399.
The Fornax-A group is considered to be falling into the main cluster for the first time.
They find that most member galaxies they explored show signs of the pre-processes in the group environment such as \HI~tails, truncated \HI~disk and \HI~deficiency.
\cite{Loni:2021qe} find that the possible members of the NGC~1365 subgroup \citep{Drinkwater:2001so} also show \HI~deficiency.

Even within the virial radius of the Fornax cluster, \cite{Iodice:2019vc} identified three groups of galaxies: the old core; the clump on the north-northwest side of the cluster; and a group of galaxies that fell in more recently.
\cite{Iodice:2019bf} find that the most bright early-type galaxies (ETGs) reside in the former two regions and there are intra-cluster baryons \citep[diffuse light and globular clusters,][]{Iodice:2017xv} that are indicative of the tidal interaction.
They discuss that the diffuse morphology of the intra-cluster light (ICL) and their relative locations to the bright ETGs suggest the gravitational interaction between galaxies rather than the galaxy-cluster interaction.

Combined with the previous studies, our results suggest that the atomic gas in the Fornax galaxies has been stripped tidally and/or by the ram pressure in a short time, and the (slightly) boosted SFEs due to the tidal interaction would accelerate the atomic- and molecular-gas depletion along with the strangulation.
In the early epoch of the cluster formation, the pre-processes in the group environment would have already reduced certain amount of the cold gas and star-formation activity in some galaxies in the core of the cluster \citep{Iodice:2019vc}.


\section{Summary}
\label{sec:summary}

We observed 64 Fornax galaxies with the ALMA Morita array in the cycle 5 and CO emission is detected from 23 out of the 64 galaxies (Figure~\ref{fig:radec_co}).
Combined with the ancillary data of stellar mass, SFR and atomic-gas mass, we investigated atomic- and molecular-gas-related properties of the massive Fornax galaxies ($M_{\rm star}>10^9$~M$_\odot$).
Our main results are as follows:
\begin{description}
\item[Fornax vs field (Section~\ref{sec:field})]
Compared to field galaxies, the Fornax galaxies have lower SFR, sSFR, and both atomic- and molecular-gas fractions and higher SFE from atomic gas, suggesting that the reduction of the gas reservoirs is more essential than the SFE reduction for the quenching in the Fornax galaxies.
Espacially, the Fornax galaxies with low $\Delta$(MS) tend to have low atomic- and molecular-gas fractions and high SFE from atomic gas compared to field galaxies.
This suggests the atomic-gas reduction processes occurred in a timescale shorter than the typical gas-depletion timescale by star formation in field galaxies.
The Fornax galaxies have a higher $R_{\rm H_2}$ median and a wider range of $R_{\rm H_2}$ compared to the field galaxies.
This suggests that the \HI~gas is more depleted than the H$_2$ gas and that there are various galaxies in different stages of the gas stripping.
However, the galaxies with \HI~measurements are biased to SF galaxies and the obtained properties related to \HI~gas would not represent the common feature of our ALMA sample.
The on-going \HI~survey with MeerKAT will shed light on this problem \citep{Serra:2019mh}.

\item[Clustocentric distance (Section~\ref{sec:distance})] 
The galaxies nearer to the cluster center have low $\mu_{\rm H_2}$ (the Spearman's rank-order correlation coefficients $r$ of $0.5$).
The $\mu_{\rm H_2}$ median of outskirt galaxies is already slightly lower than the those of the field galaxies.

\item[Local galaxy density (Section~\ref{sec:density})]
The distance to the 5-th nearest neighbour on the sky ($R_{\rm 5, sky}$) positively correlates to SFR, sSFR, and $\mu_{\rm H_2}$ ($r\sim0.4$).
In case of the distance to the 5-th nearest neighbour on the PSD ($R_{\rm 5, PSD}$), there are correlations with SFR, sSFR, and $\mu_{\rm H_2}$ with correlation coefficients of $\sim0.4$ and a negative correlation with SFE$_{\rm gas}$ with a correlation coefficient of $\sim-0.4$.
SFR and sSFR more strongly depend on $R_{\rm 5, sky}$ or $R_{\rm 5, PSD}$ than the clustocenric distance.
On the other hand, $\mu_{\rm H_2}$ dependence on the local density is expected to just reflect the dependence on the clustocentric distance given the positive correlation between $R_{\rm 5, sky}$ or $R_{\rm 5, PSD}$ with the clustocentric distance and the stronger dependence of $\mu_{\rm H_2}$ on the clustocentric distance than on $R_{\rm 5, sky}$ and $R_{\rm 5, PSD}$.

\item[Accretion phase (Section~\ref{sec:accretion})]
The product of $(|\Delta_{\rm gal}|/\sigma_{\rm Fornax})\times(R_{\rm proj}/R_{200})$ positively correlates to SFR, sSFR, and $\mu_{\rm H_2}$ ($r$ of $\sim0.4-0.6$), suggesting that galaxies that were accreted to the Fornax cluster at an earlier epoch tend to have lower star-formation activity and molecular-gas contents.
SFR, sSFR, and $\mu_{\rm H_2}$ more strongly correlates to $(|\Delta_{\rm gal}|/\sigma_{\rm Fornax})\times(R_{\rm proj}/R_{200})$ than the other environment-related quantities.

\item[SF quenching processes in Fornax (Section~\ref{sec:quenching})] 
Our results suggest that decreasing gas fractions is more important than decreasing SFEs for star-formation quenching in the Fornax galaxies with $M_{\rm star}>10^9$~M$_\odot$.
In addition, the atomic-gas is considered to be stripped tidally and by the ram pressure, which would lead to the depletion of molecular gas and eventually low star-formation activity in the galaxies.
The galaxy-cluster interaction would accelerate molecular-gas deficiency by increasing SFE along with the strangulation, while pre-processes in the group environment would have reduced the molecular-gas reservoirs of the Fornax galaxies in the early phase of the cluster formation. 
Detailed comparison of the morphology of the stellar and cold-gas components in galaxies is required to shed light on the direct effect of the tidal interaction and the ram-pressure stripping on the molecular gas contents in the Fornax galaxies.

\end{description}


Further studies with larger statistics and sample homogeneity, as well as high resolution for cold gas, SFR and stellar components, would be required to determine the dominant quenching process in the Fornax cluster.

\begin{figure*}[]
\begin{center}
\includegraphics[width=0.92\textwidth, bb=0 0 1328 768]{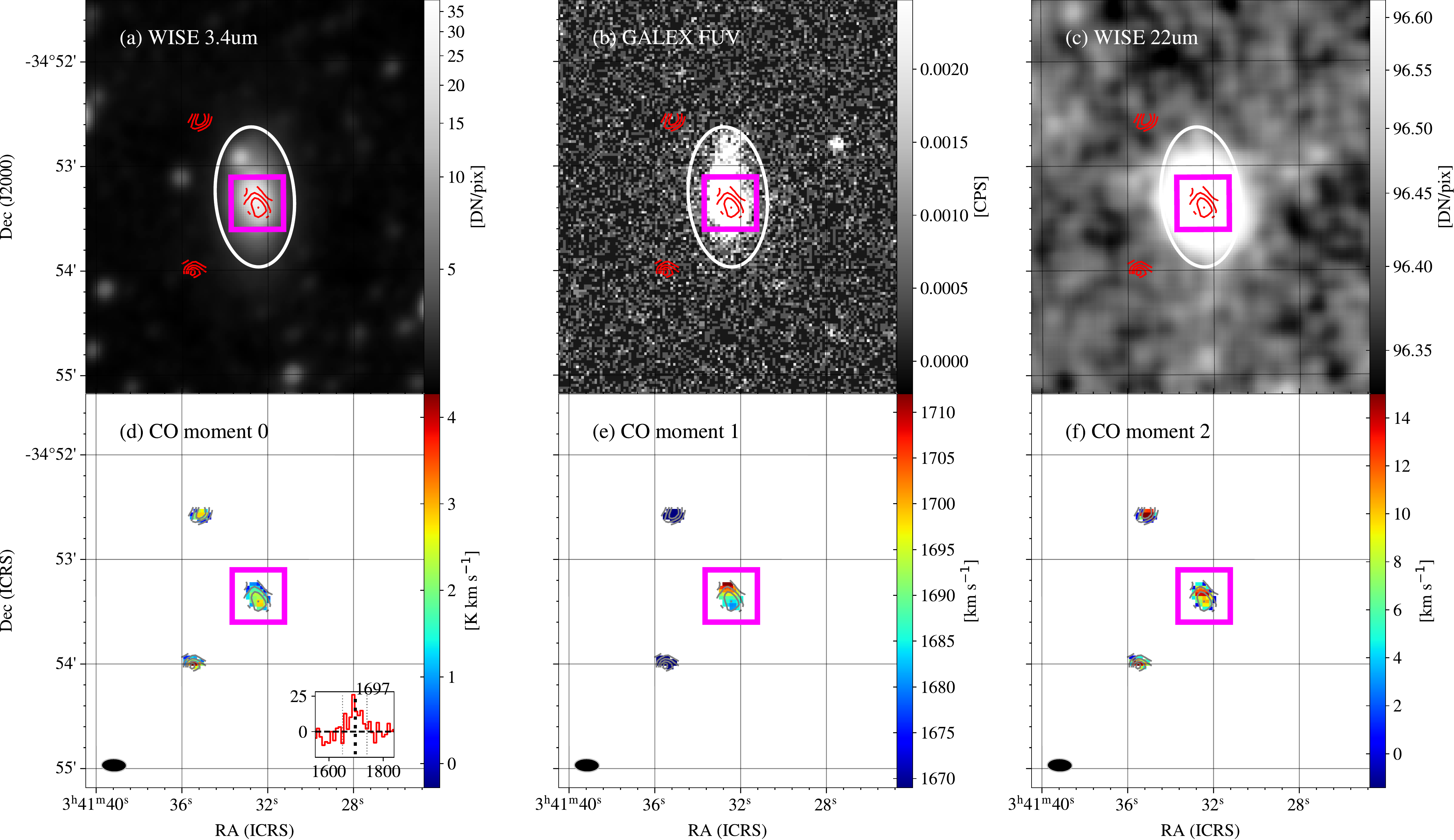}
\end{center}
\caption{
Ancillary images and CO moment maps of ESO358-51: (a) WISE $3.4 \mu$m, (b) GALEX FUV, (c) WISE $22 \mu$m, (d) CO integrated-intensity map, (e) CO velocity map, and (f) CO velocity-dispersion map. CO spectrum is also shown in the panel (d). The vertical and horizontal axes of the CO spectrum plot are the brightness temperature in the unit of mK and the velocity in the unit of km~s$^{-1}$, respectively. The black and grey dotted lines on the CO spectrum panel indicate the galaxy systemic velocity and the velocity range in which the molecular gas mass is calculated, respectively. The red and grey contours indicate CO integrated-intensity maps. The aperture ellipse for the WISE and GALEX data to calculate M$_{\rm star}$ and SFR is indicated with white solid line. The magenta line indicates the area to generate total CO spectra [shown in the panel (d)] to calculate $M_{\rm mol}$.
}
\label{fig:mommaps_ESO358-51}
\end{figure*}

\begin{figure*}[]
\begin{center}
\includegraphics[width=0.92\textwidth, bb=0 0 1328 768]{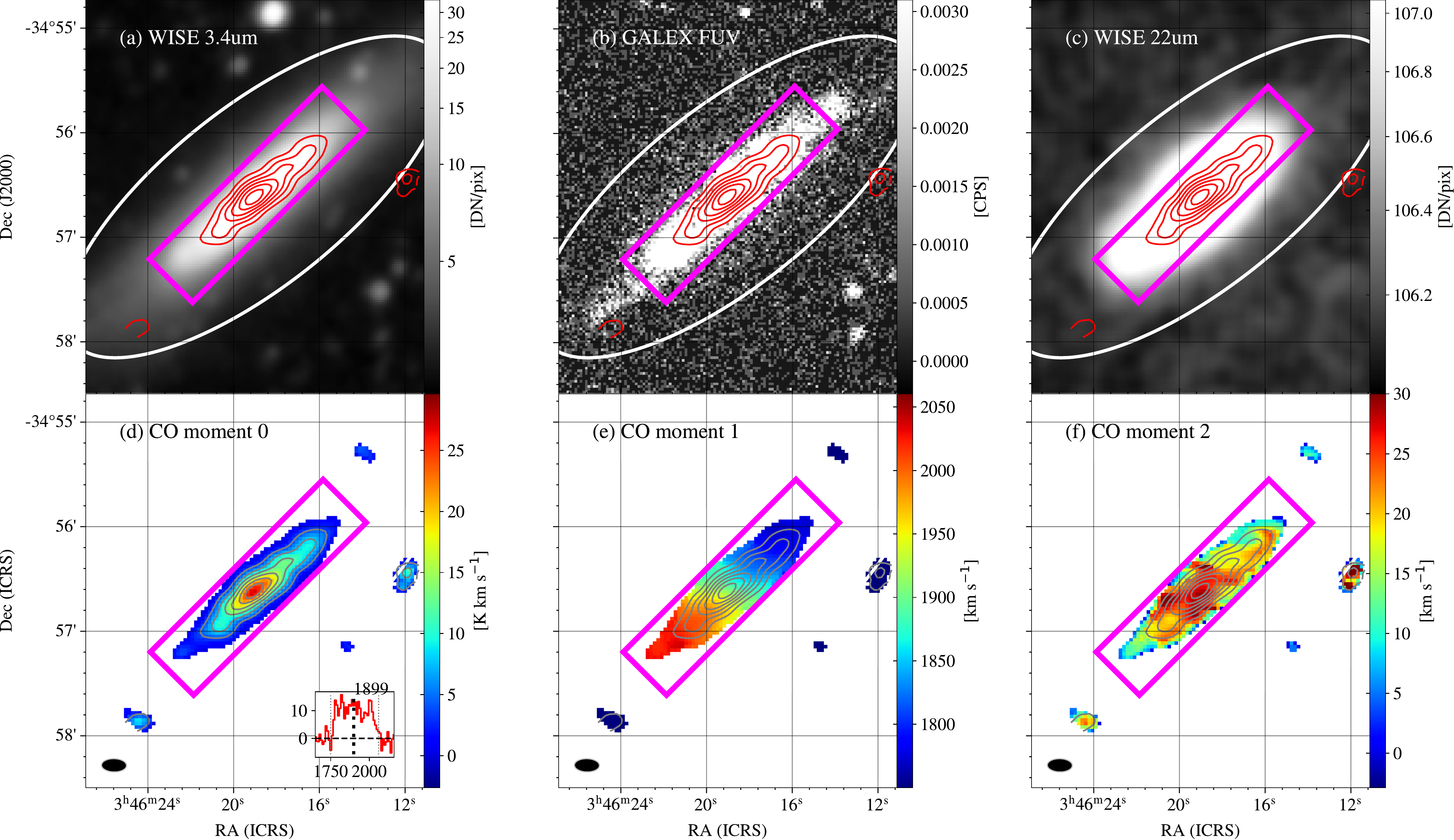}
\end{center}
\caption{
Same as Figure~\ref{fig:mommaps_ESO358-51}, but for ESO358-G063.
}
\label{fig:mommaps_ESO358-G063}
\end{figure*}

\begin{figure*}[]
\begin{center}
\includegraphics[width=0.92\textwidth, bb=0 0 1365 753]{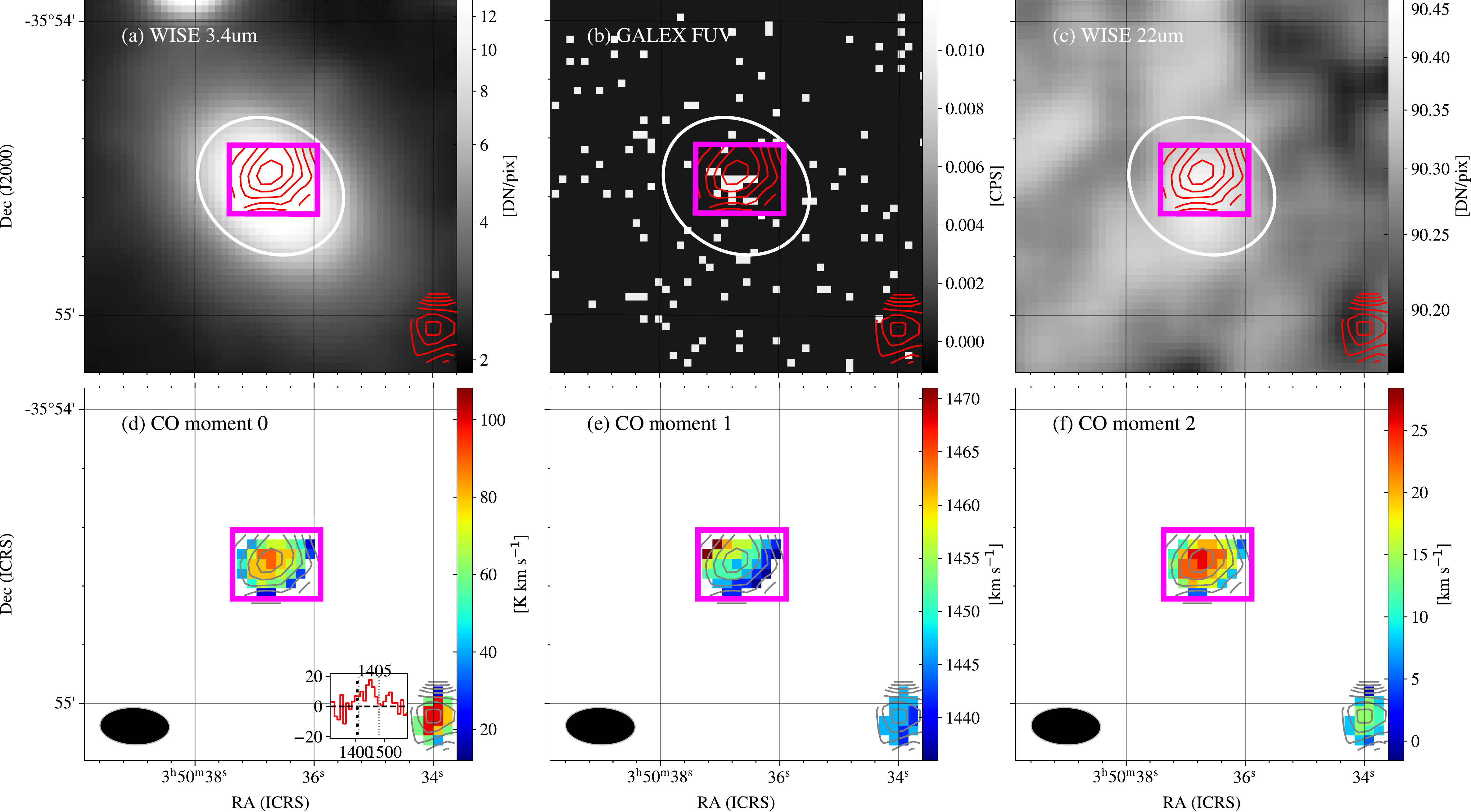}
\end{center}
\caption{
Same as Figure~\ref{fig:mommaps_ESO358-51}, but for ESO359-2.
}
\label{fig:mommaps_ESO359-2}
\end{figure*}

\begin{figure*}[]
\begin{center}
\includegraphics[width=0.92\textwidth, bb=0 0 1360 755]{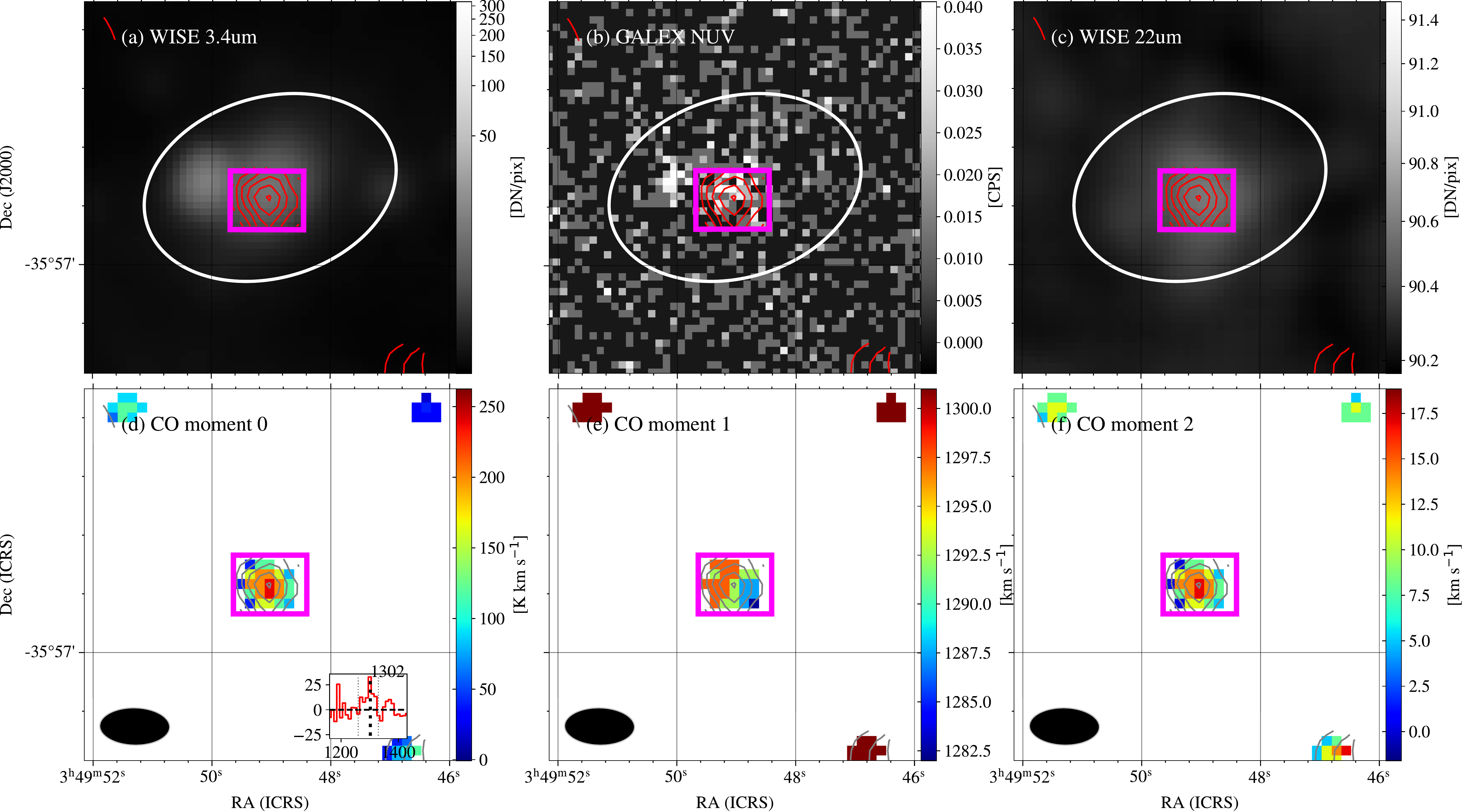}
\end{center}
\caption{
Same as Figure~\ref{fig:mommaps_ESO358-51}, but for FCC332.
}
\label{fig:mommaps_FCC332}
\end{figure*}

\begin{figure*}[]
\begin{center}
\includegraphics[width=0.92\textwidth, bb=0 0 1320 768]{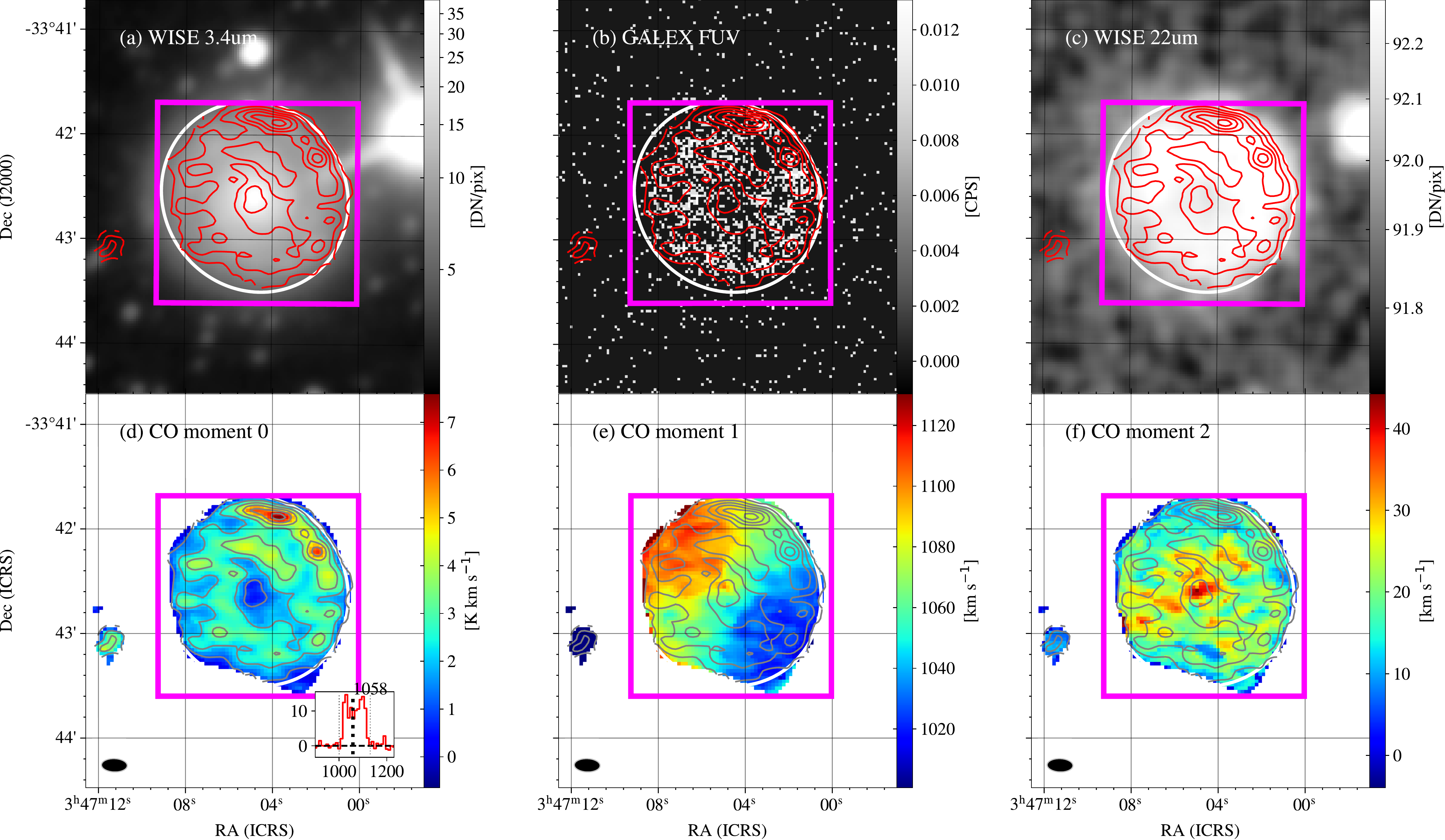}
\end{center}
\caption{
Same as Figure~\ref{fig:mommaps_ESO358-51}, but for IC1993.
}
\label{fig:mommaps_IC1993}
\end{figure*}

\begin{figure*}[]
\begin{center}
\includegraphics[width=0.92\textwidth, bb=0 0 1365 753]{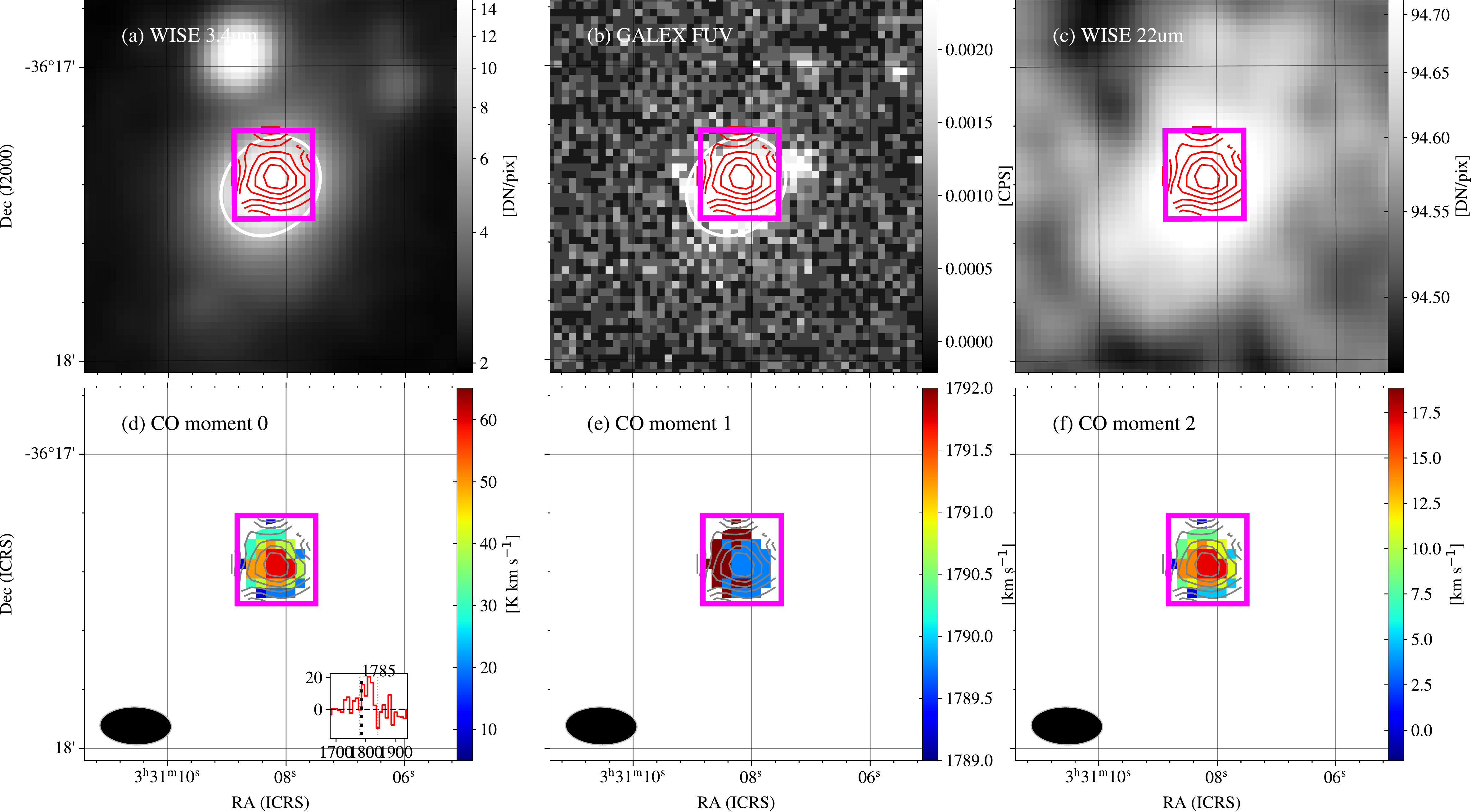}
\end{center}
\caption{
Same as Figure~\ref{fig:mommaps_ESO358-51}, but for MCG-06-08-024.
}
\label{fig:mommaps_MCG-06-08-024}
\end{figure*}

\begin{figure*}[]
\begin{center}
\includegraphics[width=0.92\textwidth, bb=0 0 1381 753]{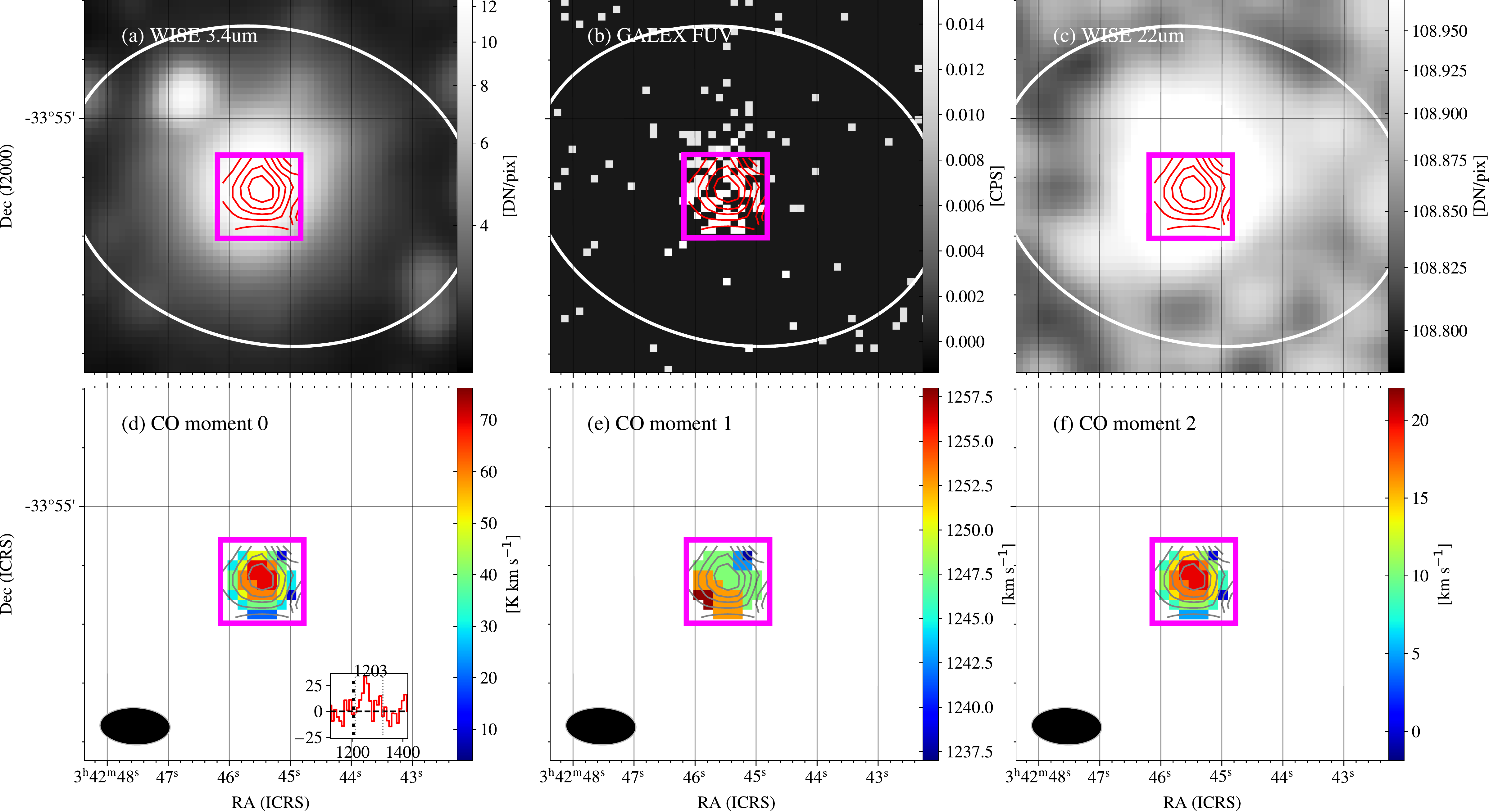}
\end{center}
\caption{
Same as Figure~\ref{fig:mommaps_ESO358-51}, but for MCG-06-09-023.
}
\label{fig:mommaps_MCG-06-09-023}
\end{figure*}

\begin{figure*}[]
\begin{center}
\includegraphics[width=0.92\textwidth, bb=0 0 1328 768]{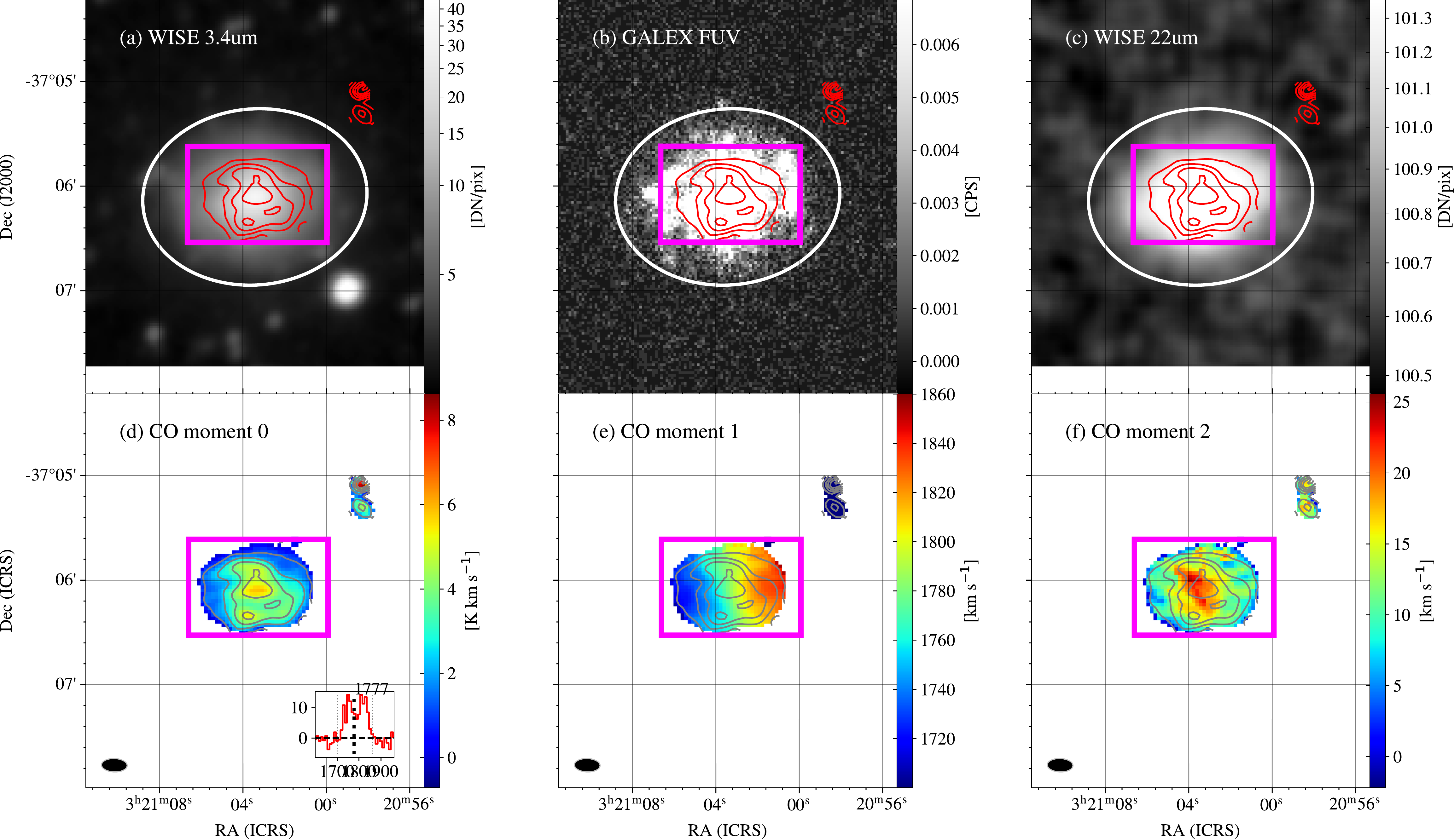}
\end{center}
\caption{
Same as Figure~\ref{fig:mommaps_ESO358-51}, but for NGC1310.
}
\label{fig:mommaps_NGC1310}
\end{figure*}

\begin{figure*}[]
\begin{center}
\includegraphics[width=0.92\textwidth, bb=0 0 1320 773]{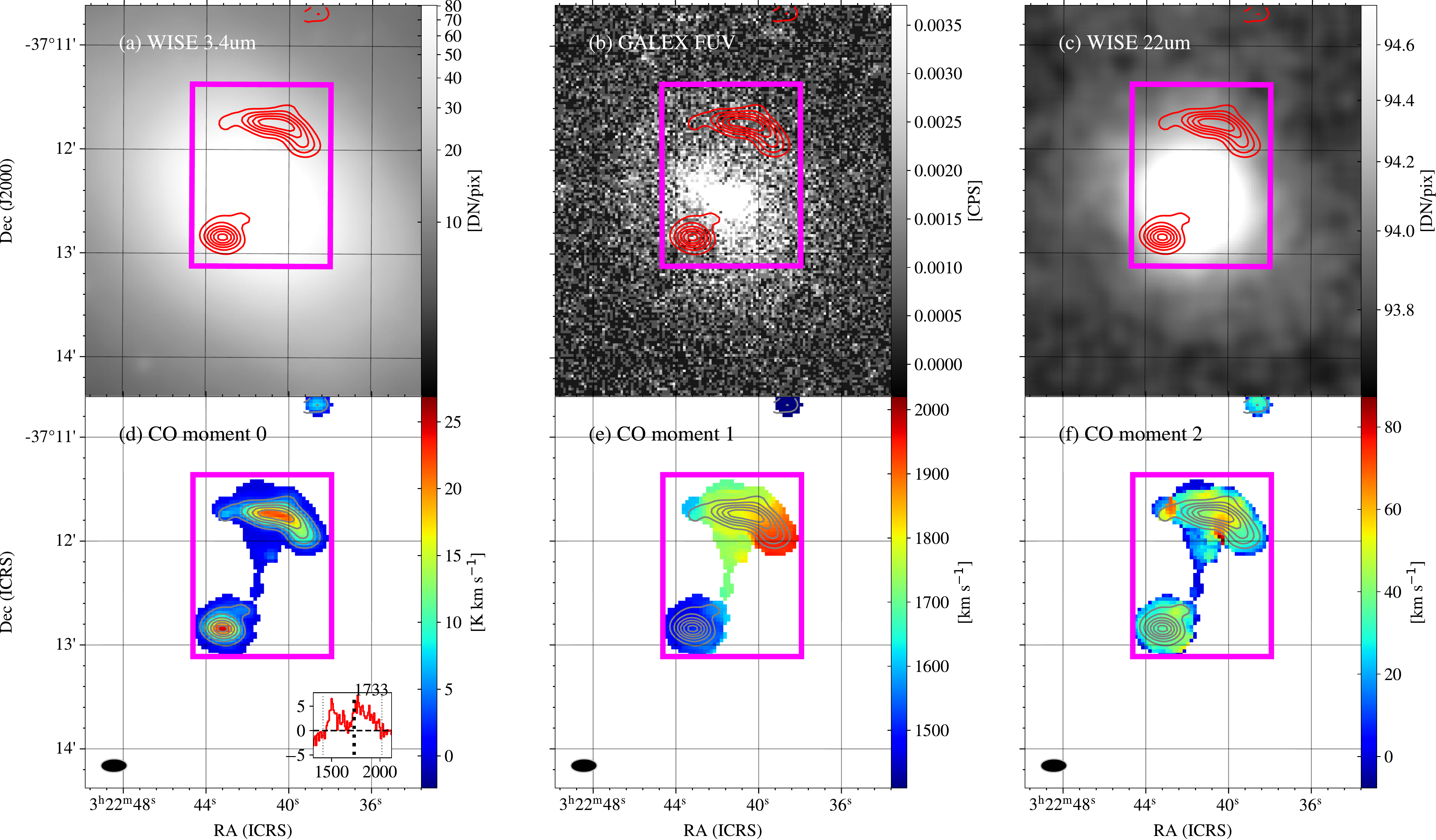}
\end{center}
\caption{
Same as Figure~\ref{fig:mommaps_ESO358-51}, but for NGC1316.
}
\label{fig:mommaps_NGC1316}
\end{figure*}

\begin{figure*}[]
\begin{center}
\includegraphics[width=0.92\textwidth, bb=0 0 1344 768]{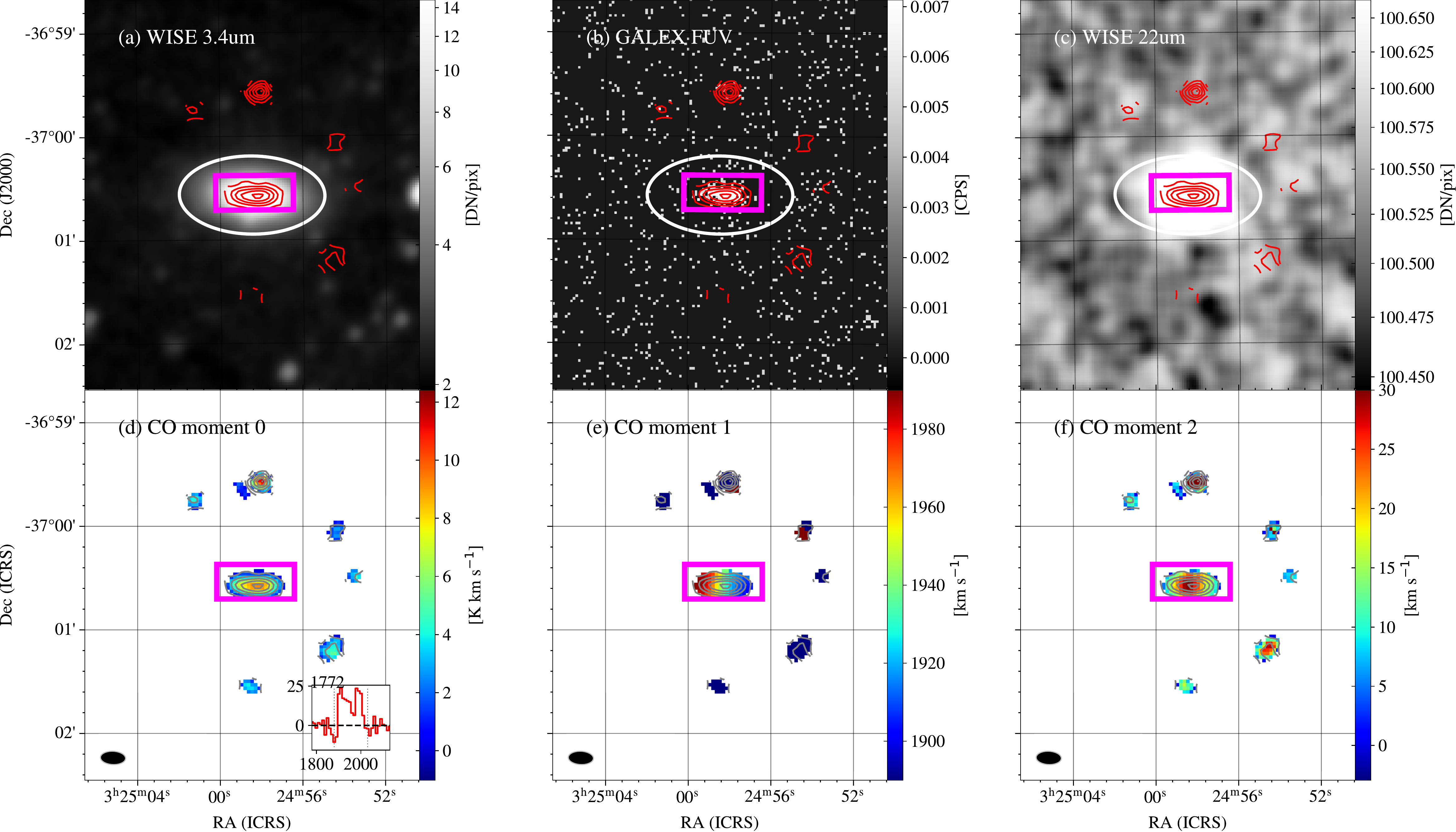}
\end{center}
\caption{
Same as Figure~\ref{fig:mommaps_ESO358-51}, but for NGC1316C.
}
\label{fig:mommaps_NGC1316C}
\end{figure*}

\begin{figure*}[]
\begin{center}
\includegraphics[width=0.92\textwidth, bb=0 0 1320 772]{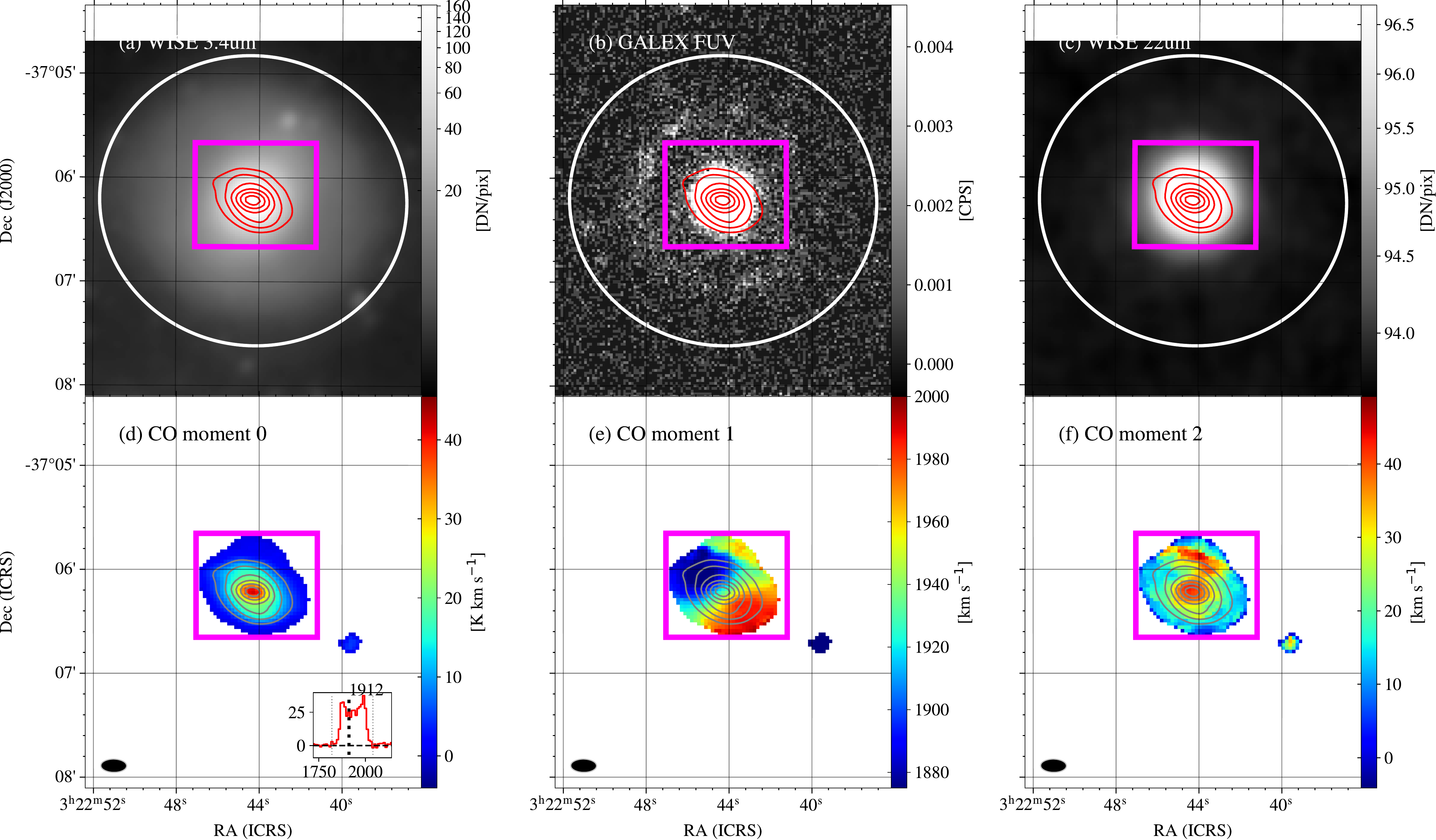}
\end{center}
\caption{
Same as Figure~\ref{fig:mommaps_ESO358-51}, but for NGC1317.
}
\label{fig:mommaps_NGC1317}
\end{figure*}

\begin{figure*}[]
\begin{center}
\includegraphics[width=0.92\textwidth, bb=0 0 1328 773]{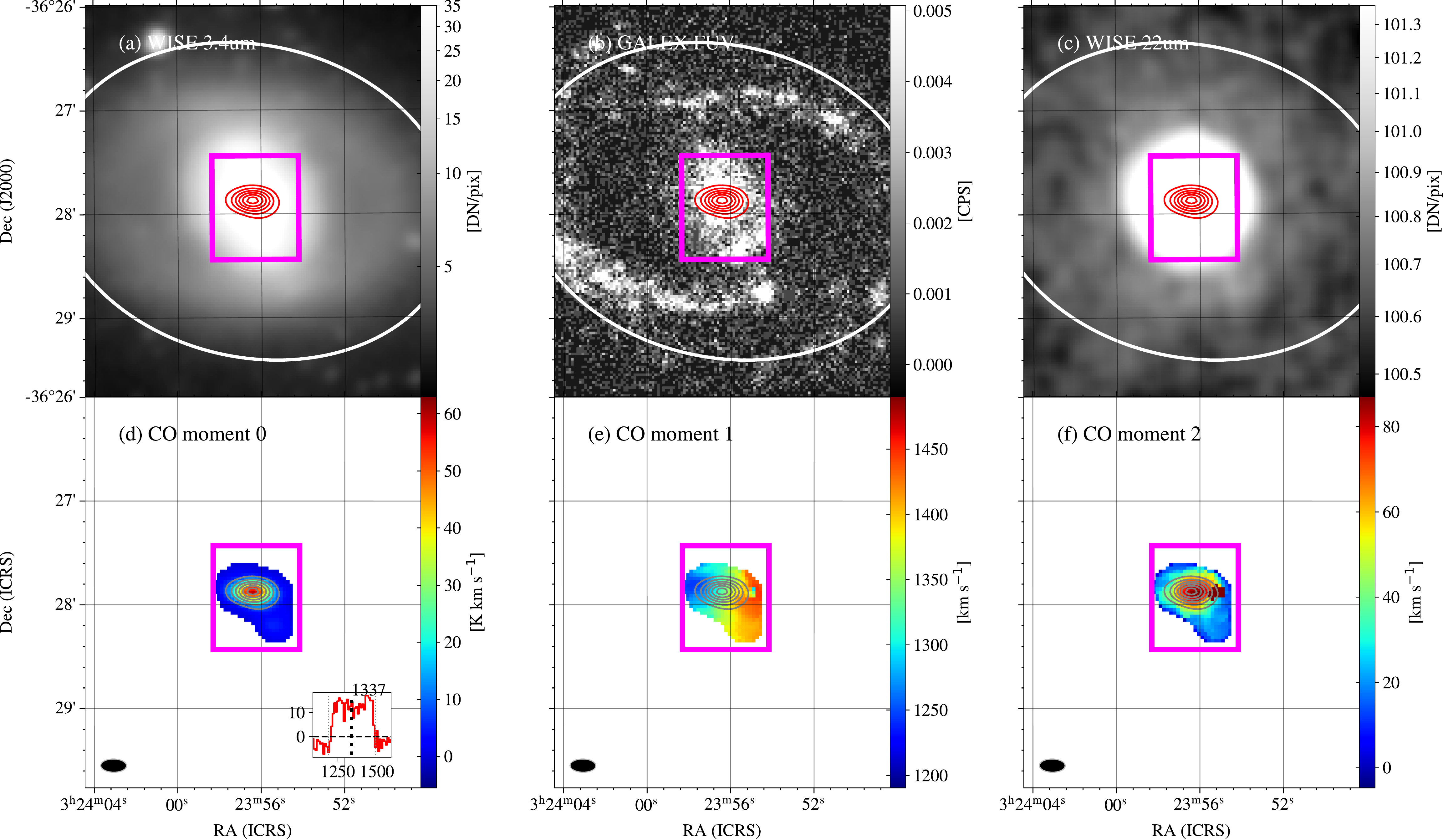}
\end{center}
\caption{
Same as Figure~\ref{fig:mommaps_ESO358-51}, but for NGC1326.
}
\label{fig:mommaps_NGC1326}
\end{figure*}

\begin{figure*}[]
\begin{center}
\includegraphics[width=0.92\textwidth, bb=0 0 1320 768]{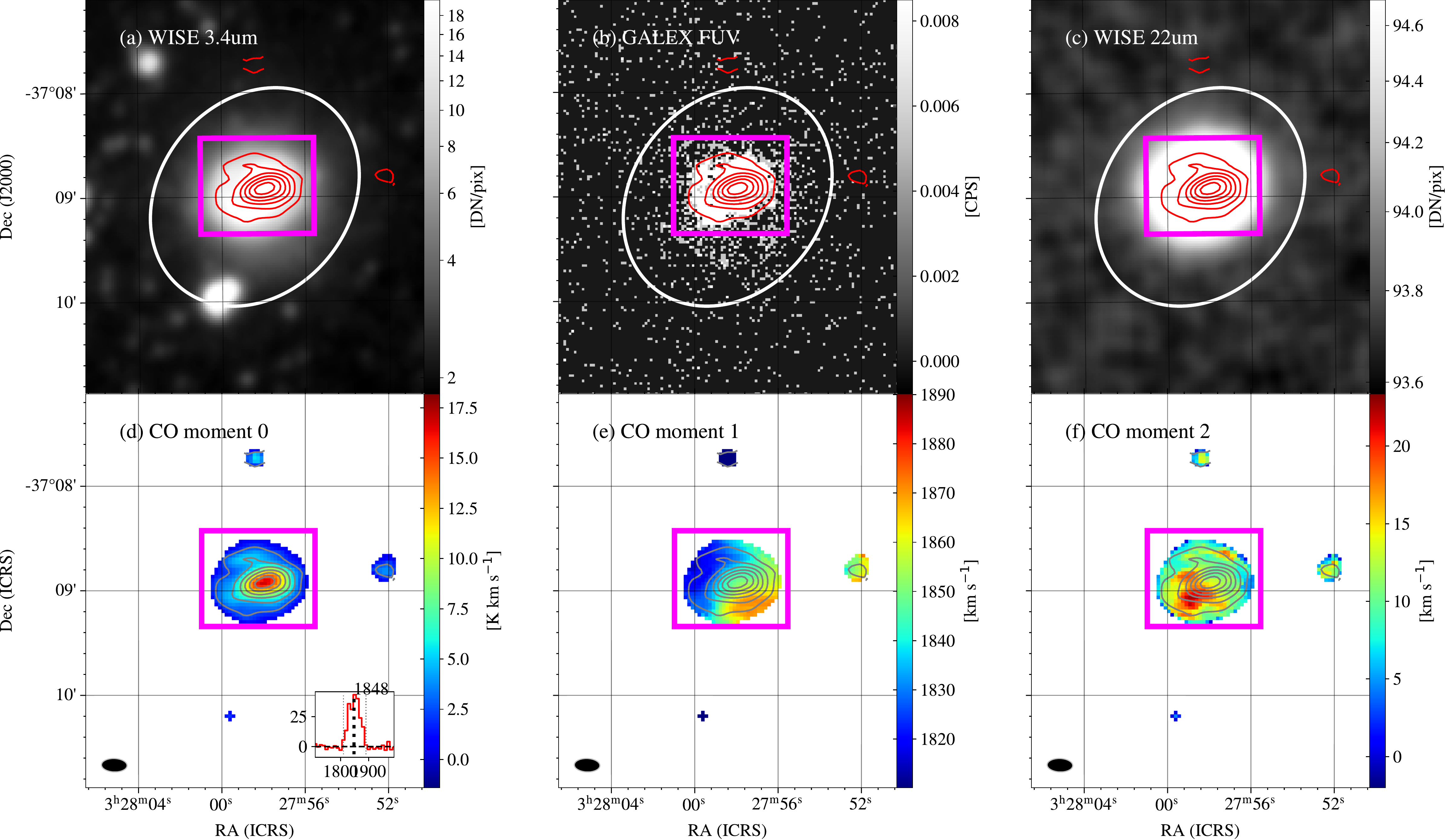}
\end{center}
\caption{
Same as Figure~\ref{fig:mommaps_ESO358-51}, but for NGC1341.
}
\label{fig:mommaps_NGC1341}
\end{figure*}

\begin{figure*}[]
\begin{center}
\includegraphics[width=0.92\textwidth, bb=0 0 1334 770]{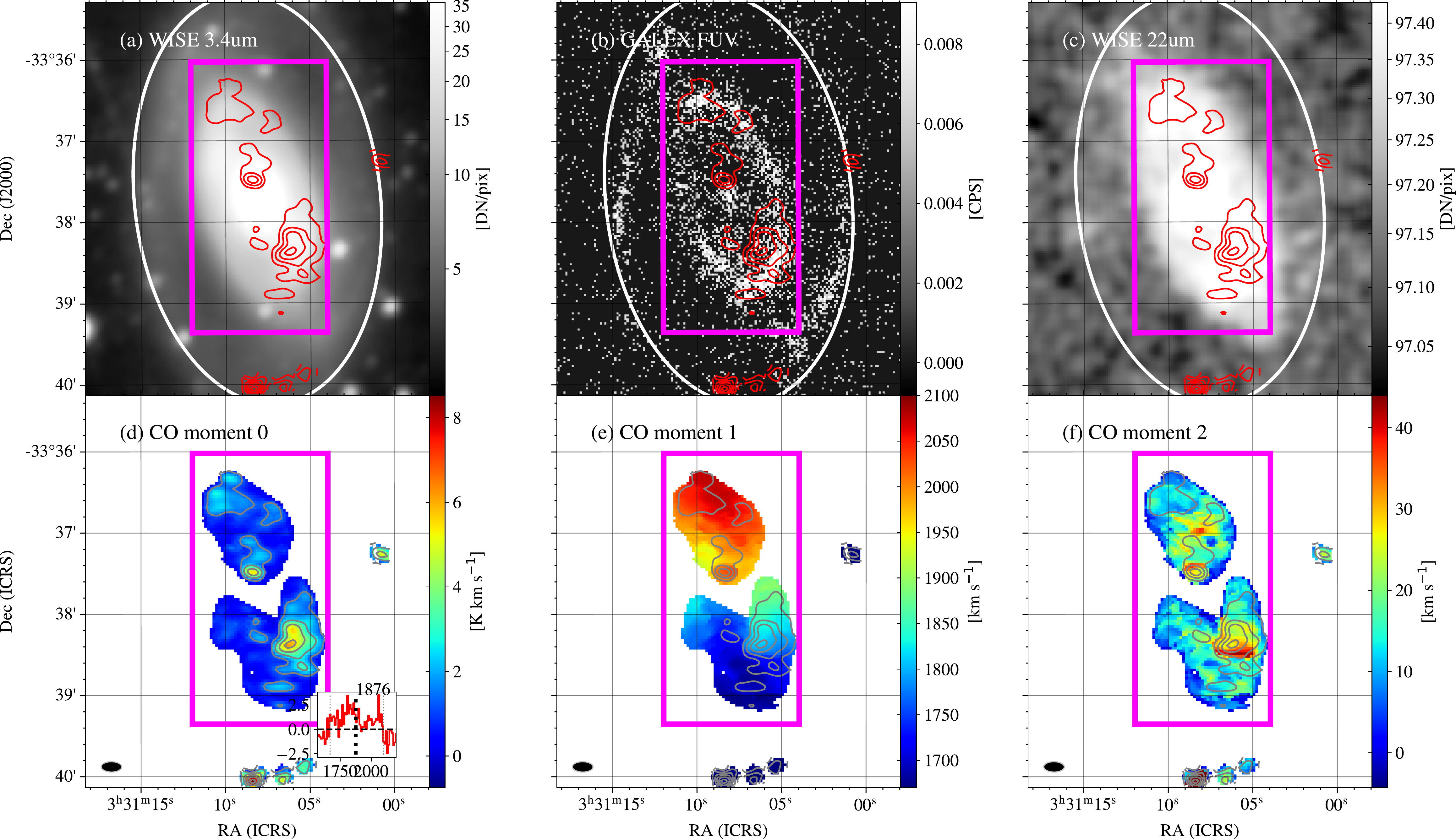}
\end{center}
\caption{
Same as Figure~\ref{fig:mommaps_ESO358-51}, but for NGC1350.
}
\label{fig:mommaps_NGC1350}
\end{figure*}

\begin{figure*}[]
\begin{center}
\includegraphics[width=0.92\textwidth, bb=0 0 1328 768]{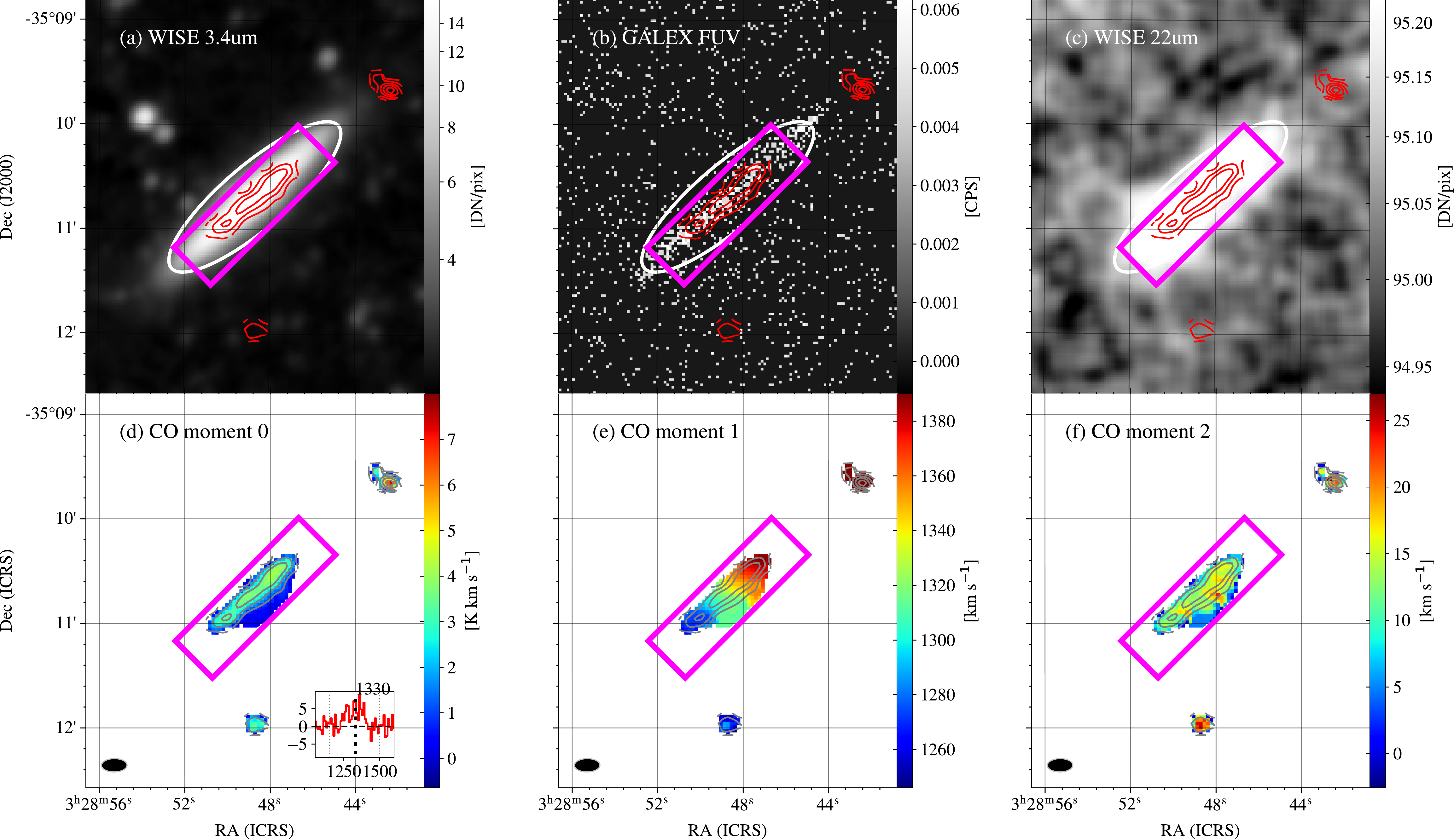}
\end{center}
\caption{
Same as Figure~\ref{fig:mommaps_ESO358-51}, but for NGC1351A.
}
\label{fig:mommaps_NGC1351A}
\end{figure*}

\begin{figure*}[]
\begin{center}
\includegraphics[width=0.92\textwidth, bb=0 0 1280 771]{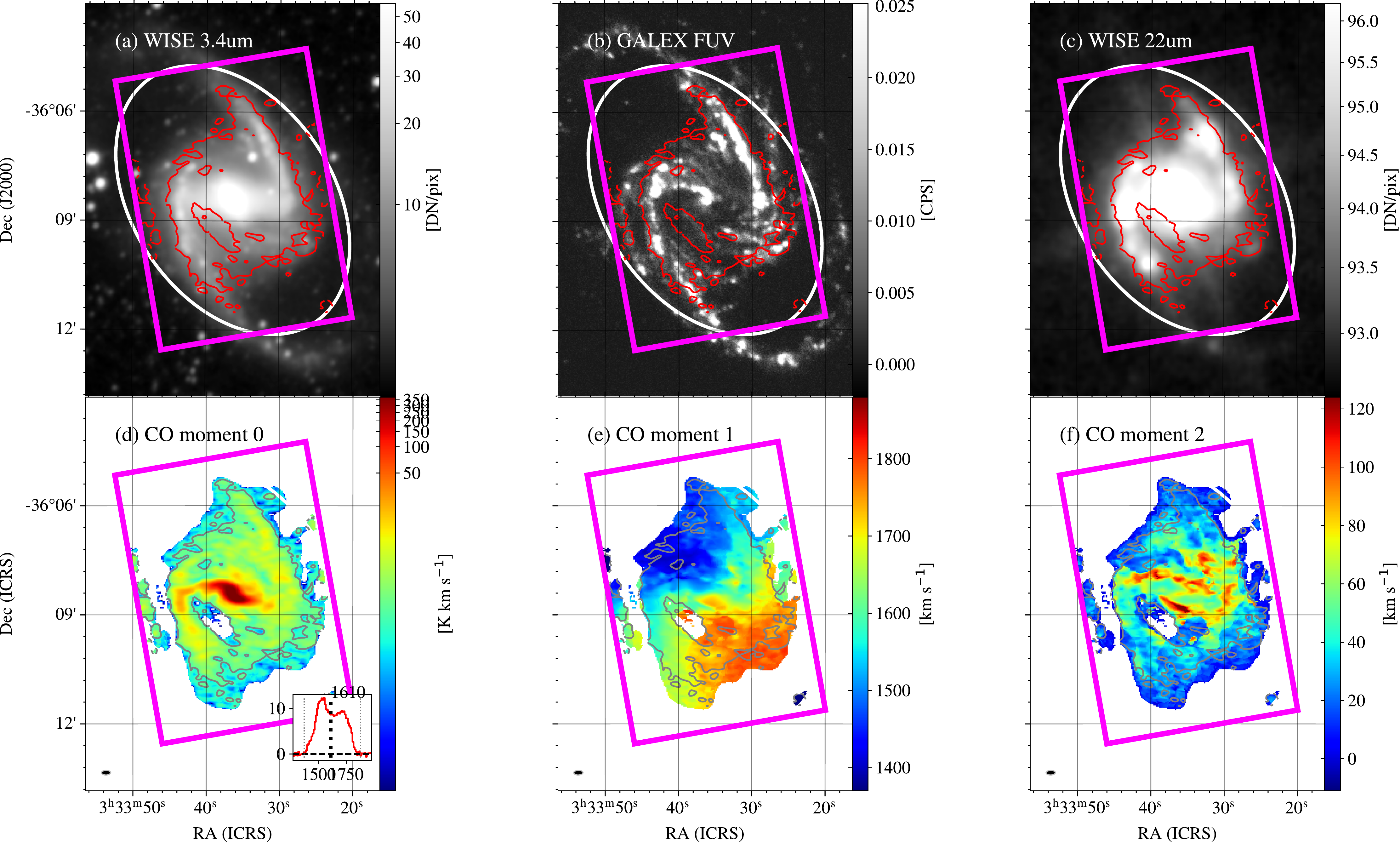}
\end{center}
\caption{
Same as Figure~\ref{fig:mommaps_ESO358-51}, but for NGC1365.
}
\label{fig:mommaps_NGC1365}
\end{figure*}

\begin{figure*}[]
\begin{center}
\includegraphics[width=0.92\textwidth, bb=0 0 1365 753]{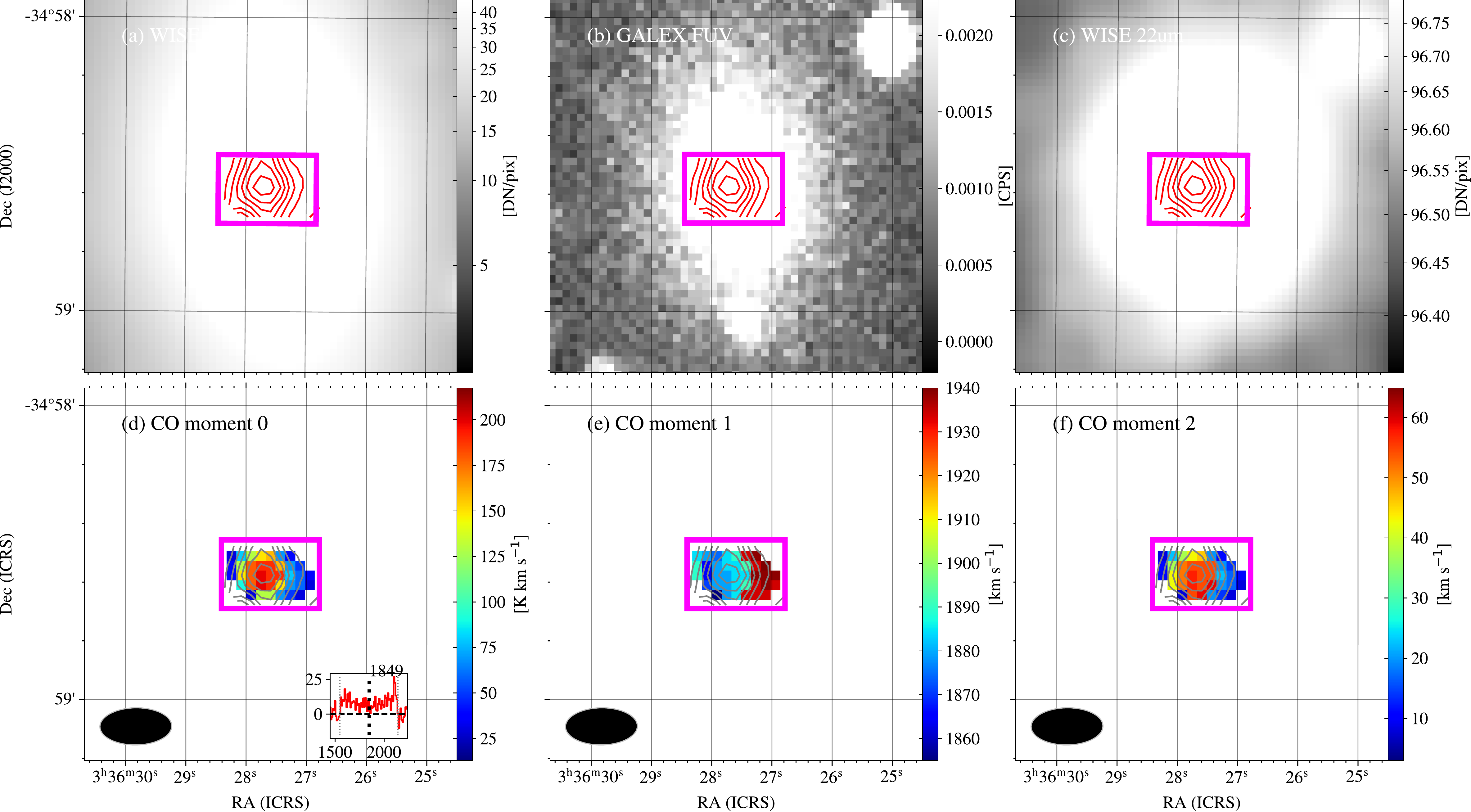}
\end{center}
\caption{
Same as Figure~\ref{fig:mommaps_ESO358-51}, but for NGC1380.
}
\label{fig:mommaps_NGC1380}
\end{figure*}

\begin{figure*}[]
\begin{center}
\includegraphics[width=0.92\textwidth, bb=0 0 1320 773]{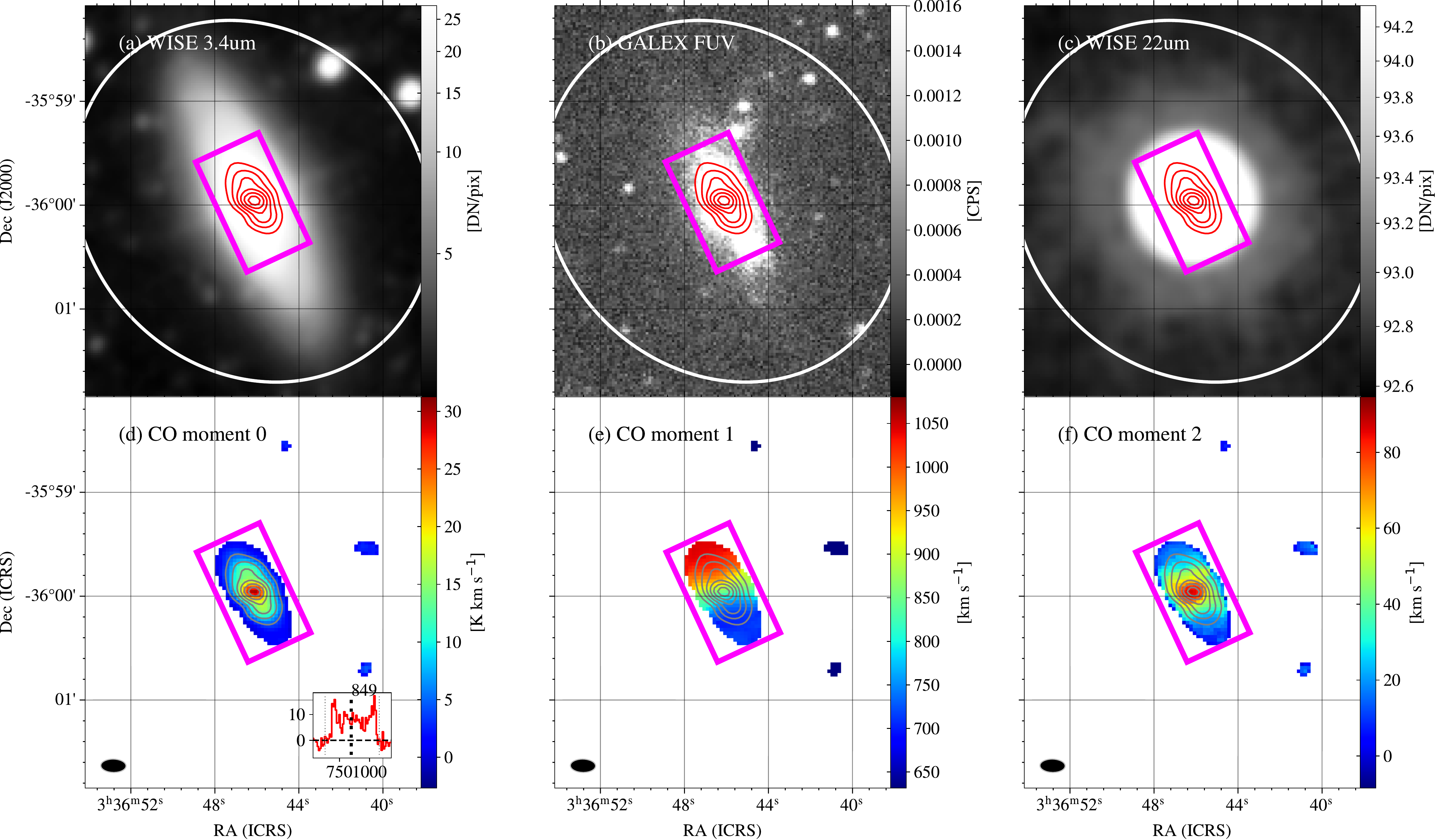}
\end{center}
\caption{
Same as Figure~\ref{fig:mommaps_ESO358-51}, but for NGC1386.
}
\label{fig:mommaps_NGC1386}
\end{figure*}

\begin{figure*}[]
\begin{center}
\includegraphics[width=0.92\textwidth, bb=0 0 1328 768]{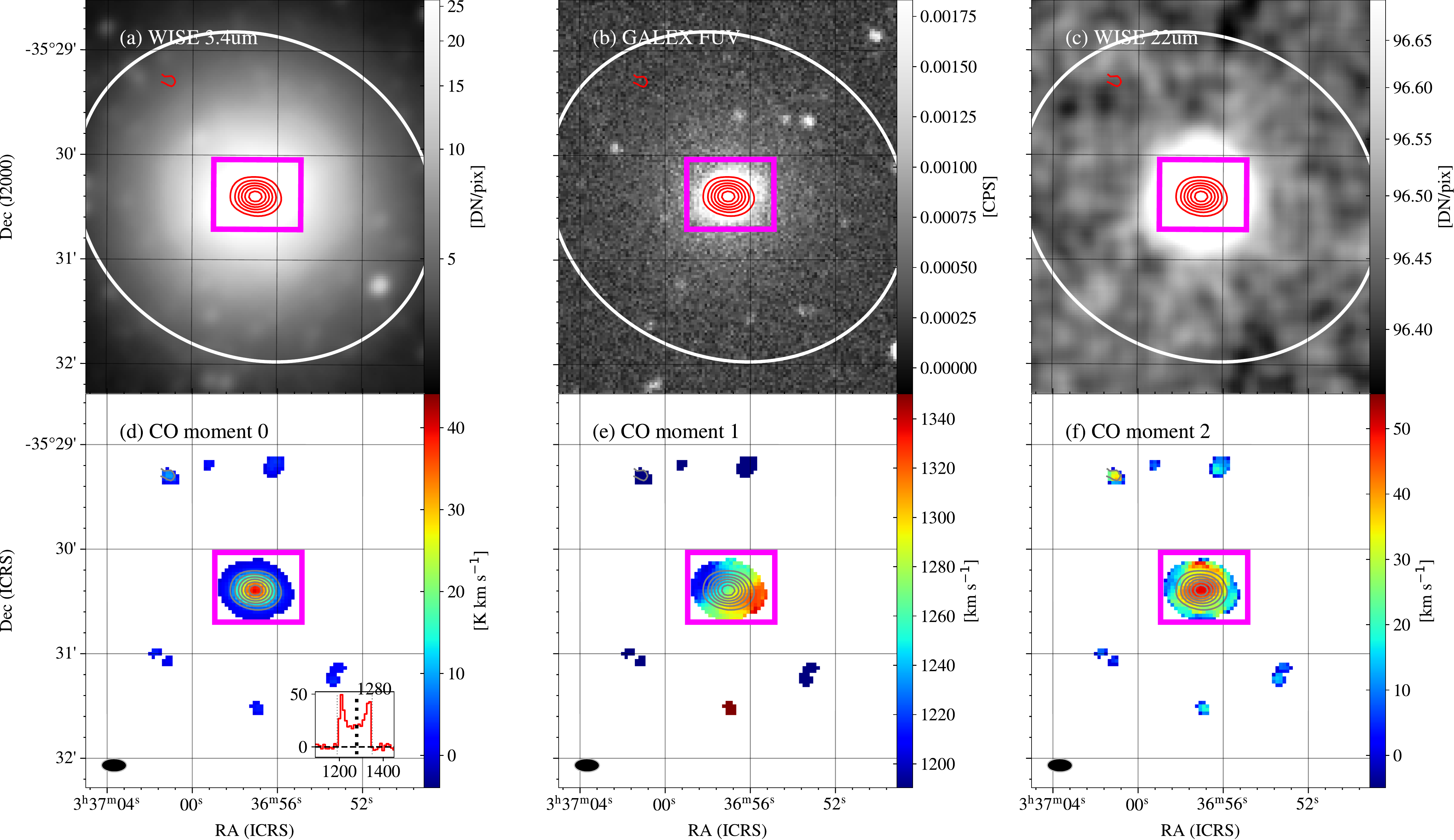}
\end{center}
\caption{
Same as Figure~\ref{fig:mommaps_ESO358-51}, but for NGC1387.
}
\label{fig:mommaps_NGC1387}
\end{figure*}

\begin{figure*}[]
\begin{center}
\includegraphics[width=0.92\textwidth, bb=0 0 1328 769]{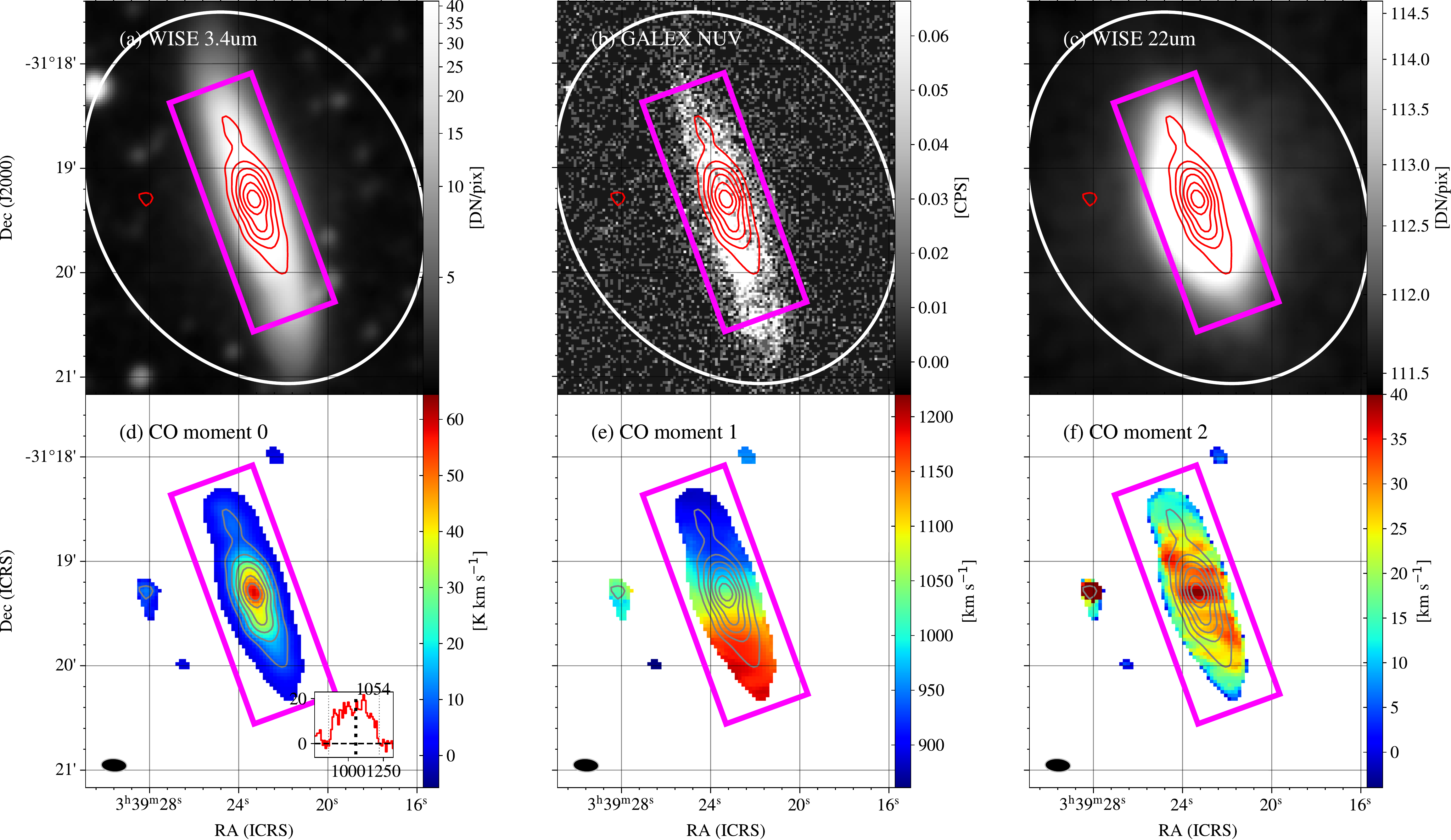}
\end{center}
\caption{
Same as Figure~\ref{fig:mommaps_ESO358-51}, but for NGC1406.
}
\label{fig:mommaps_NGC1406}
\end{figure*}

\begin{figure*}[]
\begin{center}
\includegraphics[width=0.92\textwidth, bb=0 0 1316 768]{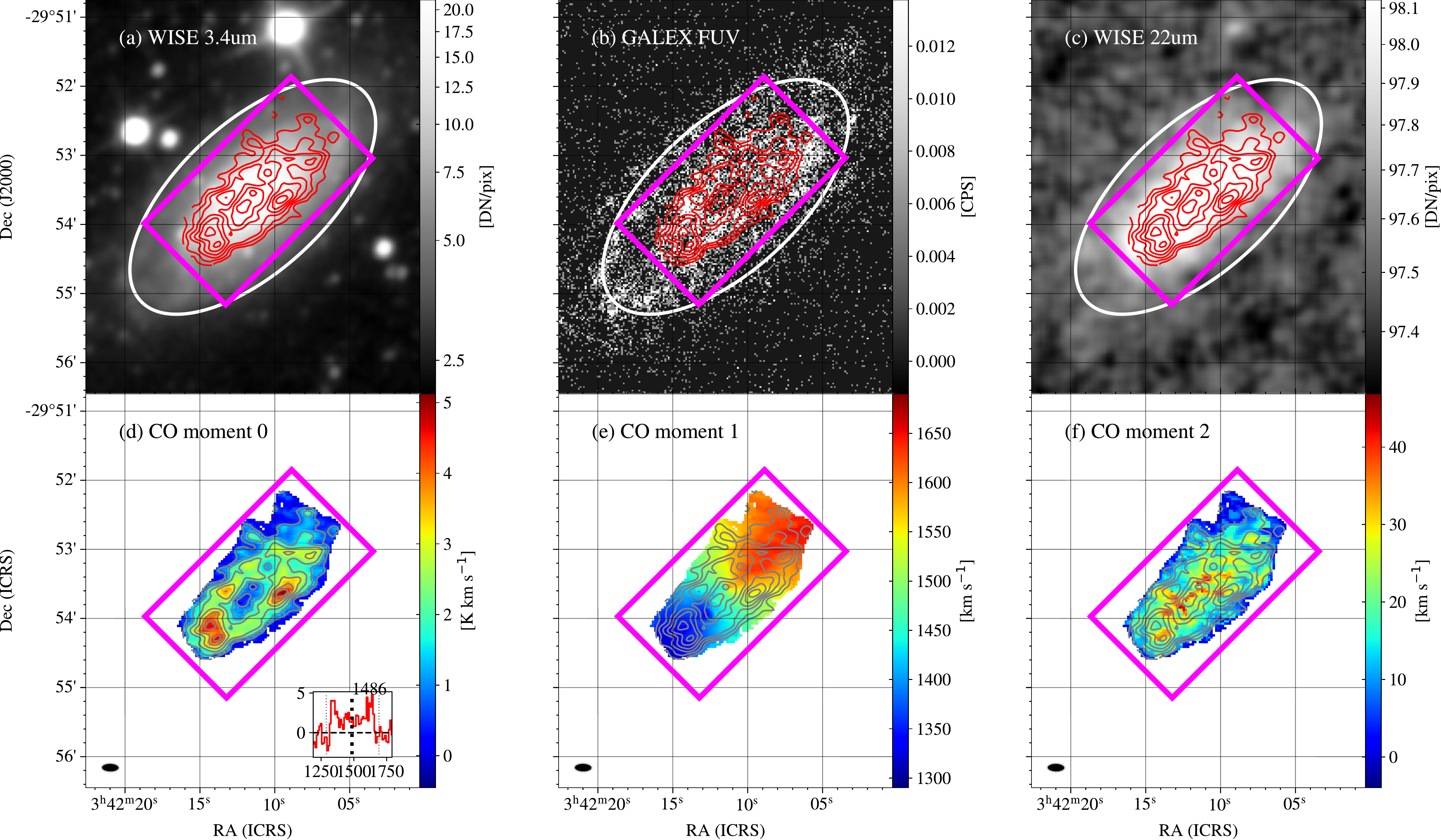}
\end{center}
\caption{
Same as Figure~\ref{fig:mommaps_ESO358-51}, but for NGC1425.
}
\label{fig:mommaps_NGC1425}
\end{figure*}

\begin{figure*}[]
\begin{center}
\includegraphics[width=0.92\textwidth, bb=0 0 1328 768]{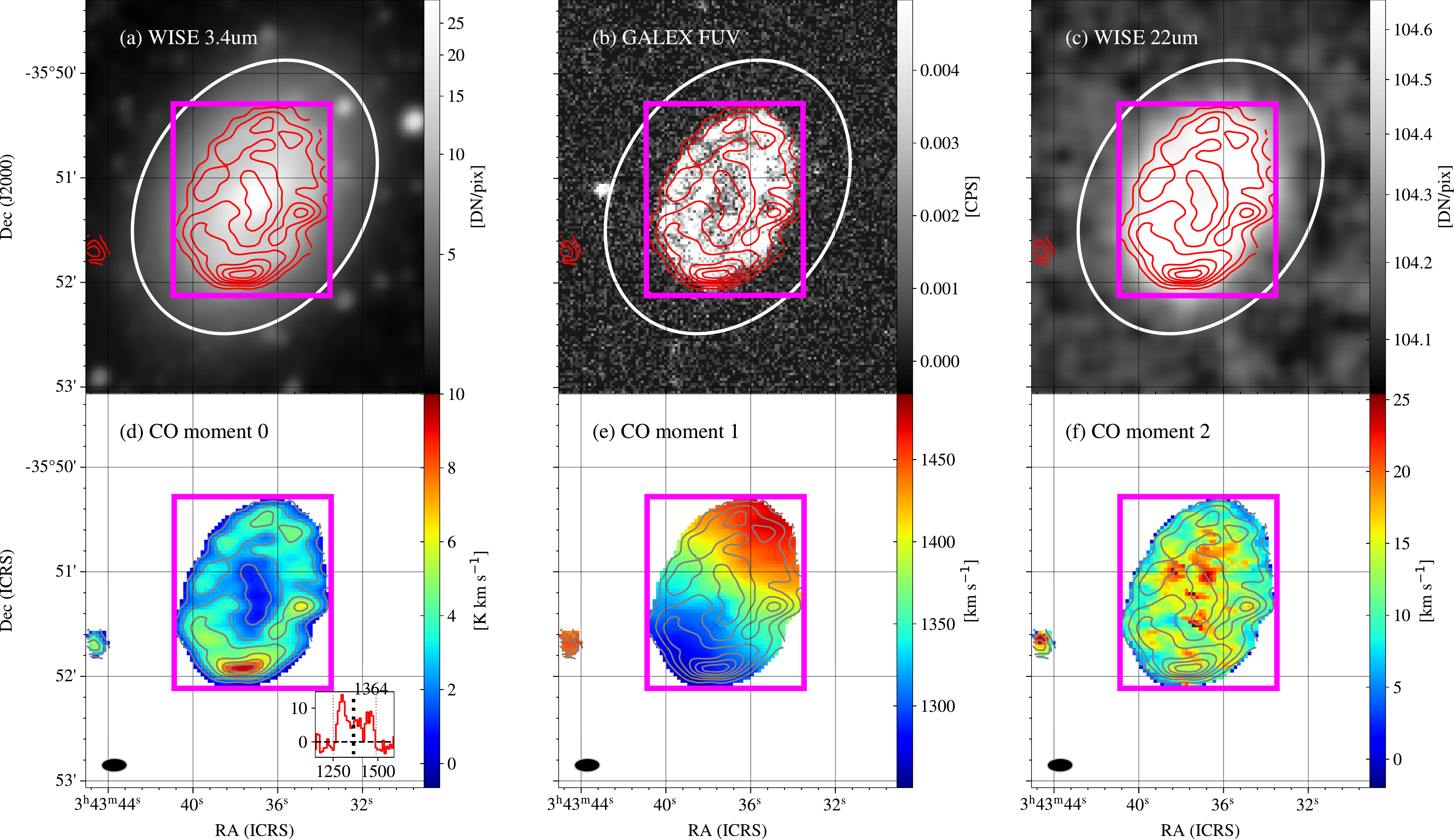}
\end{center}
\caption{
Same as Figure~\ref{fig:mommaps_ESO358-51}, but for NGC1436.
}
\label{fig:mommaps_NGC1436}
\end{figure*}

\begin{figure*}[]
\begin{center}
\includegraphics[width=0.92\textwidth, bb=0 0 1336 768]{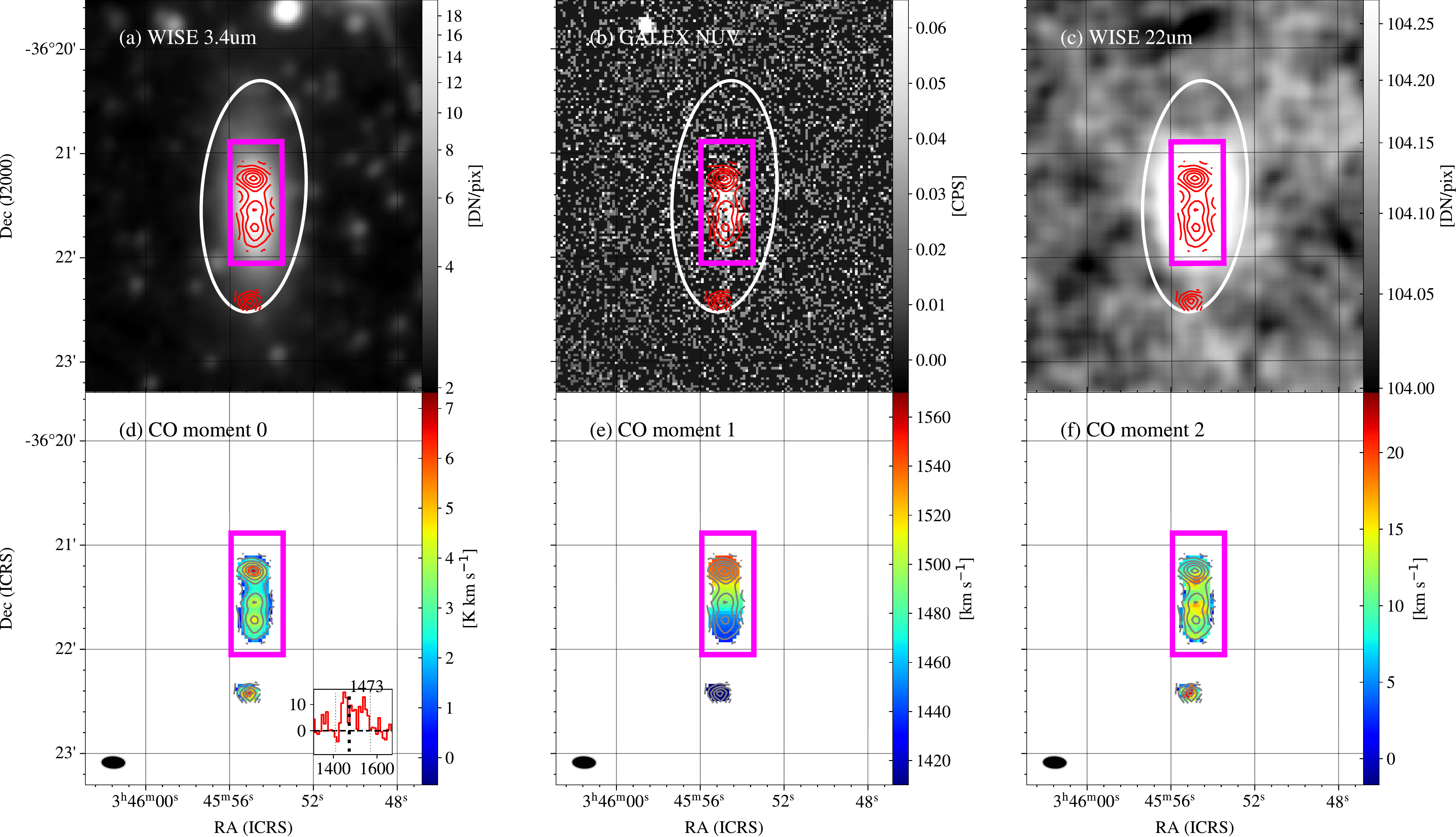}
\end{center}
\caption{
Same as Figure~\ref{fig:mommaps_ESO358-51}, but for NGC1437B.
}
\label{fig:mommaps_NGC1437B}
\end{figure*}

\newpage
\startlongtable
\begin{longrotatetable}
\begin{deluxetable*}{lcccccccc}
\tablecaption{Parameters for ALMA data analysis \label{tab:alma}}
\tablehead{
\tableline\tableline
\colhead{Name} & \colhead{SB ID\tablenotemark{a}} & \colhead{$b_{\rm maj}; b_{\rm min}; b_{\rm P.A.}$} & \colhead{Region shape\tablenotemark{b}} & \colhead{Center\tablenotemark{c}} & \colhead{width/$a$\tablenotemark{d}} & \colhead{height/$b$\tablenotemark{e}} & \colhead{P.A.\tablenotemark{f}} & \colhead{Velocity range\tablenotemark{g}}\\ 
\nocolhead{} & \nocolhead{} & \colhead{$''; ''; ^\circ$} & \nocolhead{} & \colhead{deg., deg.} & \colhead{arcsec} & \colhead{arcsec} & \colhead{deg.} & \colhead{km~s$^{-1}$}
}
\startdata
ESO302-14 & $1$ & $14.5; 7.7; 87.1$ & Ellipse & $57.920417, -38.452222$ & $35.0$ & $17.2$ & $67.9$ & $701-1001$ \\ 
ESO302-9 & $1$ & $14.8; 7.7; 87.8$ & Ellipse & $56.892167, -38.576528$ & $70.5$ & $27.8$ & $128.6$ & $821-1120$ \\ 
ESO357-25 & $1$ & $14.6; 7.7; 87.4$ & Ellipse & $50.90525, -35.778389$ & $27.9$ & $10.4$ & $117.2$ & $1650-1950$ \\ 
ESO358-16 & $2$ & $14.4; 7.6; 89.1$ & Ellipse & $53.288327, -35.718527$ & $21.1$ & $8.4$ & $66.2$ & $1540-1840$ \\ 
ESO358-20 & $1$ & $14.7; 7.7; 86.8$ & Ellipse & $53.738958, -32.639833$ & $40.4$ & $22.6$ & $55.0$ & $1590-1890$ \\ 
ESO358-5 & $1$ & $14.8; 7.6; 87.5$ & Ellipse & $51.819208, -33.486361$ & $43.4$ & $31.3$ & $12.5$ & $1450-1750$ \\ 
ESO358-51 & $1+2$ & $14.3; 7.6; 88.6$ & Rectangle & $55.385833, -34.88972188888889$ & $30.0$ & $30.0$ & $0.0$ & $1650-1740$ \\ 
ESO358-G015 & $2$ & $14.4; 7.6; 88.9$ & Ellipse & $53.278542, -34.808111$ & $27.1$ & $17.6$ & $84.4$ & $1210-1510$ \\ 
ESO358-G063 & $2$ & $14.4; 7.6; 88.9$ & Rectangle & $56.579208, -34.943556$ & $140.0$ & $35.0$ & $45.0$ & $1750-2060$ \\ 
ESO359-2 & $2$ & $14.2; 7.6; 88.1$ & Rectangle & $57.653319777777774, -35.909333$ & $18.0$ & $14.0$ & $0.0$ & $1410-1480$ \\ 
ESO359-3 & $2$ & $14.6; 7.7; 85.8$ & Ellipse & $58.003833, -33.467639$ & $56.0$ & $29.1$ & $44.0$ & $1400-1700$ \\ 
ESO359-5 & $2$ & $14.6; 7.6; 87.6$ & Ellipse & $58.514244, -36.063653$ & $27.0$ & $17.3$ & $45.3$ & $1210-1510$ \\ 
FCC102 & $2$ & $14.7; 7.7; -89.1$ & Ellipse & $53.04474, -36.220809$ & $19.1$ & $11.7$ & $59.5$ & $1550-1850$ \\ 
FCC117 & $2$ & $14.6; 7.8; 88.5$ & Ellipse & $53.31106, -37.819638$ & $6.5$ & $6.5$ & $45.0$ & $1340-1640$ \\ 
FCC120 & $2$ & $14.6; 7.7; -89.5$ & Ellipse & $53.392593, -36.605914$ & $21.4$ & $8.2$ & $132.9$ & $731-1031$ \\ 
FCC177 & $2$ & $14.6; 7.7; -89.2$ & Ellipse & $54.197875, -34.739611$ & $63.2$ & $35.7$ & $82.1$ & $1390-1690$ \\ 
FCC198 & $2$ & $14.5; 7.7; 88.5$ & Ellipse & $54.427823, -37.208339$ & $6.5$ & $6.5$ & $45.0$ & $1340-1640$ \\ 
FCC206 & $2$ & $14.5; 7.7; 88.5$ & Ellipse & $54.556202, -37.29018$ & $49.2$ & $26.2$ & $-24.3$ & $1230-1530$ \\ 
FCC207 & $2$ & $14.4; 7.6; 89.0$ & Ellipse & $54.580292, -35.129083$ & $15.0$ & $11.5$ & $-17.6$ & $1250-1550$ \\ 
FCC261 & $2$ & $14.4; 7.6; 88.5$ & Ellipse & $55.339708, -33.769222$ & $64.2$ & $22.2$ & $49.2$ & $1320-1620$ \\ 
FCC302 & $2$ & $14.4; 7.6; 88.8$ & Ellipse & $56.300593, -35.570906$ & $46.3$ & $9.7$ & $13.1$ & $632-931$ \\ 
FCC306 & $2$ & $14.5; 7.8; 88.0$ & Ellipse & $56.439158, -36.34653$ & $18.8$ & $13.9$ & $-35.4$ & $711-1011$ \\ 
FCC316 & $2$ & $14.3; 7.6; 88.6$ & Ellipse & $56.756333, -36.437472$ & $17.7$ & $10.0$ & $35.0$ & $1370-1670$ \\ 
FCC32 & $2$ & $14.7; 7.7; -88.9$ & Ellipse & $51.2185, -35.435444$ & $22.0$ & $18.9$ & $72.8$ & $1150-1450$ \\ 
FCC332 & $2$ & $14.3; 7.6; 88.8$ & Rectangle & $57.45494444444444, -35.94669411111111$ & $15.0$ & $12.0$ & $0.0$ & $1260-1330$ \\ 
FCC44 & $2$ & $14.7; 7.7; -89.1$ & Ellipse & $51.531017, -35.127491$ & $11.8$ & $8.3$ & $33.6$ & $1080-1380$ \\ 
IC1993 & $1$ & $14.8; 7.6; 88.0$ & Rectangle & $56.770042, -33.71124988888889$ & $115.0$ & $115.0$ & $0.0$ & $1001-1130$ \\ 
IC335 & $1$ & $14.7; 7.6; 87.2$ & Ellipse & $53.879333, -34.447056$ & $79.7$ & $35.9$ & $4.2$ & $1440-1740$ \\ 
MCG-06-08-024 & $2$ & $14.6; 7.7; 88.4$ & Rectangle & $52.78469477777777, -36.28986122222223$ & $16.0$ & $18.0$ & $0.0$ & $1780-1840$ \\ 
MCG-06-09-023 & $2$ & $14.3; 7.6; 87.9$ & Rectangle & $55.690083, -33.92147222222222$ & $17.0$ & $17.0$ & $0.0$ & $1210-1320$ \\ 
NGC1310 & $1$ & $14.8; 7.7; 88.0$ & Rectangle & $50.264292, -37.101694$ & $80.0$ & $55.0$ & $0.0$ & $1700-1860$ \\ 
NGC1316 & $1$ & $15.2; 7.7; -89.3$ & Rectangle & $50.6727, -37.2045$ & $80.0$ & $105.0$ & $0.0$ & $1410-2020$ \\ 
NGC1316C & $1$ & $14.6; 7.7; 87.4$ & Rectangle & $51.243583, -37.009472$ & $45.0$ & $20.0$ & $0.0$ & $1880-2030$ \\ 
NGC1317 & $1$ & $14.9; 7.6; 89.2$ & Rectangle & $50.684527, -37.10313244444444$ & $70.0$ & $60.0$ & $0.0$ & $1820-2040$ \\ 
NGC1326 & $1$ & $14.9; 7.7; 89.2$ & Rectangle & $50.985, -36.46605588888889$ & $50.0$ & $60.0$ & $0.0$ & $1190-1490$ \\ 
NGC1326A & $1$ & $14.7; 7.7; 88.0$ & Ellipse & $51.285405, -36.363949$ & $62.3$ & $27.7$ & $12.7$ & $1650-1950$ \\ 
NGC1326B & $1$ & $14.9; 7.7; 88.9$ & Ellipse & $51.334782, -36.384954$ & $91.9$ & $29.8$ & $32.7$ & $831-1130$ \\ 
NGC1336 & $1$ & $14.6; 7.7; 87.5$ & Ellipse & $51.634125, -35.713556$ & $76.2$ & $52.5$ & $109.2$ & $1240-1540$ \\ 
NGC1339 & $1$ & $14.8; 7.6; 87.1$ & Ellipse & $52.027417, -32.286111$ & $75.2$ & $59.2$ & $49.9$ & $1220-1520$ \\ 
NGC1341 & $1$ & $14.7; 7.7; 87.9$ & Rectangle & $51.993417, -37.14861111111111$ & $65.0$ & $55.0$ & $0.0$ & $1810-1890$ \\ 
NGC1350 & $1$ & $15.4; 7.7; -89.2$ & Rectangle & $52.783833, -33.628639$ & $100.0$ & $200.0$ & $0.0$ & $1670-2100$ \\ 
NGC1351 & $1$ & $14.8; 7.6; 87.7$ & Ellipse & $52.64575, -34.853944$ & $110.1$ & $77.8$ & $53.3$ & $1340-1640$ \\ 
NGC1351A & $2$ & $14.7; 7.6; -89.4$ & Rectangle & $52.20355555555556, -35.17980566666667$ & $100.0$ & $30.0$ & $45.0$ & $1150-1500$ \\ 
NGC1365 & $2$ & $15.5; 7.6; -88.6$ & Rectangle & $53.401548, -36.140402$ & $320.0$ & $450.0$ & $10.0$ & $1370-1880$ \\ 
NGC1374 & $1$ & $14.8; 7.6; 88.1$ & Ellipse & $53.819125, -35.22625$ & $78.3$ & $74.5$ & $63.0$ & $1120-1420$ \\ 
NGC1375 & $1$ & $14.6; 7.6; 87.1$ & Ellipse & $53.820083, -35.265667$ & $54.1$ & $27.0$ & $0.2$ & $572-871$ \\ 
NGC1379 & $1$ & $14.6; 7.6; 87.6$ & Ellipse & $54.016458, -35.441194$ & $90.3$ & $82.7$ & $-4.2$ & $1150-1450$ \\ 
NGC1380 & $2$ & $14.7; 7.6; -89.3$ & Rectangle & $54.11566244444444, -34.97678055555556$ & $20.0$ & $14.0$ & $0.0$ & $1550-2141$ \\ 
NGC1381 & $1$ & $14.7; 7.6; 87.5$ & Ellipse & $54.132, -35.295194$ & $81.8$ & $41.3$ & $51.8$ & $1550-1850$ \\ 
NGC1386 & $2$ & $14.6; 7.7; 88.8$ & Rectangle & $54.19298055555556, -35.99996355555556$ & $40.0$ & $70.0$ & $25.0$ & $632-1080$ \\ 
NGC1387 & $2$ & $14.4; 7.6; 89.2$ & Rectangle & $54.23775, -35.506639$ & $50.0$ & $40.0$ & $0.0$ & $1190-1350$ \\ 
NGC1399 & $1$ & $14.8; 7.6; 88.5$ & Ellipse & $54.620941, -35.450657$ & $271.1$ & $225.9$ & $-15.5$ & $1250-1550$ \\ 
NGC1404 & $1$ & $14.7; 7.6; 87.6$ & Ellipse & $54.716333, -35.594389$ & $115.0$ & $88.6$ & $72.4$ & $1770-2070$ \\ 
NGC1406 & $1$ & $14.7; 7.6; 86.7$ & Rectangle & $54.84791633333333, -31.32252811111111$ & $50.0$ & $140.0$ & $20.0$ & $861-1220$ \\ 
NGC1419 & $1$ & $14.5; 7.6; 86.8$ & Ellipse & $55.175458, -37.510833$ & $53.0$ & $43.3$ & $-7.4$ & $1420-1720$ \\ 
NGC1425 & $1$ & $15.3; 7.5; 88.9$ & Rectangle & $55.54668088888889, -29.892221888888887$ & $180.0$ & $100.0$ & $45.0$ & $1290-1690$ \\ 
NGC1427 & $1$ & $14.7; 7.6; 87.6$ & Ellipse & $55.580933, -35.392563$ & $125.0$ & $106.8$ & $-19.1$ & $1210-1510$ \\ 
NGC1427A & $2$ & $14.4; 7.6; 89.2$ & Ellipse & $55.03875, -35.624444$ & $87.5$ & $55.6$ & $-18.7$ & $1850-2151$ \\ 
NGC1436 & $2$ & $14.7; 7.7; -89.0$ & Rectangle & $55.90561111111111, -35.853861333333334$ & $90.0$ & $110.0$ & $0.0$ & $1250-1490$ \\ 
NGC1437A & $2$ & $14.6; 7.7; 88.5$ & Ellipse & $55.75914, -36.273374$ & $52.7$ & $44.4$ & $17.5$ & $711-1011$ \\ 
NGC1437B & $2$ & $14.2; 7.7; 87.9$ & Rectangle & $56.478542, -36.35836088888889$ & $30.0$ & $70.0$ & $0.0$ & $1410-1570$ \\ 
NGC1484 & $1$ & $14.6; 7.6; 87.0$ & Ellipse & $58.583875, -36.968889$ & $72.2$ & $24.7$ & $-13.9$ & $871-1170$ \\ 
PGC012625 & $1$ & $14.8; 7.7; 89.0$ & Ellipse & $50.52, -37.5875$ & $34.0$ & $17.0$ & $100.0$ & $1450-1750$ \\ 
WOMBAT I & $1$ & $14.7; 7.7; 87.6$ & Ellipse & $55.245185, -38.855266$ & $23.5$ & $19.5$ & $37.4$ & $662-961$ \\ 
\enddata
\tablenotetext{a}{See Table~\ref{tab:sbs}.} \tablenotetext{b}{The region shape to generate the galactic CO spectra for calculating total molecular gas mass of galaxies. Rectangle and Ellipse are adopted for the CO-detected and CO-non-detected galaxies.}  \tablenotetext{c}{The region center for the galactic CO spectra.} \tablenotetext{d}{The width and semi-major axis for the region shapes of Rectangle and Ellipse, respectively.} \tablenotetext{e}{The height and semi-minor axis for the region shapes of Rectangle and Ellipse, respectively.} \tablenotetext{f}{The position angle of the region calculated from the North in a counterclockwise direction.} \tablenotetext{g}{The velocity range to calculate the CO integrated intensity. The range was determied by eye for the CO detected sources. $\sim300$~km~s$^{-1}$ is adopted for the CO-non-detected sources.}
\end{deluxetable*}
\end{longrotatetable}

\startlongtable
\begin{deluxetable*}{lccc}
\tablecaption{Parameters for CAAPR$^{a}$ \label{tab:caapr}}
\tablehead{
\tableline\tableline
\colhead{Name} & \colhead{Semi-major} & \colhead{Semi-minor} & \colhead{P.A.}\\ 
\nocolhead{} & \colhead{arcsec} & \colhead{arcsec} & \colhead{deg.}
}
\startdata
ESO302-14 & $70.0$ & $34.5$ & $67.9$ \\ 
ESO302-9 & $141.0$ & $55.6$ & $128.6$ \\ 
ESO357-25 & $55.8$ & $20.8$ & $117.2$ \\ 
ESO358-16 & $42.2$ & $16.8$ & $66.2$ \\ 
ESO358-20 & $80.8$ & $45.2$ & $55.0$ \\ 
ESO358-5 & $86.8$ & $62.6$ & $12.5$ \\ 
ESO358-51 & $80.5$ & $45.2$ & $94.5$ \\ 
ESO358-G015 & $54.2$ & $35.1$ & $84.4$ \\ 
ESO358-G063 & $266.7$ & $102.3$ & $38.6$ \\ 
ESO359-2 & $31.4$ & $26.3$ & $-35.0$ \\ 
ESO359-3 & $111.9$ & $58.2$ & $44.0$ \\ 
ESO359-5 & $54.0$ & $34.7$ & $45.3$ \\ 
FCC102 & $38.1$ & $23.4$ & $59.5$ \\ 
FCC117 & $13.0$ & $13.0$ & $45.0$ \\ 
FCC120 & $42.9$ & $16.4$ & $132.9$ \\ 
FCC177 & $126.5$ & $71.4$ & $82.1$ \\ 
FCC198 & $13.0$ & $13.0$ & $45.0$ \\ 
FCC206 & $98.5$ & $52.4$ & $-24.3$ \\ 
FCC207 & $29.9$ & $23.0$ & $-17.6$ \\ 
FCC261 & $128.4$ & $44.4$ & $49.2$ \\ 
FCC302 & $92.6$ & $19.3$ & $13.1$ \\ 
FCC306 & $37.6$ & $27.8$ & $-35.4$ \\ 
FCC316 & $35.5$ & $20.0$ & $35.0$ \\ 
FCC32 & $44.0$ & $37.9$ & $72.8$ \\ 
FCC332 & $52.6$ & $36.9$ & $16.4$ \\ 
FCC44 & $23.6$ & $16.6$ & $33.6$ \\ 
IC1993 & $112.0$ & $104.5$ & $126.4$ \\ 
IC335 & $159.5$ & $71.7$ & $4.2$ \\ 
MCG-06-08-024 & $22.0$ & $18.3$ & $45.0$ \\ 
MCG-06-09-023 & $84.5$ & $64.3$ & $-12.5$ \\ 
NGC1310 & $128.8$ & $101.0$ & $5.1$ \\ 
NGC1316 & $797.4$ & $557.0$ & $-34.9$ \\ 
NGC1316C & $84.4$ & $45.2$ & $-0.7$ \\ 
NGC1317 & $177.5$ & $167.3$ & $-8.5$ \\ 
NGC1326 & $236.6$ & $180.2$ & $-13.0$ \\ 
NGC1326A & $124.5$ & $55.4$ & $12.7$ \\ 
NGC1326B & $183.8$ & $59.7$ & $32.7$ \\ 
NGC1336 & $152.4$ & $105.0$ & $109.2$ \\ 
NGC1339 & $150.4$ & $118.5$ & $49.9$ \\ 
NGC1341 & $134.5$ & $109.4$ & $51.4$ \\ 
NGC1350 & $304.1$ & $181.5$ & $96.5$ \\ 
NGC1351 & $220.3$ & $155.6$ & $53.3$ \\ 
NGC1351A & $125.9$ & $36.5$ & $40.5$ \\ 
NGC1365 & $486.1$ & $330.9$ & $124.0$ \\ 
NGC1374 & $156.7$ & $149.0$ & $63.0$ \\ 
NGC1375 & $108.2$ & $54.1$ & $0.2$ \\ 
NGC1379 & $180.5$ & $165.4$ & $-4.2$ \\ 
NGC1380 & $329.1$ & $240.4$ & $98.6$ \\ 
NGC1381 & $163.5$ & $82.6$ & $51.8$ \\ 
NGC1386 & $221.1$ & $194.2$ & $133.2$ \\ 
NGC1387 & $211.2$ & $182.9$ & $-27.7$ \\ 
NGC1399 & $542.2$ & $451.7$ & $-15.5$ \\ 
NGC1404 & $230.0$ & $177.3$ & $72.4$ \\ 
NGC1406 & $226.0$ & $181.8$ & $123.5$ \\ 
NGC1419 & $106.0$ & $86.6$ & $-7.4$ \\ 
NGC1425 & $265.7$ & $128.2$ & $42.9$ \\ 
NGC1427 & $250.0$ & $213.7$ & $-19.1$ \\ 
NGC1427A & $175.0$ & $111.3$ & $-18.7$ \\ 
NGC1436 & $168.0$ & $127.0$ & $56.9$ \\ 
NGC1437A & $105.4$ & $88.7$ & $17.5$ \\ 
NGC1437B & $133.5$ & $59.9$ & $85.8$ \\ 
NGC1484 & $144.3$ & $49.4$ & $-13.9$ \\ 
PGC012625 & $68.0$ & $34.0$ & $100.0$ \\ 
WOMBAT I & $46.9$ & $39.0$ & $37.4$ \\ 
\enddata
\tablenotetext{a}{The photometric aperture ellipses for MCG-06-08-024 and PGC012625 are manually determined to prevent nearby stars from being included in the ellipses.}
\end{deluxetable*}

\startlongtable
\begin{deluxetable*}{lcccccccc}
\tablecaption{Derived physical quantities of sample Fornax galaxies \label{tab:derived}}
\tablehead{
\tableline\tableline
\colhead{Name} & \colhead{$\log{M_{\rm star}}$} & \colhead{$M_{\rm star}$ method\tablenotemark{a}} & \colhead{$\log{{\rm SFR}}$} & \colhead{SFR method\tablenotemark{b}} & \colhead{$\log{M_{\rm mol}}$\tablenotemark{c}} & \colhead{$\log{M_{\rm atom}}$} & \colhead{$M_{\rm atom}$ ref.\tablenotemark{d}}\\ 
\nocolhead{Name} & \colhead{$M_\odot$} & \nocolhead{} & \colhead{$M_\odot$~yr$^{-1}$} & \nocolhead{} & \colhead{$M_\odot$} & \colhead{$M_\odot$} & \nocolhead{}
}
\startdata
ESO302-14 & $7.94\pm0.16$ & $11$ & $-1.27\pm0.20$ & $11$ & $<7.54$ & $8.92\pm0.10$ & $3$ \\ 
ESO302-9 & $8.66\pm0.10$ & $11$ & $-1.12\pm0.20$ & $12$ & $<7.42$ & $9.04\pm0.08$ & $3$ \\ 
ESO357-25 & $8.70\pm0.10$ & $11$ & $-1.80\pm0.20$ & $11$ & $<7.66$ & $8.12\pm0.09$ & $3$ \\ 
ESO358-16 & $7.40\pm0.08$ & $4$ & $-2.36\pm0.02$ & $4$ & $<7.62$ & $7.98\pm0.18$ & $1$ \\ 
ESO358-20 & $8.92\pm0.13$ & $11$ & $-1.51\pm0.20$ & $11$ & $<7.64$ & $7.40\pm0.10$ & $3$ \\ 
ESO358-5 & $8.77\pm0.10$ & $11$ & $-1.38\pm0.20$ & $11$ & $<7.56$ & $8.53\pm0.19$ & $3$ \\ 
ESO358-51 & $9.15\pm0.10$ & $11$ & $-1.04\pm0.20$ & $11$ & $7.56\pm0.07$ & $8.13\pm0.12$ & $1$ \\ 
ESO358-G015 & $8.62\pm0.10$ & $11$ & $-1.91\pm0.20$ & $11$ & $<7.60$ & $8.09\pm0.14$ & $1$ \\ 
ESO358-G063 & $10.12\pm0.11$ & $11$ & $-0.29\pm0.20$ & $11$ & $8.78\pm0.02$ & $9.36\pm0.11$ & $1$ \\ 
ESO359-2 & $9.20\pm0.10$ & $12$ & $-1.95\pm0.20$ & $13$ & $6.84\pm0.12$ & $-$ & $-$ \\ 
ESO359-3 & $9.57\pm0.10$ & $11$ & $-1.25\pm0.20$ & $11$ & $<7.58$ & $8.66\pm0.11$ & $3$ \\ 
ESO359-5 & $7.98\pm0.06$ & $4$ & $-1.84\pm0.03$ & $4$ & $<7.43$ & $8.83\pm0.11$ & $3$ \\ 
FCC102 & $7.49\pm0.07$ & $4$ & $-2.34\pm0.02$ & $4$ & $<7.52$ & $7.83\pm0.12$ & $1$ \\ 
FCC117 & $6.34\pm0.29$ & $4$ & $-3.37\pm0.10$ & $4$ & $<7.68$ & $-$ & $-$ \\ 
FCC120 & $7.57\pm0.06$ & $4$ & $-2.36\pm0.02$ & $4$ & $<7.63$ & $7.74\pm0.15$ & $1$ \\ 
FCC177 & $9.84\pm0.15$ & $12$ & $-2.12\pm0.23$ & $13$ & $<7.50$ & $-$ & $-$ \\ 
FCC198 & $7.39\pm0.11$ & $1$ & $-3.35\pm0.30$ & $1$ & $<7.57$ & $-$ & $-$ \\ 
FCC206 & $8.40\pm0.07$ & $4$ & $-2.07\pm0.03$ & $4$ & $<7.59$ & $-$ & $-$ \\ 
FCC207 & $8.44\pm0.03$ & $1$ & $-2.80\pm0.24$ & $1$ & $<7.66$ & $<7.13$ & $1$ \\ 
FCC261 & $8.48\pm0.07$ & $4$ & $-1.99\pm0.03$ & $4$ & $<7.68$ & $<7.13$ & $1$ \\ 
FCC302 & $7.73\pm0.27$ & $11$ & $-1.48\pm0.20$ & $11$ & $<7.62$ & $9.17\pm0.11$ & $1$ \\ 
FCC306 & $7.89\pm0.05$ & $1$ & $-1.76\pm0.04$ & $1$ & $<7.79$ & $8.17\pm0.11$ & $1$ \\ 
FCC316 & $8.01\pm0.05$ & $1$ & $-2.35\pm0.12$ & $2$ & $<7.67$ & $-$ & $-$ \\ 
FCC32 & $8.67\pm0.03$ & $1$ & $-1.87\pm0.06$ & $2$ & $<7.47$ & $-$ & $-$ \\ 
FCC332 & $8.61\pm0.03$ & $5$ & $-2.16\pm0.04$ & $5$ & $6.82\pm0.11$ & $<7.13$ & $1$ \\ 
FCC44 & $7.76\pm0.06$ & $4$ & $-3.14\pm0.06$ & $4$ & $<7.56$ & $-$ & $-$ \\ 
IC1993 & $10.16\pm0.10$ & $11$ & $-0.66\pm0.20$ & $11$ & $8.81\pm0.01$ & $8.31\pm0.14$ & $3$ \\ 
IC335 & $9.99\pm0.10$ & $11$ & $-1.71\pm0.20$ & $11$ & $<7.61$ & $-$ & $-$ \\ 
MCG-06-08-024 & $8.77\pm0.01$ & $1$ & $-1.66\pm0.02$ & $1$ & $6.80\pm0.12$ & $-$ & $-$ \\ 
MCG-06-09-023 & $9.11\pm0.10$ & $12$ & $-1.94\pm0.20$ & $13$ & $7.16\pm0.09$ & $-$ & $-$ \\ 
NGC1310 & $9.65\pm0.10$ & $11$ & $-0.60\pm0.20$ & $11$ & $8.36\pm0.02$ & $8.70\pm0.01$ & $2$ \\ 
NGC1316 & $11.57\pm0.10$ & $11$ & $0.02\pm0.20$ & $11$ & $8.74\pm0.03$ & $7.85\pm0.04$ & $2$ \\ 
NGC1316C & $9.33\pm0.10$ & $11$ & $-1.49\pm0.20$ & $11$ & $7.83\pm0.04$ & $7.25\pm0.07$ & $2$ \\ 
NGC1317 & $10.59\pm0.10$ & $11$ & $-0.34\pm0.20$ & $11$ & $8.87\pm0.01$ & $8.46\pm0.00$ & $2$ \\ 
NGC1326 & $10.66\pm0.10$ & $11$ & $0.06\pm0.20$ & $11$ & $8.62\pm0.02$ & $9.38\pm0.01$ & $2$ \\ 
NGC1326A & $8.84\pm0.10$ & $11$ & $-0.98\pm0.20$ & $11$ & $<7.53$ & $9.17\pm0.03$ & $2$ \\ 
NGC1326B & $9.06\pm0.11$ & $11$ & $-0.56\pm0.20$ & $11$ & $<7.67$ & $9.67\pm0.01$ & $2$ \\ 
NGC1336 & $9.78\pm0.10$ & $11$ & $-1.63\pm0.20$ & $12$ & $<7.56$ & $-$ & $-$ \\ 
NGC1339 & $10.17\pm0.10$ & $11$ & $-1.61\pm0.20$ & $11$ & $<7.63$ & $-$ & $-$ \\ 
NGC1341 & $9.58\pm0.10$ & $11$ & $-0.32\pm0.20$ & $11$ & $8.48\pm0.01$ & $8.48\pm0.16$ & $3$ \\ 
NGC1350 & $10.77\pm0.10$ & $11$ & $-0.46\pm0.20$ & $11$ & $8.59\pm0.06$ & $9.13\pm0.16$ & $3$ \\ 
NGC1351 & $10.27\pm0.10$ & $11$ & $-1.41\pm0.20$ & $11$ & $<7.56$ & $-$ & $-$ \\ 
NGC1351A & $9.55\pm0.11$ & $11$ & $-1.27\pm0.20$ & $11$ & $7.94\pm0.08$ & $8.83\pm0.12$ & $1$ \\ 
NGC1365 & $10.84\pm0.10$ & $11$ & $1.24\pm0.20$ & $11$ & $10.30\pm0.00$ & $10.31\pm0.12$ & $1$ \\ 
NGC1374 & $10.42\pm0.10$ & $11$ & $-1.44\pm0.20$ & $11$ & $<7.53$ & $-$ & $-$ \\ 
NGC1375 & $9.84\pm0.10$ & $11$ & $-1.88\pm0.20$ & $11$ & $<7.74$ & $-$ & $-$ \\ 
NGC1379 & $10.41\pm0.10$ & $11$ & $-1.27\pm0.20$ & $11$ & $<7.49$ & $-$ & $-$ \\ 
NGC1380 & $10.92\pm0.10$ & $11$ & $-0.81\pm0.20$ & $11$ & $7.77\pm0.03$ & $<7.83$ & $1$ \\ 
NGC1381 & $10.28\pm0.10$ & $11$ & $-1.69\pm0.20$ & $11$ & $<7.56$ & $-$ & $-$ \\ 
NGC1386 & $10.25\pm0.10$ & $11$ & $0.26\pm0.20$ & $11$ & $8.58\pm0.02$ & $<7.83$ & $1$ \\ 
NGC1387 & $10.70\pm0.10$ & $11$ & $-0.65\pm0.20$ & $11$ & $8.53\pm0.01$ & $<7.61$ & $1$ \\ 
NGC1399 & $11.31\pm0.10$ & $11$ & $-0.27\pm0.20$ & $11$ & $<7.50$ & $-$ & $-$ \\ 
NGC1404 & $11.00\pm0.10$ & $11$ & $-0.74\pm0.20$ & $11$ & $<7.46$ & $-$ & $-$ \\ 
NGC1406 & $10.12\pm0.10$ & $11$ & $0.22\pm0.20$ & $12$ & $9.13\pm0.01$ & $9.35\pm0.09$ & $3$ \\ 
NGC1419 & $9.73\pm0.10$ & $11$ & $-2.27\pm0.21$ & $11$ & $<7.62$ & $-$ & $-$ \\ 
NGC1425 & $10.36\pm0.10$ & $11$ & $-0.13\pm0.20$ & $11$ & $8.70\pm0.03$ & $9.52\pm0.13$ & $3$ \\ 
NGC1427 & $10.46\pm0.10$ & $11$ & $-1.46\pm0.20$ & $11$ & $<7.57$ & $-$ & $-$ \\ 
NGC1427A & $9.14\pm0.17$ & $11$ & $-0.89\pm0.20$ & $11$ & $<7.50$ & $9.46\pm0.12$ & $1$ \\ 
NGC1436 & $10.16\pm0.10$ & $11$ & $-0.64\pm0.20$ & $11$ & $8.78\pm0.02$ & $7.91\pm0.20$ & $1$ \\ 
NGC1437A & $8.75\pm0.10$ & $11$ & $-1.07\pm0.20$ & $11$ & $<7.45$ & $8.91\pm0.10$ & $1$ \\ 
NGC1437B & $9.29\pm0.11$ & $11$ & $-0.64\pm0.20$ & $12$ & $7.97\pm0.05$ & $8.51\pm0.13$ & $1$ \\ 
NGC1484 & $9.38\pm0.02$ & $1$ & $-1.07\pm0.02$ & $1$ & $<7.51$ & $8.94\pm0.04$ & $3$ \\ 
PGC012625 & $7.45\pm0.27$ & $1$ & $-2.46\pm0.30$ & $1$ & $<7.67$ & $8.52\pm0.13$ & $3$ \\ 
WOMBAT I & $8.10\pm0.04$ & $4$ & $-1.74\pm0.02$ & $4$ & $<7.71$ & $8.33\pm0.13$ & $3$ \\ 
\enddata
\tablenotetext{a}{Double-digit and single-digit numbers indicate that $M_{\rm star}$ values from \cite{Leroy:2019cu} and those derived in this study, respectively. For the values from \cite{Leroy:2019cu}: 11, SSFRLIKE; 12, W4W1; 13, FIXEDUL. See Table~6 in \cite{Leroy:2019cu} for details. For the values derived in this study, we determine M/L with sSFR values estimated as follows: 1, as 11 with SFR derived using FUV+W4; 2, as 11 with SFR derived using NUV+W4; 3, as 12 with W4-to-W1 color; 4, 5, and 6, with equation 24 in \cite{Leroy:2019cu} using FUV-to-W1 color, NUV-to-W1 color, and W3-to-W1 color, respectively.} \tablenotetext{b}{As a note for $a$, double-digit and single-digit numbers indicate that SFR values from \cite{Leroy:2019cu} and those derived in this study, respectively. For the values from \cite{Leroy:2019cu}, SFR using: 11, FUV+W4; 12, NUV+W4; 13, W4. See Table~7 in \cite{Leroy:2019cu} for details. For the values derived in this study, we determine SFR using: 1, FUV+W4; 2, NUV+W4; 3, W4; 4, FUV; 5, NUV, and 6, W3.} \tablenotetext{c}{The constant CO-to-H$_2$ conversion factor of 4.36~$M_\odot$~(K~km~s$^{-1}$~pc$^2$) is adopted for all the galaxies. \tablenotetext{d}{Refereces for the $M_{\rm atom}$ data, 1: \cite{Loni:2021qe}; 2: \cite{Kleiner:2021bx}; 3: Hyper-LEDA}}
\end{deluxetable*}

\startlongtable
\begin{deluxetable*}{lccccc}
\tablecaption{Basic properties of sample Fornax galaxies \label{tab:basic}}
\tablehead{
\tableline\tableline
\colhead{Name} & \colhead{FCC\#} & \colhead{R.A.} & \colhead{Decl.} & \colhead{V (kLSR.)} & \colhead{Morphology}\\ 
\nocolhead{Name} & \nocolhead{FCC\#} & \colhead{deg.} & \colhead{deg.} & \colhead{km~s$^{-1}$} & \nocolhead{Morphology}
}
\startdata
ESO302-14 & $-$ & $57.920417$ & $-38.452222$ & $854$ & IB(s)m \\ 
ESO302-9 & $322$ & $56.892167$ & $-38.576528$ & $970$ & SB(s)dm? \\ 
ESO357-25 & $26$ & $50.90525$ & $-35.778389$ & $1806$ & SAB0$^{\wedge}$0? \\ 
ESO358-16 & $115$ & $53.288327$ & $-35.718527$ & $1700$ & Sc \\ 
ESO358-20 & $139$ & $53.738958$ & $-32.639833$ & $1752$ & IB(s)m? pec \\ 
ESO358-5 & $53$ & $51.819208$ & $-33.486361$ & $1611$ & SAB(s)m pec? \\ 
ESO358-51 & $263$ & $55.385833$ & $-34.888333$ & $1707$ & S0/a? \\ 
ESO358-G015 & $113$ & $53.278542$ & $-34.808111$ & $1371$ & SBm? pec \\ 
ESO358-G063 & $312$ & $56.579208$ & $-34.943556$ & $1911$ & I0? \\ 
ESO359-2 & $335$ & $57.653042$ & $-35.909333$ & $1412$ & SB0$^{\wedge}$-? \\ 
ESO359-3 & $338$ & $58.003833$ & $-33.467639$ & $1557$ & Sab? edge-on \\ 
ESO359-5 & $-$ & $58.514244$ & $-36.063653$ & $1370$ & dwarf \\ 
FCC102 & $102$ & $53.04474$ & $-36.220809$ & $1706$ & Irr \\ 
FCC117 & $117$ & $53.31106$ & $-37.819638$ & $1500$ & $-$ \\ 
FCC120 & $120$ & $53.392593$ & $-36.605914$ & $887$ & S \\ 
FCC177 & $177$ & $54.197875$ & $-34.739611$ & $1544$ & S0$^{\wedge}$0? edge-on \\ 
FCC198 & $198$ & $54.427823$ & $-37.208339$ & $1500$ & $-$ \\ 
FCC206 & $206$ & $54.556202$ & $-37.29018$ & $1385$ & dwarf \\ 
FCC207 & $207$ & $54.580292$ & $-35.129083$ & $1403$ & S0 \\ 
FCC261 & $261$ & $55.339708$ & $-33.769222$ & $1475$ & $-$ \\ 
FCC302 & $302$ & $56.300593$ & $-35.570906$ & $786$ & IB(s)m? edge-on \\ 
FCC306 & $306$ & $56.439158$ & $-36.34653$ & $868$ & S \\ 
FCC316 & $316$ & $56.756333$ & $-36.437472$ & $1529$ & dwarf \\ 
FCC32 & $32$ & $51.2185$ & $-35.435444$ & $1301$ & S0 \\ 
FCC332 & $332$ & $57.45425$ & $-35.945583$ & $1308$ & S0 \\ 
FCC44 & $44$ & $51.531017$ & $-35.127491$ & $1232$ & $-$ \\ 
IC1993 & $315$ & $56.770042$ & $-33.709861$ & $1062$ & (R')SAB(rs)b \\ 
IC335 & $153$ & $53.879333$ & $-34.447056$ & $1602$ & S0 edge-on \\ 
MCG-06-08-024 & $90$ & $52.784417$ & $-36.290139$ & $1796$ & SA0$^{\wedge}$-? \\ 
MCG-06-09-023 & $282$ & $55.688833$ & $-33.9205$ & $1208$ & S0 \\ 
NGC1310 & $-$ & $50.264292$ & $-37.101694$ & $1788$ & SA(s)c? \\ 
NGC1316 & $21$ & $50.673825$ & $-37.208227$ & $1743$ & SAB0$^{\wedge}$0(s) pec \\ 
NGC1316C & $33$ & $51.243583$ & $-37.009472$ & $1783$ & (R')SA0$^{\wedge}$0? \\ 
NGC1317 & $22$ & $50.684527$ & $-37.103688$ & $1924$ & SAB(r)a \\ 
NGC1326 & $29$ & $50.985$ & $-36.464667$ & $1343$ & (R)SB0$^{\wedge}$+(r) \\ 
NGC1326A & $37$ & $51.285405$ & $-36.363949$ & $1814$ & SB(s)m? \\ 
NGC1326B & $39$ & $51.334782$ & $-36.384954$ & $982$ & SB(s)m? edge-on \\ 
NGC1336 & $47$ & $51.634125$ & $-35.713556$ & $1401$ & SA0$^{\wedge}$- \\ 
NGC1339 & $63$ & $52.027417$ & $-32.286111$ & $1375$ & cD pec? \\ 
NGC1341 & $62$ & $51.993417$ & $-37.15$ & $1859$ & SAB(s)ab \\ 
NGC1350 & $88$ & $52.783833$ & $-33.628639$ & $1888$ & (R')SB(r)ab \\ 
NGC1351 & $83$ & $52.64575$ & $-34.853944$ & $1497$ & SA0$^{\wedge}$- pec? \\ 
NGC1351A & $67$ & $52.203$ & $-35.178139$ & $1336$ & SB(rs)bc? edge-on \\ 
NGC1365 & $121$ & $53.401548$ & $-36.140402$ & $1619$ & SB(s)b \\ 
NGC1374 & $147$ & $53.819125$ & $-35.22625$ & $1277$ & E \\ 
NGC1375 & $148$ & $53.820083$ & $-35.265667$ & $723$ & S0 \\ 
NGC1379 & $161$ & $54.016458$ & $-35.441194$ & $1307$ & E \\ 
NGC1380 & $167$ & $54.114968$ & $-34.976225$ & $1860$ & SA0 \\ 
NGC1381 & $170$ & $54.132$ & $-35.295194$ & $1707$ & SA0? edge-on \\ 
NGC1386 & $179$ & $54.192425$ & $-35.999408$ & $851$ & SB0$^{\wedge}$+(s) \\ 
NGC1387 & $184$ & $54.23775$ & $-35.506639$ & $1285$ & SAB0$^{\wedge}$-(s) \\ 
NGC1399 & $213$ & $54.620941$ & $-35.450657$ & $1408$ & E1 pec \\ 
NGC1404 & $219$ & $54.716333$ & $-35.594389$ & $1930$ & E2 \\ 
NGC1406 & $-$ & $54.847083$ & $-31.321417$ & $1058$ & SB(s)bc? edge-on \\ 
NGC1419 & $249$ & $55.175458$ & $-37.510833$ & $1576$ & dE3 \\ 
NGC1425 & $-$ & $55.547792$ & $-29.893333$ & $1493$ & SA(s)b \\ 
NGC1427 & $276$ & $55.580933$ & $-35.392563$ & $1371$ & E4 \\ 
NGC1427A & $235$ & $55.03875$ & $-35.624444$ & $2011$ & dE2 \\ 
NGC1436 & $290$ & $55.9045$ & $-35.853028$ & $1370$ & ScII \\ 
NGC1437A & $285$ & $55.75914$ & $-36.273374$ & $868$ & SdIII? \\ 
NGC1437B & $308$ & $56.478542$ & $-36.356972$ & $1480$ & Sd(onedge) \\ 
NGC1484 & $-$ & $58.583875$ & $-36.968889$ & $1022$ & SB(s)b? \\ 
PGC012625 & $-$ & $50.52$ & $-37.5875$ & $1609$ & IB(s)m? \\ 
WOMBAT I & $-$ & $55.245185$ & $-38.855266$ & $812$ & S \\ 
\enddata
\end{deluxetable*}

\acknowledgments
We thank the anonymous referee for her/his constructive comments, which improved the manuscript.
KMM thank Hidenobu Yajima for the discussion on the dominant environmental effect on Fornax cluster galaxies. 
This work was supported by JSPS KAKENHI Grant Numbers 16H02158, 19J40004, 19H01931, 19H05076, 20H05861, 21H01128 and 21H04496.
This work has also been supported in part by the Sumitomo Foundation Fiscal 2018 Grant for Basic Science Research Projects (180923), and the Collaboration Funding of the Institute of Statistical Mathematics ``New Development of the Studies on Galaxy Evolution with a Method of Data Science''.
FMM acknowledges funding from the European Research Council (ERC) under the European Union's Horizon 2020 research and innovation programme (grant agreement no. 679627; project name FORNAX).
BQF was supported by the Australian Research Council Centre of Excellence for All Sky Astrophysics in 3 Dimensions (ASTRO 3D), through project number CE170100013.
BL acknowledges the support from the Korea Astronomy and Space Science Institute grant funded by the Korea government (MSIT) (Project No. 2022-1- 840-05).
This paper makes use of the following ALMA data: ADS/JAO.ALMA\#2017.1.00129.S. ALMA is a partnership of ESO (representing its member states), NSF (USA) and NINS (Japan), together with NRC (Canada), MOST and ASIAA (Taiwan), and KASI (Republic of Korea), in cooperation with the Republic of Chile. The Joint ALMA Observatory is operated by ESO, AUI/NRAO and NAOJ.
Data analysis was carried out on the Multi-wavelength Data Analysis System operated by the Astronomy Data Center (ADC), National Astronomical Observatory of Japan.
This research has made use of the NASA/IPAC Extragalactic Database (NED), which is operated by the Jet Propulsion Laboratory, California Institute of Technology, under contract with the National Aeronautics and Space Administration.

%

\vspace{5mm}
\facilities{GALEX, WISE, ALMA, ATCA, MeerKAT, Parkes}


\software{astropy \citep{Astropy-Collaboration:2013uf},
	APLpy \citep{Robitaille:2012lz}
	}



\appendix


\section{Appendix information\label{sec:supplementaryplots}}

\begin{figure*}[]
\begin{center}
\includegraphics[width=\textwidth, bb=0 0 1867 939]{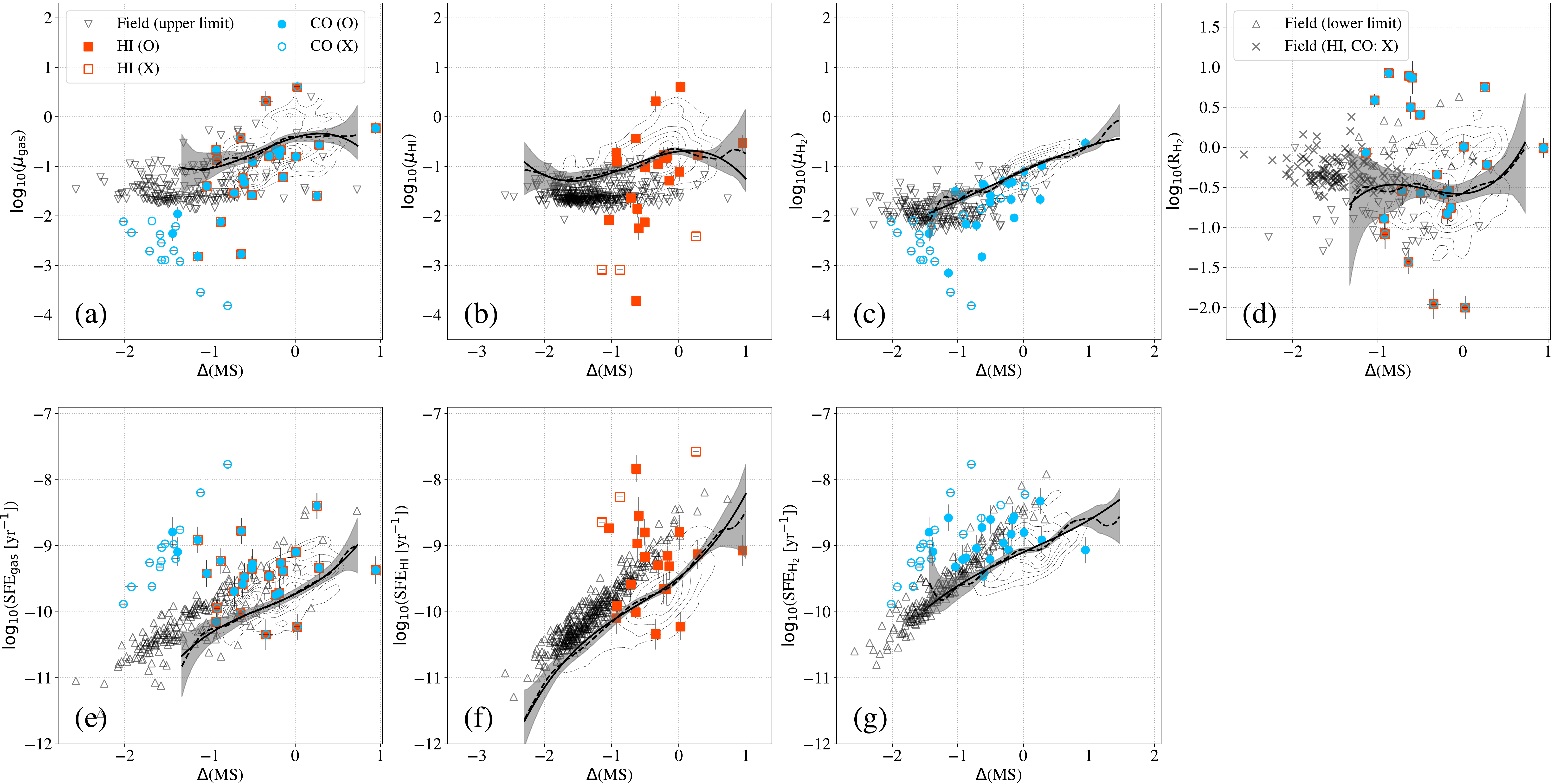}
\end{center}
\caption{
Comparison between the Fornax galaxies and field galaxies with $M_{\rm star}>10^9$~M$_\odot$: relationships between the offsets from the main sequence of star-forming galaxies with gas fractions, $R_{\rm H_2}$, and SFEs.
Symbols and lines are same as in Figure~\ref{fig:comp_field}.
Our Fornax sample has lower gas fractions and higher $R_{\rm H_2}$ and SFEs than the field galaxies at $\Delta({\rm MS})\sim-1$.
}
\label{fig:comp_field4}
\end{figure*}

\begin{figure*}[]
\begin{center}
\includegraphics[width=.8\textwidth, bb=0 0 968 495]{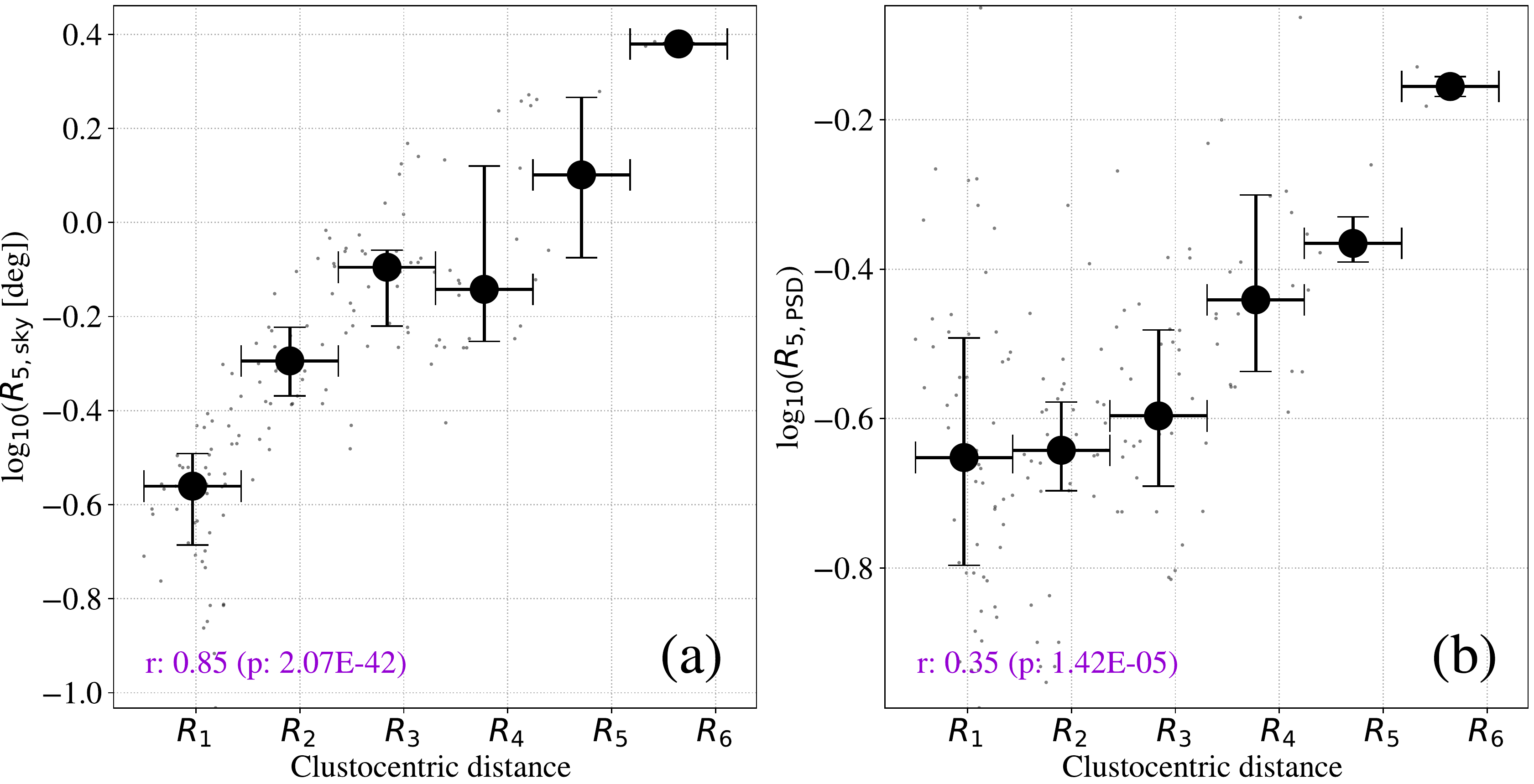}
\includegraphics[width=.8\textwidth, bb=0 0 968 495]{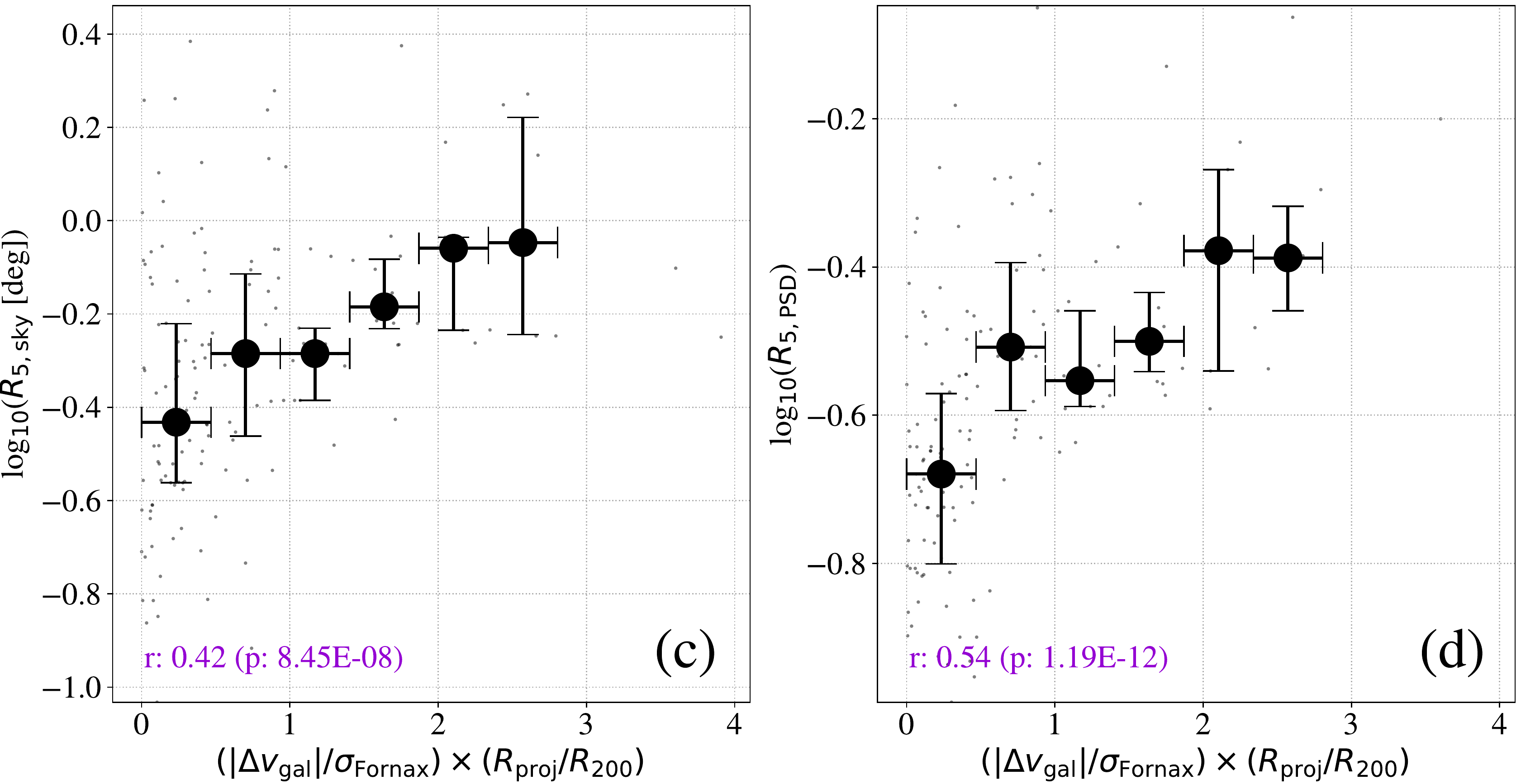}
\end{center}
\caption{
The relationship between the clustocentric distance with (a) $R_{\rm 5, sky}$ and (b) $R_{\rm 5, PSD}$, and between the accretion phase with (c) $R_{\rm 5, sky}$ and (d) $R_{\rm 5, PSD}$.
The median values for each bin of the clustocentric distance are indicated with black filled circle.
The upper and lower caps indicate the 3rd and 1st quartiles of $R_{\rm 5, sky}$ or $R_{\rm 5, PSD}$ in each clustocentric-distance or accretion phase bin, respectively.
The Spearman's rank-order correlation coefficient $r$ and the $p$-value are shown on the lower left corner of each panel in purple.
}
\label{fig:dist_r5}
\end{figure*}

\begin{figure*}[]
\begin{center}
\includegraphics[width=150mm, bb=0 0 1422 1390]{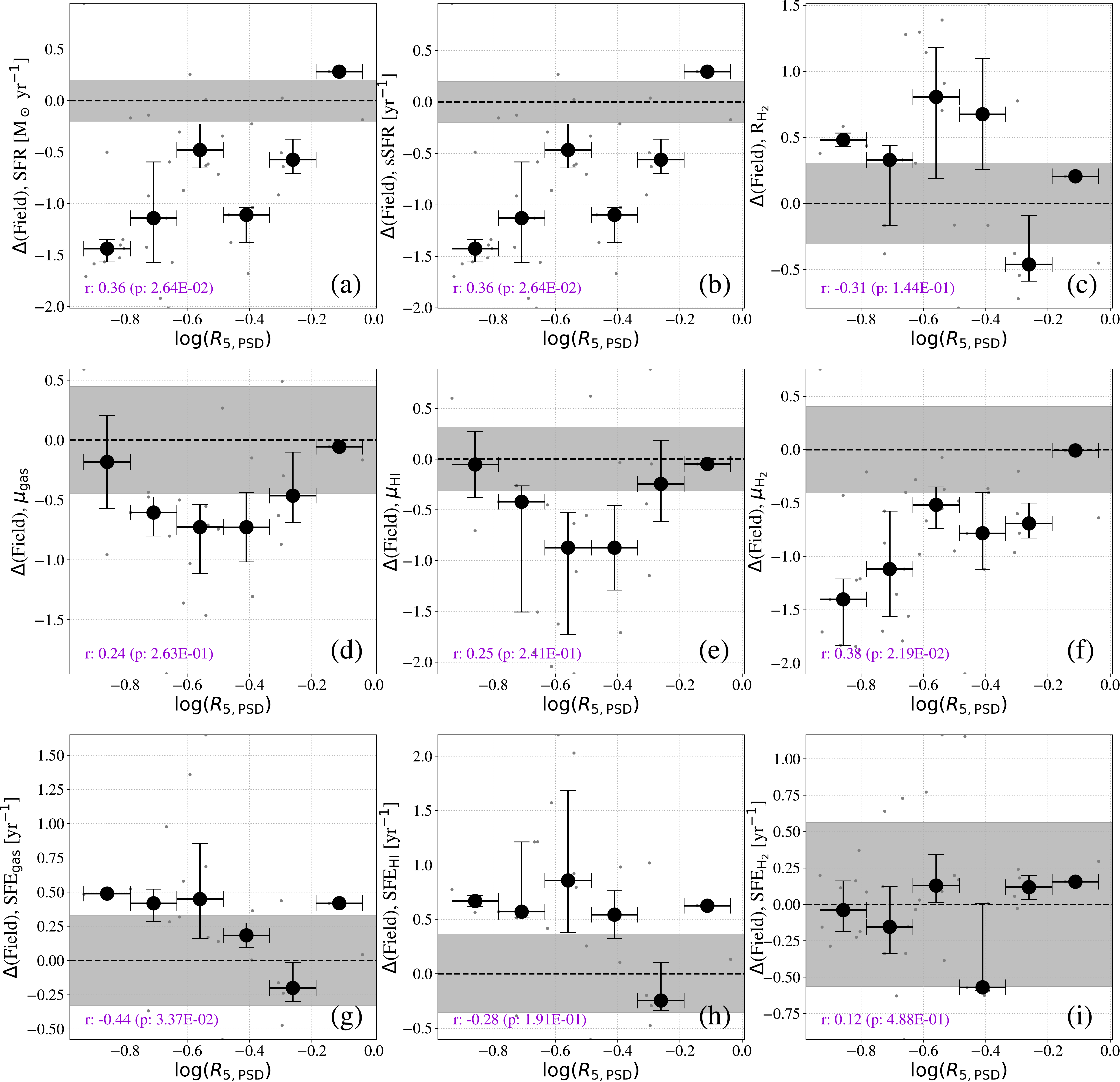}
\end{center}
\caption{
The relationships between $R_{\rm 5, PSD}$ with the key quantities [$\Delta({\rm Field})$ values].
Symbols are same as in Figure~\ref{fig:radial_wul}.
$R_{\rm 5,PSD}$ positively correlates to $\mu_{\rm H_2}$ and negatively correlates to SFE$_{\rm gas}$.
}
\label{fig:numberdensity}
\end{figure*}

\begin{figure*}[]
\begin{center}
\includegraphics[width=1.\textwidth, bb=0 0 1586 1829]{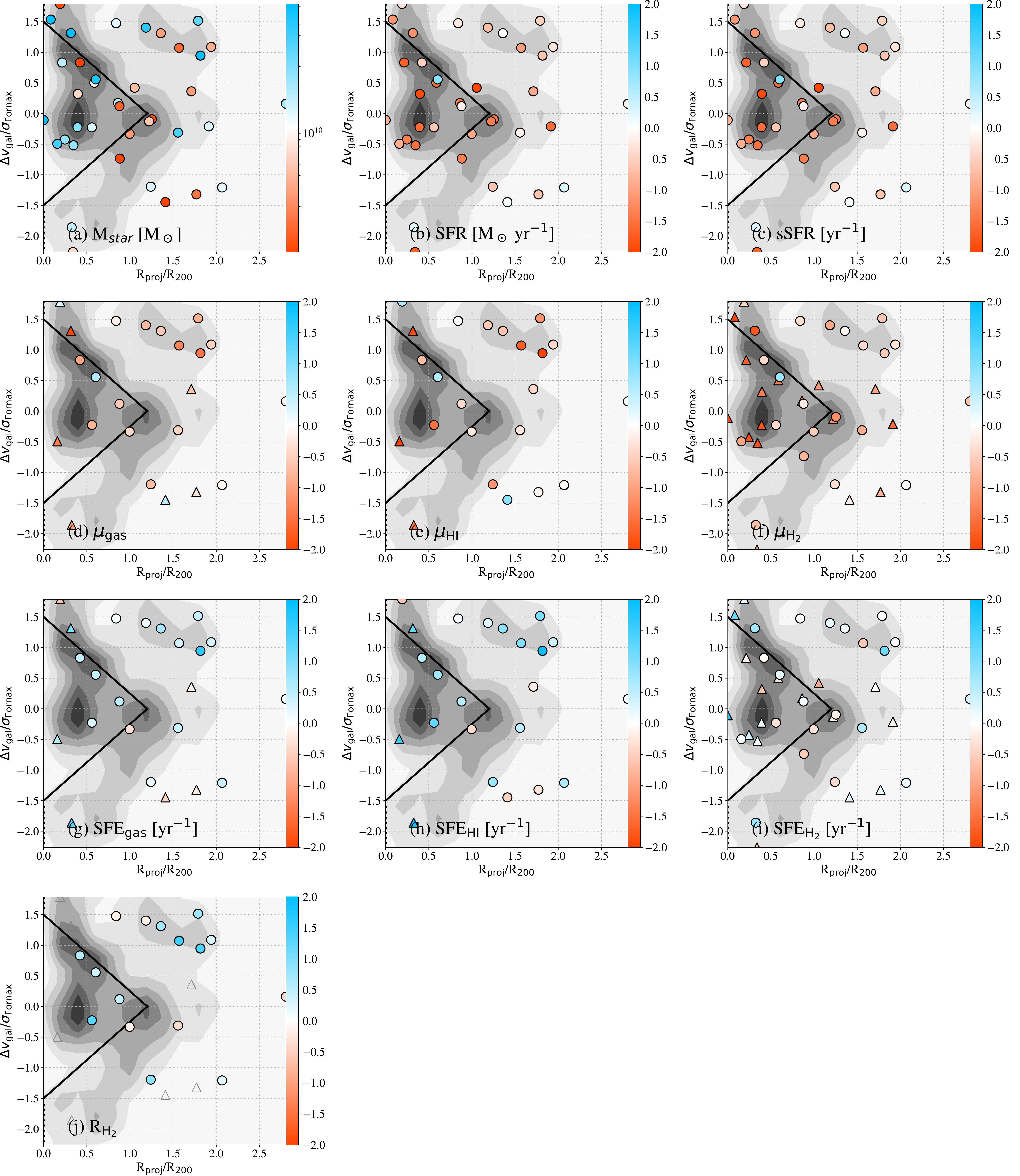}
\end{center}
\caption{
Phase-space diagram of our sample galaxies in the Fornax cluster, color-coded for the key quantities defined in section~\ref{sec:coldgassfproperties}.
}
\label{fig:psd_all}
\end{figure*}

In this sections, supplementary plots are presented:
Figure\ref{fig:comp_field4} shows the dependence of the gas fractions, $R_{\rm H_2}$, and SFE on the distance from the main-sequence of star-forming galaxies;
Figure~\ref{fig:dist_r5} shows the relationships between the clustocentric distance with $R_{\rm 5,sky}$ and $R_{\rm 5,PSD}$, and the relationships between the accretion phase with $R_{\rm 5,sky}$ and $R_{\rm 5,PSD}$;
Figure~\ref{fig:numberdensity} shows the relationship between the $R_{\rm 5,PSD}$ with the key quantities;
Figure~\ref{fig:hist_c2_inclp} compares the key quantities of the VIR and the RIF galaxies;
Figure~\ref{fig:psd_all} shows the PSD color-coded for the key quantities explored in this study.


\bibliographystyle{aasjournal}
\bibliography{myref_moro}



\end{document}